\documentclass[a4paper,11pt]{article}
\usepackage{amssymb}
\usepackage{graphicx}
\usepackage{amsmath}

\DeclareFontFamily{U}{rsf}{}
\DeclareFontShape{U}{rsf}{m}{n}{
  <5> <6> rsfs5 <7> <8> <9> rsfs7 <10-> rsfs10}{}
\DeclareMathAlphabet\Scr{U}{rsf}{m}{n} \makeatletter
\@addtoreset{equation}{section} \makeatother

\def\be{\begin{equation}}
\def\ee{\end{equation}}
\def\ba{\begin{array}}
\def\ea{\end{array}}

\newcommand{\bea}{\begin{eqnarray}}
\newcommand{\eea}{\end{eqnarray}}

\begin{document}

\begin{titlepage}

\begin{flushright}
CERN-PH-TH/2012-154\\
UCB-PTH-12/09\\
\end{flushright}

\vskip 2.5 cm
\begin{center}  {\Huge{\bf      Jordan Pairs, $E_{6}$ and U-Duality\\ \vskip 0.2 cm in Five Dimensions
    }}

\vskip 1.5 cm

{\Large{\bf Sergio Ferrara$^{1,2}$}, {\bf Alessio Marrani$^1$}, and {\bf Bruno Zumino$^{3,4}$}}

\vskip 1.0 cm

$^1${\sl Physics Department, Theory Unit, CERN,  CH 1211, Geneva 23, Switzerland\\
\texttt{sergio.ferrara@cern.ch}\\
\texttt{alessio.marrani@cern.ch}}\\

\vskip 0.5 cm

$^2${\sl INFN - Laboratori Nazionali di Frascati, Via Enrico Fermi
40, 00044 Frascati, Italy}\\

\vskip 0.5 cm

$^3${ \sl Department of Physics, University of California, Berkeley, CA 94720-8162, USA}

\vskip 0.5 cm

$^4${ \sl Lawrence Berkeley National Laboratory, Theory Group, Berkeley, CA 94720-8162, USA\\
\texttt{zumino@thsrv.lbl.gov}}

       \vskip 0.5 cm

\end{center}

 \vskip 3.5 cm

\begin{abstract}
By exploiting the \textit{Jordan pair} structure of $U$-duality Lie algebras
in $D=3$ and the relation to the \textit{super-Ehlers} symmetry in $D=5$, we
elucidate the massless multiplet structure of the spectrum of a broad class
of $D=5$ supergravity theories. Both \textit{simple} and \textit{semi-simple}%
, Euclidean rank-$3$ Jordan algebras are considered. Theories sharing the
same bosonic sector but with different supersymmetrizations are also
analyzed.
\end{abstract}

\vspace{24pt} \end{titlepage}


\newpage \tableofcontents 

\section{\label{Intro}Introduction}

In recent past, exceptional Lie groups with their various real forms have
been shown to play a major role in order to exploit several dynamical
properties of supergravity theories in different dimensions.

Their relevance was highlighted by the seminal work of Cremmer and Julia
\cite{CJ-1}, in which the existence of an exceptional electric-magnetic
duality symmetry in $\mathcal{N}=8$, $D=4$ supergravity was established,
based on the maximally non-compact (\textit{split}) form $E_{7(7)}$ of the
exceptional group $E_{7}$.

Further advances were pionereed by G\"{u}naydin, Sierra and Townsend \cite%
{GST}, which established the close relation between exceptional Lie groups
occurring in $D=5$ supergravity theories and Jordan algebras. In particular,
different real forms of $E_{6}$ (namey, the split form $E_{6(6)}$ for
maximal $\mathcal{N}=8$ supergravity and the minimally non-compact form $%
E_{6(-26)}$ for exceptional minimal Maxwell-Einstein $\mathcal{N}=2$
supergravity) made their appearance as \textit{reduced structure} symmetries
of the corresponding rank-$3$ Euclidean Jordan algebras. These latter are
characterized by a cubic norm, which is directly related to \textit{real
special geometry} of the $D=5$ vector multiplets' scalar manifold (see also
\textit{e.g.} \cite{dWVVP}, and Refs. therein) and to the Bekenstein-Hawking
extremal $D=5$ black hole entropy, when the Jordan algebra elements are
identified with the black hole charges (see \textit{e.g.} \cite%
{FG-D=5,CFMZ1-D=5}, and Refs. therein).

The $U$-duality\footnote{%
Here $U$-duality is referred to as the \textquotedblleft
continuous\textquotedblright\ symmetries of \cite{CJ-1}. Their discrete
versions are the $U$-duality non-perturbative string theory symmetries
introduced by Hull and Townsend \cite{HT-1}.} symmetry of maximal
supergravity in $D$ space-time dimensions is given by the so-called \textit{%
Cremmer-Julia sequence} $E_{11-D\left( 11-D\right) }$ \cite%
{Cremmer-Lects,Julia-Lects}, which for $D>5$ yields classical groups. This sequence occurs in (at least)
two sets of group embeddings for maximal supergravity in $3<D\leqslant 11$:
\begin{eqnarray}
E_{11-D\left( 11-D\right) } &\supset &E_{10-D\left( 10-D\right) }\times
SO(1,1);  \label{emb-N=16} \\
E_{8(8)} &\supset &E_{11-D\left( 11-D\right) }\times SL(D-2,\mathbb{R}).
\label{embbb}
\end{eqnarray}

(\ref{emb-N=16}) describes the (maximal and symmetric) embedding of the $U$%
-duality group in $D+1$ dimensions into the corresponding $U$-duality group
in $D$ dimensions; the commuting $SO(1,1)$ factor pertains to the
compactification radius in the Kaluza-Klein $\left( D+1\right) \rightarrow D$
dimensional reduction.

On the other hand, (\ref{embbb}) describes the (generally not maximal nor
symmetric) embedding of the $U$-duality group in $D$ dimensions into the $D=3
$ $U$-duality group $E_{8(8)}$; the commuting $SL(D-2,\mathbb{R})$ factor
can physically be interpreted as the \textit{Ehlers group} in $D$
dimensions. Some time ago \cite{GS}, it was shown that $SL(D-2,\mathbb{R})$
is a symmetry of $D$-dimensional Einstein gravity, provided that the theory
is formulated in the light-cone gauge. The embedding (\ref{embbb}) can be
obtained by suitable manipulations of the extended Dynkin diagram (also
called Dynkin diagram of the affine untwisted Kac-Moody algebra) of $E_{8(8)}
$, as discussed \textit{e.g.} in \cite{Julia-Lects} and in \cite{Keu-1}.

The present investigation is devoted to the study of the embedding (\ref%
{embbb}) in $D=5$:%
\begin{equation}
E_{8(8)}\supset E_{6(6)}\times SL(3,\mathbb{R}),  \label{pre-1}
\end{equation}%
which, as recently pointed out in \cite{Truini-1}, is related to the
so-called \textit{Jordan pairs}. As also recently discussed in \cite%
{super-Ehlers-1}, for non-maximal supersymmetry, (\ref{pre-1}) is
generalized as%
\begin{equation}
G_{N}^{3}\supset G_{N}^{5}\times SL(3,\mathbb{R}),  \label{pre-2}
\end{equation}%
where $G_{N}^{3}$ and $G_{N}^{5}$ respectively are the $D=3$ and $D=5$ $U$%
-duality groups of the theory with $2N$ supersymmetries. For instance, in
the minimal ($N=4$, corresponding to $\mathcal{N}=2$ supercharges in $D=5$)
exceptional supergravity \cite{GST}, (\ref{pre-2}) specifies to a different
non-compact, real form of (\ref{pre-1}), namely:%
\begin{equation}
E_{8(-24)}\supset E_{6(-26)}\times SL(3,\mathbb{R}).  \label{pre-3}
\end{equation}%
As mentioned above and also recently analyzed in \cite{super-Ehlers-1}, the $%
SL(3,\mathbb{R})$ appearing in (\ref{pre-1}) and (\ref{pre-2}) can
physically be interpreted as the \textit{Ehlers group} in $D=5$.

Interestingly enough, the supermultiplet structure of the underlying theory
enjoys a natural explanation in terms of the \textit{Jordan pair} embedding (%
\ref{pre-2}), if one considers the corresponding embedding of the maximal
compact subgroups, which may largely differ depending on the relevant
non-compact, real form. For example, in the maximal and minimal exceptional
cases, the maximal compact level of (\ref{pre-1}) and of (\ref{pre-3})
respectively yields%
\begin{eqnarray}
SO(16) &\supset &Usp(8)\times SU(2)_{J};  \label{pre-1-mcs} \\
E_{7(-133)}\times SU(2) &\supset &F_{4(-52)}\times SU(2)_{J},
\label{pre-3-mcs}
\end{eqnarray}%
where the $SU(2)_{J}$ on the r.h.s. (maximal compact subgroup of the Ehlers $%
SL(3,\mathbb{R})$) is the massless spin (helicity) group in $D=5$.\medskip\

The plan of the paper is as follows.

In Sec. \ref{Simple-General}, we introduce the $q$-parametrized sequence of
exceptional Lie algebras, and its decomposition in terms of \textit{Jordan
pairs}, by starting from the treatment of the compact case recently given in
\cite{Truini-1}, and pointing out the relation to \textit{simple}, Euclidean
rank-$3$ Jordan algebras. Suitable non-compact, real forms, relevant for
application to the \textit{super-Ehlers} symmetry \cite{super-Ehlers-1} in $%
D=5$ supergravity, are considered.

In particular, Sec. \ref{q=8-Sec} deals with \textit{maximal} supergravity,
related to $\mathfrak{J}_{3}^{\mathbb{O}_{s}}$ (Subsec. \ref{J3-Os-Subsec}),
and with \textit{minimal} exceptional magical supergravity, related to $%
\mathfrak{J}_{3}^{\mathbb{O}}$ \cite{GST}. In the latter case, considered in
Subsec. \ref{J3-O-Subsec}, the existence of a $D$-independent hypersector is
crucial to recover the massless multiplet structure \textit{via}
representation theory.

The interesting case of a pair of supergravity theory sharing the same
bosonic sector, but with a different fermionic sector and thus with
different supersymmetry properties, is considered in Sec. \ref{q=4-J3-H-Sec}%
; namely, $\mathcal{N}=6$ \textit{\textquotedblleft pure"} theory \textit{%
versus} minimal ($\mathcal{N}=2$) matter-coupled Maxwell-Einstein
quaternionic supergravity, both related to $\mathfrak{J}_{3}^{\mathbb{H}}$
\cite{GST}.

Sec. \ref{Simple-List} lists the \textit{Jordan pair} embeddings for all
\textit{simple}, Euclidean rank-$3$ Jordan algebras, also including the
non-generic case of the $D=5$ uplift of the so-called $T^{3}$ model ($q=-2/3$%
).

\textit{Semi-simple}, Euclidean rank-$3$ Jordan algebras are then analyzed
in Sec. \ref{Semi-Simple-General}, focussing on the two infinite classes
relevant for minimal and half-maximal supergravity in $D=5$ (the former
class includes the $D=5$ uplift of the so-called $STU$ model, which is
separately analyzed in Subsec. \ref{STU}).

Within the \textit{semi-simple} framework, a pair of theories with the same
bosonic sector but different supersymmetry features (namely minimal $%
\mathfrak{J}_{3}^{2,6}$-related theory \textit{versus} half-maximal $%
\mathfrak{J}_{3}^{6,2}$-related theory, both matter coupled) is then
analyzed in detail in Sec. \ref{Semi-Simple-Twin}.

Final observations and remarks are given in the concluding Sec. \ref%
{Conclusion}.

\section{\label{Simple-General}\textit{Jordan Pairs} : the \textit{Simple}
Case}

We start by briefly recalling that a \textit{Jordan algebra} $\mathfrak{J}$
\cite{Jordan,JVNW} is a vector space defined over a ground field $\mathbb{F}$
equipped with a bilinear product $\circ $ satisfying
\begin{eqnarray}
X\circ Y &=&Y\circ X; \\
X^{2}\circ \left( X\circ Y\right) &=&X\circ \left( X^{2}\circ Y\right)
,~~\forall X,Y\in \mathfrak{J}.  \notag
\end{eqnarray}%
The Jordan algebras relevant for the present investigation are \textit{rank-}%
$3$ Jordan algebras $\mathfrak{J}_{3}$ over $\mathbb{F}=\mathbb{R}$, which
come equipped with a cubic norm%
\begin{eqnarray}
N &:&\mathfrak{J}\rightarrow \mathbb{R},  \notag \\
N\left( \lambda X\right) &=&\lambda ^{3}N\left( X\right) ,~\forall \lambda
\in \mathbb{R},X\in \mathfrak{J}.
\end{eqnarray}

As an example, we anticipate that for both the rank-$3$ Jordan algebras $%
\mathfrak{J}_{3}^{\mathbb{O}}$ and $\mathfrak{J}_{3}^{\mathbb{O}_{s}}$
treated in Sec. \ref{q=8-Sec}, the relevant vector space is the
representation space $\mathbf{27}$ pertaining to the fundamental irrep. of $%
E_{6(-26)}$ resp. $E_{6(6)}$, and the cubic norm $N$ is realized in terms of
he completely symmetric invariant rank-$3$ tensor $d_{IJK}$ in the $\mathbf{%
27}$ ($I,J,K=1,...,27$):%
\begin{eqnarray}
\left( \mathbf{27}\times \mathbf{27}\times \mathbf{27}\right) _{s} &\ni
&\exists !\mathbf{1}\equiv d_{IJK}; \\
N\left( X\right) &\equiv &d_{IJK}X^{I}X^{J}X^{K}.
\end{eqnarray}

There is a general prescription for constructing rank-$3$ Jordan algebras,
due to Freudenthal, Springer and Tits \cite{Springer,McCrimmon-pre,McCrimmon}%
, for which all the properties of the Jordan algebra are essentially
determined by the cubic norm $N$ (for a sketch of the construction see also
\cite{Duff-Freud}).\medskip

The $q$-parametrized sequence of $U$-duality (non--compact, real) Lie
algebras $\mathfrak{L}^{q}$ in $D=3$ (Lorentzian) space-time dimensions can
be characterized as follows:%
\begin{equation}
\mathfrak{L}^{q}=\mathfrak{sl}\left( 3,\mathbb{R}\right) \oplus \mathfrak{str%
}_{0}\left( \mathfrak{J}_{3}^{q}\right) \oplus \mathbf{3}\times \mathfrak{J}%
_{3}^{q}\oplus \mathbf{3}^{\prime }\times \mathfrak{J}_{3}^{q\prime },
\label{1}
\end{equation}%
where $\mathfrak{J}_{3}^{q}$ is a rank-$3$ Euclidean \textit{simple} Jordan
algebra; for the cases $q=8,4,2,1$, the parameter $q$ is defined as $q\equiv
$dim$_{\mathbb{R}}\mathbb{A}$, with $\mathbb{A}$ denoting one of the four
normed division algebras $\mathbb{A}=\mathbb{O},\mathbb{H},\mathbb{C},%
\mathbb{R}$ (from the famous \textquotedblleft 1,2,4,8" Hurwitz's Theorem;
see \textit{e.g.} \cite{McCrimmon}), respectively. $\mathfrak{J}_{3}^{q}$
fits into a $\left( 3q+3\right) $-dimensional irrep. of the \textit{reduced
structure} Lie algebra $\mathfrak{str}_{0}\left( \mathfrak{J}_{3}^{q}\right)
$, which is nothing but the $D=5$ $U$-duality Lie algebra. Also,
\begin{equation}
\mathfrak{L}^{q}=\mathfrak{qconf}\left( \mathfrak{J}_{3}^{q}\right)
\end{equation}%
is the \textit{quasi-conformal} algebra of $\mathfrak{J}_{3}^{q}$ \cite%
{GKN,GP-04}, \textit{i.e.} the $U$-duality Lie algebra in $D=3$ (see \textit{%
e.g.} \cite{G-Lects,Small-Orbits-Phys} for an introduction to the
application of Jordan algebras and their symmetries in supergravity\footnote{%
In these theories, the $U$-duality Lie algebra in $D=4$ is given by $%
\mathfrak{conf}\left( \mathfrak{J}_{3}\right) =\mathfrak{aut}\left(
\mathfrak{F}\left( \mathfrak{J}_{3}\right) \right) $, where $\mathfrak{F}%
\left( \mathfrak{J}_{3}\right) $ denotes the \textit{Freudenthal triple
system} constructed over $\mathfrak{J}_{3}$.}, and lists of Refs.).

(\ref{1}) is a suitable non-compact, real version of the decomposition of
the compact Lie algebras (subscript \textquotedblleft $c$" stands for
\textit{compact}) \cite{Truini-1}%
\begin{equation}
\mathfrak{L}_{c}^{q}=\mathfrak{su}\left( 3\right) \oplus \mathfrak{str}%
_{0,c}\left( \mathfrak{J}_{3}^{q}\right) \oplus \mathbf{3}\times \mathfrak{J}%
_{3}^{q}\oplus \overline{\mathbf{3}}\times \overline{\mathfrak{J}_{3}^{q}},
\label{2}
\end{equation}%
with the various cases given by the following Table 1:
\begin{table}[h]
\begin{center}
\begin{tabular}{|c||c|c|c|c|c|c|c|}
\hline
$q$ & $8$ & $4$ & $2$ & $1$ & $0$ & $-2/3$ & $-1$ \\ \hline
\rule[-1mm]{0mm}{6mm} $\mathfrak{L}_{c}^{q}$ & $\mathfrak{e}_{8\left(
-248\right) }$ & $\mathfrak{e}_{7\left( -133\right) }$ & $\mathfrak{e}%
_{6\left( -78\right) }$ & $\mathfrak{f}_{4\left( -52\right) }$ & $\mathfrak{%
so}\left( 8\right)$ & $\mathfrak{g}_{2\left( -14\right) }$ & $\mathfrak{su}%
\left( 3\right)$ \\ \hline
\rule[-1mm]{0mm}{6mm} $\mathfrak{str}_{0,c}$ & $\mathfrak{e}_{6\left(
-78\right) }$ & $\mathfrak{su}\left( 6\right)$ & $\mathfrak{su}\left(
3\right) \oplus \mathfrak{su}\left( 3\right)$ & $\mathfrak{su}\left(
3\right) $ & $\mathfrak{u}\left( 1\right) \oplus \mathfrak{u}\left( 1\right)$
& $-$ & $-$ \\ \hline
\end{tabular}%
\end{center}
\par
\label{grouptheorytable}
\end{table}
The sequence $\mathfrak{L}_{c}^{q}$ is usually named \textit{%
\textquotedblleft exceptional sequence"} (or \textit{\textquotedblleft
exceptional series"}; see \textit{e.g.} \cite{LM-1}, and Refs. therein).

At Lie group level, the algebraic decompositions (\ref{1}) and (\ref{2}) are
Cartan decompositions respectively pertaining to the following maximal
non-symmetric embeddings:%
\begin{eqnarray}
QConf\left( \mathfrak{J}_{3}^{q}\right) &\supset &SL\left( 3,\mathbb{R}%
\right) \times Str_{0}\left( \mathfrak{J}_{3}^{q}\right) ;  \label{emb-1} \\
QConf_{c}\left( \mathfrak{J}_{3}^{q}\right) &\supset &SU\left( 3\right)
\times Str_{0,c}\left( \mathfrak{J}_{3}^{q}\right) .  \label{emb-2}
\end{eqnarray}%
The non-semi-simple part of the r.h.s. of (\ref{1}) and (\ref{2}) is given
by a pair of triplets of Jordan algebras, which is usually named \textit{%
\textquotedblleft Jordan pair"} (for a recent application in the compact
case and a list of Refs., see \textit{e.g.} \cite{Truini-1}).

(Suitable real, non-compact forms of) all exceptional Lie algebras can be
characterized as \textit{quasi-conformal} algebras\footnote{%
The case $q=-1$ is trivial ($\mathfrak{su}(3)=\mathfrak{su}(3)$), and it
corresponds to \textit{\textquotedblleft pure"} $\mathcal{N}=2$, $D=(3,1)$
supergravity; therefore, it does not admit an uplift to five dimensions, and
it will henceforth not be considered. Moreover, $\mathfrak{su}(2)$ might be
considered as $q=-4/3$ element of the sequence in the second row of Table 1,
as well. However, this is a limit case of the \textquotedblleft exceptional"
sequence reported in Table 1, not pertaining to Jordan pairs nor to
supergravity in $D=3$ dimensions, and thus we will disregard it.} of
suitable Euclidean simple Jordan algebras of rank $3$. Moreover, in the next
Secs. we will consider the extension of \textit{Jordan pairs} to \textit{%
semi-simple} Euclidean Jordan algebras of rank $3$ of relevance for
supergravity theories (to which the case of $\mathfrak{so}\left( 8\right) $,
$q=0$ belongs).\medskip

As recently analyzed in \cite{super-Ehlers-1}, the $SL(3,\mathbb{R})$
appearing in (\ref{emb-1}) can physically be interpreted as the \textit{%
Ehlers group} in $D=5$. Three decades ago, it was shown \cite{GS} that the $%
D $-dimensional \textit{Ehlers group} $SL(D-2,\mathbb{R})$ is a symmetry of $%
D$-dimensional Einstein gravity, provided that the theory is formulated in
the light-cone gauge. For any $D\geqslant 4$-dimensional Lorentzian
space-time, this results enables to identify the graviton degrees of freedom
with the Riemannian coset%
\begin{equation}
\mathcal{M}_{grav}=\frac{SL\left( D-2,\mathbb{R}\right) _{\text{Ehlers}}}{%
SO\left( D-2\right) _{J}},  \label{Ehlers-M}
\end{equation}%
even if the action of the theory is not simply the sigma model action on
this coset (with the exception of $D=3$). In $D=5$, this statement reduces
to the well known fact that the massless graviton described by the
Einstein-Hilbert action with five degrees of freedom allows for an
enhancement of the \textit{massless} spin subgroup $SO\left( 3\right)
_{J}\sim SU(2)_{J}$ of the Lorentz group in $D=5$ (Lorentzian) space-time
dimensions the \textit{non-compact} Ehlers group :%
\begin{equation}
SU(2)_{J}\rightarrow SL(3,\mathbb{R})_{\text{Ehlers}}.
\end{equation}

As studied \textit{e.g.} in \cite{Br-1,Br-2,Br-3,Keu-1,Keu-2}, in $\mathcal{N%
}$-extended supergravity theories in $D$ dimensions, the Ehlers group enjoys
an interesting interplay with the $U$-duality symmetry $G_{\mathcal{N}}^{D}$%
; algebraically, it can be defined as the commutant of $G_{\mathcal{N}}^{D}$
itself inside the $D=3$ $U$-duality $G_{N}^{3}$:%
\begin{equation}
G_{\mathcal{N}}^{3}\supset G_{\mathcal{N}}^{D}\times SL(D-2,\mathbb{R})_{%
\text{Ehler}}.  \label{emb}
\end{equation}%
In \cite{super-Ehlers-1}, the direct product $G_{\mathcal{N}}^{D}\times
SL(D-2,\mathbb{R})$ was dubbed \textit{super-Ehlers group}, and it was
conjectured to be a manifest \textit{off-shell} symmetry in the Hamiltonian
light-cone formulation of the $\mathcal{N}$-extended supergravity theory.

In $D=5$, the specification of (\ref{emb}) for all $\mathcal{N}>2$ theories,
as well as for a broad class of $\mathcal{N}=2$ models, is given by (\ref%
{emb-1}) itself. This latter is a non-symmetric embedding, but it is however
\textit{maximal}; thus, no further \textit{\textquotedblleft enhancement"}
of the super-Ehlers symmetry into some larger symmetry occurs\footnote{%
However, enhancement to infinite-dimensional Lie algebras, along the lines
of \cite{Sagnotti-1}, should occur.}, as instead is the case in $D=10$ type
IIB supergravity and other theories \cite{super-Ehlers-1}.\medskip

The present paper is devoted to the detailed analysis of suitable
non-compact real form of \textit{Jordan pairs}, and elucidation of their
relevance for the algebraic definition of the super-Ehlers symmetry in $D=5$
supergravity theories, as well as for the determination of the multiplet
structure of the massless spectrum.

\section{\label{q=8-Sec}$q=8$}

Let us consider the case $q=8$. From (\ref{2}) and Table 1, it corresponds to%
\begin{equation}
\mathfrak{e}_{8\left( -248\right) }=\mathfrak{su}\left( 3\right) \oplus
\mathfrak{e}_{6\left( -78\right) }\oplus \mathbf{3}\times \mathbf{27}\oplus
\overline{\mathbf{3}}\times \overline{\mathbf{27}},
\end{equation}%
or, at compact group level:%
\begin{eqnarray}
E_{8\left( -248\right) } &\supset &SU\left( 3\right) \times E_{6\left(
-78\right) }; \\
\mathbf{248} &=&\left( \mathbf{8},\mathbf{1}\right) +\left( \mathbf{1},%
\mathbf{78}\right) +\left( \mathbf{3},\mathbf{27}\right) +\left( \overline{%
\mathbf{3}},\overline{\mathbf{27}}\right) ,
\end{eqnarray}%
where $\mathbf{27}$ is the fundamental irrep. of $E_{6\left( -78\right) }$.

By confining ourselves to \textit{Euclidean} rank-$3$ simple Jordan
algebras, two possibility arise (recall (\ref{pre-1}) and (\ref{pre-3})):%
\begin{equation}
q=8:\left\{
\begin{array}{l}
\mathfrak{J}_{3}^{\mathbb{O}}:\left\{
\begin{array}{l}
\mathfrak{e}_{8\left( -24\right) }=\mathfrak{sl}\left( 3,\mathbb{R}\right)
\oplus \mathfrak{e}_{6\left( -26\right) }\oplus \mathbf{3}\times \mathbf{27}%
\oplus \mathbf{3}^{\prime }\times \mathbf{27}^{\prime }, \\
~ \\
E_{8\left( -24\right) }\supset SL\left( 3,\mathbb{R}\right) \times
E_{6\left( -26\right) };%
\end{array}%
\right. \\
\\
\mathfrak{J}_{3}^{\mathbb{O}_{s}}:\left\{
\begin{array}{l}
\mathfrak{e}_{8\left( 8\right) }=\mathfrak{sl}\left( 3,\mathbb{R}\right)
\oplus \mathfrak{e}_{6\left( 6\right) }\oplus \mathbf{3}\times \mathbf{27}%
\oplus \mathbf{3}^{\prime }\times \mathbf{27}^{\prime }, \\
~ \\
E_{8\left( 8\right) }\supset SL\left( 3,\mathbb{R}\right) \times E_{6\left(
6\right) },%
\end{array}%
\right.%
\end{array}%
\right.  \label{q=8}
\end{equation}%
where $E_{8\left( -24\right) }$ and $E_{8\left( 8\right) }$ are the two only
real, non-compact forms of $E_{8}$, namely the minimally non-compact and the
maximally non-compact (\textit{split}) one.

\subsection{\label{J3-Os-Subsec}$\mathfrak{J}_{3}^{\mathbb{O}_{s}}$}

Let us start by considering the split case. This has an interpretation as
\textit{maximal} supergravity ($32$ supersymmetries) \cite{GST}; this is a
\textit{\textquotedblleft pure"} theory, in which no matter coupling is
allowed, and only the gravity multiplet exists.

The maximal compact subalgebra (\textit{mcs}) of $\mathfrak{qconf}\left(
\mathfrak{J}_{3}^{\mathbb{O}_{s}}\right) =\mathfrak{e}_{8\left( 8\right) }$
and $\mathfrak{str}_{0}\left( \mathfrak{J}_{3}^{\mathbb{O}_{s}}\right) =%
\mathfrak{e}_{6\left( 6\right) }$ respectively reads%
\begin{equation}
mcs\left( \mathfrak{e}_{8\left( 8\right) }\right) =\mathfrak{so}\left(
16\right) ;~mcs\left( \mathfrak{e}_{6\left( 6\right) }\right) =\mathfrak{usp}%
\left( 8\right) ,  \label{mcs}
\end{equation}%
and the corresponding relevant maximal non-symmetric embedding is (recall (%
\ref{pre-1-mcs}))%
\begin{eqnarray}
\mathfrak{so}\left( 16\right) &=&\mathfrak{su}(2)\oplus \mathfrak{usp}\left(
8\right) \oplus \mathbf{3}\times \mathbf{27;}  \label{embb-1} \\
SO\left( 16\right) &\supset &SU\left( 2\right) \times USp\left( 8\right) ;
\label{embb-2} \\
\mathbf{120} &=&\left( \mathbf{3},\mathbf{1}\right) +\left( \mathbf{1},%
\mathbf{36}\right) +\left( \mathbf{3},\mathbf{27}\right) ,
\end{eqnarray}%
where $\mathbf{27}$ is the rank-$2$ antisymmetric skew-traceless irrep. of $%
USp\left( 8\right) $. Note that $SO\left( 16\right) $ and $USp\left(
8\right) $ are the $\mathcal{R}$-symmetry of $\mathcal{N}=16$, $D=3$ \cite%
{MS,Tollsten} and of $\mathcal{N}=8$, $D=5$ \cite{CSS} maximal supergravity,
respectively.

On the other hand, the branchings corresponding to the maximal symmetric
embeddings (\ref{mcs}) read%
\begin{eqnarray}
E_{8\left( 8\right) } &\supset &SO\left( 16\right) :\mathbf{248}=\mathbf{120}%
+\mathbf{128,}  \label{a-bis} \\
E_{6\left( 6\right) } &\supset &USp\left( 8\right) :\mathbf{78}=\mathbf{36}+%
\mathbf{42.}  \label{a-biss}
\end{eqnarray}%
In (\ref{a-bis}), $\mathbf{128}$ is one of the two chiral spinor irreps. of $%
SO\left( 16\right) $, in which the generators of the rank-$8$ symmetric
scalar coset $\frac{E_{8\left( 8\right) }}{SO\left( 16\right) }$ of $%
\mathcal{N}=16$, $D=3$ maximal supergravity sit. In (\ref{a-biss}), $\mathbf{%
42}$ is the rank-$4$ antisymmetric skew-traceless self-real irrep. of $%
USp\left( 8\right) $, in which the generators of the rank-$6$ symmetric
scalar coset $\frac{E_{6\left( 6\right) }}{USp\left( 8\right) }$ of $%
\mathcal{N}=8$, $D=5$ maximal supergravity sit.

Thus, under (\ref{embb-1})-(\ref{embb-2}), it is worth considering also the
following branchings:%
\begin{eqnarray}
SO\left( 16\right) &\supset &SU\left( 2\right) \times USp\left( 8\right) ;
\label{embb-3} \\
\mathbf{16} &=&\left( \mathbf{2},\mathbf{8}\right) ;  \notag \\
\mathbf{128} &=&\left( \mathbf{5},\mathbf{1}\right) +\left( \mathbf{3},%
\mathbf{27}\right) +\left( \mathbf{1},\mathbf{42}\right) ;  \label{b-1} \\
\mathbf{128}^{\prime } &=&\left( \mathbf{4},\mathbf{8}\right) +\left(
\mathbf{2},\mathbf{48}\right) ,  \label{f-1}
\end{eqnarray}%
where $\mathbf{48}$ is the rank-$3$ antisymmetric skew-traceless irrep. of $%
USp\left( 8\right) $, and $\mathbf{128}^{\prime }$ is the other chiral
spinor irrep. of $SO\left( 16\right) $, conjugate to $\mathbf{128}$.\medskip

Some remarks are in order.

\begin{enumerate}
\item The branchings (\ref{f-1}) and (\ref{b-1}) suggests the identification
of the $SU\left( 2\right) $ on the right-hand side of (\ref{embb-2}) (or (%
\ref{embb-3})) as the \textit{spin group} for massless particles (as
understood throughout the present investigation) in $D=5$ space-time
dimensions:%
\begin{equation}
SU\left( 2\right) \equiv SU\left( 2\right) _{J}.  \label{spin-SU(2)}
\end{equation}%
Indeed, decomposition (\ref{b-1}) corresponds to the massless bosonic
spectrum of $\mathcal{N}=8$, $D=5$ maximal supergravity ($128$ states): $1$
spin-$2$ field (graviton), $27$ spin-$1$ fields (graviphotons), and $42$
spin-$0$ fields (real scalars). On the other hand, decomposition (\ref{f-1})
yields the corresponding massless fermionic spectrum ($128$ states): $8$
spin-$3/2$ fields (gravitinos) and $48$ spin-$1/2$ fields (dilatinos). Thus,
at the level of massless spectrum, the action of supersymmetry amounts to
the following exchange of irreps.:%
\begin{equation}
SO\left( 16\right) :\underset{B}{\mathbf{128}}~\longleftrightarrow ~\underset%
{F}{\mathbf{128}^{\prime }}.  \label{susy-1}
\end{equation}

\item As denoted by the subscript \textquotedblleft $P$" \ in (\ref%
{principal-1}), the spin group $SU\left( 2\right) _{J}$ commuting with $%
USp\left( 8\right) $ inside $SO\left( 16\right) $ (recall (\ref{embb-2}) or (%
\ref{embb-3})) is the Kostant \textit{\textquotedblleft principal"} $SU(2)$
\cite{K-SU(2)} maximally embedded into the $SL\left( 3,\mathbb{R}\right) $
Ehlers group, which occurs in the embedding $E_{8\left( 8\right) }\supset
SL\left( 3,\mathbb{R}\right) \times E_{6\left( 6\right) }$ pertaining to $%
\mathfrak{J}_{3}^{\mathbb{O}_{s}}$ in (\ref{q=8}):%
\begin{equation}
SL\left( 3,\mathbb{R}\right) \supset _{P}SU\left( 2\right) _{J}:\mathbf{3}=%
\mathbf{3},~\mathbf{8}=\mathbf{3}+\mathbf{5}.  \label{principal-1}
\end{equation}%
Due to the isomorphisms $SU\left( 2\right) \sim SO\left( 3\right) $ and to
the split nature of the non-compact, real form $SL\left( 3,\mathbb{R}\right)
$ of $SU(3)$, the maximal embedding (\ref{principal-1}) is \textit{symmetric}
(whereas generally the principal $SU(2)$ embedding is non-symmetric).
Therefore, consistent with its physical interpretation as Ehlers group in $%
D=5$ \cite{GS} (see also \textit{e.g.} App. of \cite{Compere}), the \textit{%
split} form $SL\left( 3,\mathbb{R}\right) $ of the Jordan-pair $SU(3)$
maximally enhances the massless spin group $SU(2)_{J}$ in $D=5$, as given by
the principal embedding (\ref{principal-1}).

\item As a consequence of the $\mathfrak{J}_{3}^{\mathbb{O}_{s}}$-related
embedding in (\ref{q=8}) and of the embedding (\ref{embb-2}) (or (\ref%
{embb-3})), the following (maximal, non-symmetric) manifold embedding holds:%
\begin{equation}
\frac{E_{8\left( 8\right) }}{SO\left( 16\right) }\supset \frac{E_{6\left(
6\right) }}{USp\left( 8\right) }\times \frac{SL\left( 3,\mathbb{R}\right) _{%
\text{Ehlers}}}{SU\left( 2\right) _{J}}.  \label{manifold-embedding}
\end{equation}%
This has the trivial interpretation of embedding of the scalar manifold of ($%
\mathcal{N}=8$) maximal $D=5$ theory into the scalar manifold of the
corresponding ($\mathcal{N}=16$) maximal theory in $D=3$, obtained \textit{%
e.g.} by two consecutive space-like Kaluza-Klein dimensional reductions. By
recalling (\ref{Ehlers-M}), the maximal symmetric rank-$2$ $5$-dimensional
coset%
\begin{equation}
\frac{SL\left( 3,\mathbb{R}\right) _{\text{Ehlers}}}{SU\left( 2\right) _{J}}%
\sim \frac{SL\left( 3,\mathbb{R}\right) }{SO\left( 3\right) }  \label{gg}
\end{equation}%
in the r.h.s. of (\ref{manifold-embedding}) is associated to the massless
graviton degrees of freedom in $D=5$ Lorentzian space-time dimension.
Indeed, it is nothing but the $D=5$ case of the coset (\ref{Ehlers-M}).

\item Decompositions (\ref{b-1}) and (\ref{f-1}) of $\mathbf{128}$ and its
conjugate $\mathbf{128}^{\prime }$ under the embedding (\ref{embb-3}), which
are consistent with the space-time spin-statistics, are not the usual ones,
as reported \textit{e.g.} in \cite{Slansky} and \cite{Patera}. As
investigated in \cite{Dynkin-1} and \cite{Minchenko} (see also \cite%
{Dynkin-2} and, for a recent discussion, \cite{super-Ehlers-1}), in the Lie
algebra $\mathfrak{so}(2n)$ ($n\in \mathbb{N}$) there are pairs of
subalgebras which are \textit{inequivalent}, namely which are not mapped one
into the other by the conjugation by an element of $\mathfrak{so}(2n)$
itself. They are however \textit{linearly equivalent}, \textit{i.e.} in
every representation of $\mathfrak{so}(2n)$ they are mapped one into the
other by a suitable implementation of the outer $\mathfrak{so}(2n)$%
-automorphism. Clearly, a (semi-)spinor irrep. of $\mathfrak{so}(2n)$
branches differently into each of such two linearly-equivalent and
inequivalent subalgebras. The cases relevant in the present investigation
are obtained by setting $n=4$ and $n=3$ in the following maximal
non-symmetric embedding pattern\footnote{%
For the first application of such an embedding in supersymmetry, see \textit{%
e.g.} \cite{FSaZ}.}%
\begin{equation}
\begin{array}{l}
SO\left( 4n\right) \supset SU\left( 2\right) \times USp\left( 2n\right) ; \\
\mathbf{4n}=(\mathbf{2},\mathbf{2n}).%
\end{array}
\label{Emb}
\end{equation}%
From a Theorem due to Dynkin \cite{Dynkin-1,Dynkin-2}, this embedding is
nothing but a consequence of the \textit{self-conjugacy} of the
bi-fundamental $(\mathbf{2},\mathbf{2n})$ irrep. of $SU\left( 2\right)
\times USp\left( 2n\right) $:%
\begin{eqnarray}
(\mathbf{2},\mathbf{2n})\times _{s}(\mathbf{2},\mathbf{2n}) &=&(\mathbf{2}%
\times _{s}\mathbf{2,2n}\times _{s}\mathbf{2n})+(\mathbf{2}\times _{a}%
\mathbf{2,2n}\times _{a}\mathbf{2n})  \notag \\
&=&\left( \mathbf{Adj}_{SU(2)},\mathbf{Adj}_{USp(2n)}\right) +\left( \mathbf{%
1},\mathbf{\Lambda }_{0}^{2}\right) +\left( \mathbf{1},\mathbf{1}\right) ; \\
(\mathbf{2},\mathbf{2n})\times _{a}(\mathbf{2},\mathbf{2n}) &=&(\mathbf{2}%
\times _{s}\mathbf{2,2n}\times _{a}\mathbf{2n})+(\mathbf{2}\times _{a}%
\mathbf{2,2n}\times _{s}\mathbf{2n})  \notag \\
&=&\left( \mathbf{Adj}_{SU(2)},\mathbf{1}\right) +\left( \mathbf{1},\mathbf{%
Adj}_{USp(2n)}\right) +\left( \mathbf{Adj}_{SU(2)},\mathbf{\Lambda }%
_{0}^{2}\right) ,  \label{second}
\end{eqnarray}%
where $\mathbf{\Lambda }_{0}^{2}$ is the rank-$2$ antisymmetric
skew-traceless irrep. of $USp\left( 2n\right) $ (of total real dimension $%
2n^{2}-n-1$). In general, two non-equivalent (but linearly equivalent) $%
\mathfrak{usp}(2n)$ subalgebras of $\mathfrak{so}(4n)$ exist (distinguished
by a \textquotedblleft $+$" or \textquotedblleft $-$" subscript), under
which the (semi-)spinor irreps. of $\mathfrak{so}(4n)$ branch in different
way. In particular, the case $n=4$ of (\ref{Emb}) splits into a
\textquotedblleft standard" embedding (as \textit{e.g.} reported in \cite%
{Slansky} and in \cite{Patera}) pertains to, say, $USp\left( 8\right) _{+}$,
and it reads%
\begin{eqnarray}
SO\left( 16\right) &\supset &SU\left( 2\right) \times USp\left( 8\right)
_{+}; \\
\mathbf{128} &=&\left( \mathbf{4},\mathbf{8}\right) +\left( \mathbf{2},%
\mathbf{48}\right) ;  \label{b-1-no} \\
\mathbf{128}^{\prime } &=&\left( \mathbf{5},\mathbf{1}\right) +\left(
\mathbf{3},\mathbf{27}\right) +\left( \mathbf{1},\mathbf{42}\right) ,
\label{f-1-no}
\end{eqnarray}%
as well as into a \textquotedblleft non-standard" embedding, pertaining to $%
USp\left( 8\right) _{-}$, which is given by (\ref{embb-3})-(\ref{f-1}). As
discussed above, this latter is relevant (for consistency of spin-statistics
assignments in Lorentzian space-time) to $J_{3}^{\mathbb{O}_{s}}$, and thus
to maximal supergravity in $D=5$. It is immediate to realize that the role
of the conjugate semi-spinor irreps. $\mathbf{128}$ and $\mathbf{128}%
^{\prime }$ of $SO(16)$ is interchanged in the \textquotedblleft standard"
and \textquotedblleft non-standard" embeddings, or equivalently, when
decomposed with respect to the maximal (singular) subalgebras $USp\left(
8\right) _{+}$ and $USp\left( 8\right) _{-}$.

\item As recently analyzed in \cite{super-Ehlers-1}, to each (not
necessarily maximal nor symmetric) embedding (\ref{emb}) one can associate a
pseudo-Riemannian and a Riemannian compact coset, respectively:%
\begin{eqnarray}
M_{\mathcal{N}}^{D} &\equiv &\frac{G_{\mathcal{N}}^{3}}{G_{\mathcal{N}%
}^{D}\times SL\left( D-2,\mathbb{R}\right) _{\text{Ehlers}}};  \label{M} \\
\widehat{M}_{\mathcal{N}}^{D} &\equiv &\frac{mcs\left( G_{\mathcal{N}%
}^{3}\right) }{mcs\left( G_{\mathcal{N}}^{D}\right) \times SO\left(
D-2\right) _{J}},  \label{M-hat}
\end{eqnarray}%
where here \textquotedblleft $mcs$" stands for maximal compact subgroup. In
all theories of supergravity with symmetric scalar manifolds (as considered
in \cite{super-Ehlers-1}), the cosets $M_{\mathcal{N}}^{D}$ (\ref{M}) all
have \textit{vanishing character}, namely they have the same number of
compact and non-compact generators, which in turn equals the real dimension
of $\widehat{M}_{\mathcal{N}}^{D}$ (\ref{M-hat}):%
\begin{equation}
c\left( M_{\mathcal{N}}^{D}\right) =nc\left( M_{\mathcal{N}}^{D}\right) =%
\text{dim}_{\mathbb{R}}\left( \widehat{M}_{\mathcal{N}}^{D}\right) .
\label{c=nc}
\end{equation}%
In \cite{super-Ehlers-1}, this property was related to Poincar\'{e} duality
and to the symmetry of the cohomology of $M_{\mathcal{N}}^{D}$ under the
action of the Hodge involution. In $\mathcal{N}=8$, $D=5$ supergravity, (\ref%
{M})-(\ref{c=nc}) respectively specify to%
\begin{eqnarray}
M_{\mathcal{N}=8}^{5} &\equiv &\frac{E_{8(8)}}{E_{6(6)}\times SL\left( 3,%
\mathbb{R}\right) _{\text{Ehlers}}};  \label{M-1} \\
\widehat{M}_{\mathcal{N}=8}^{5} &\equiv &\frac{SO(16)}{USp(8)\times SO\left(
3\right) _{J}};  \label{M-hat-1} \\
c\left( M_{\mathcal{N}=8}^{5}\right) &=&nc\left( M_{\mathcal{N}%
=8}^{5}\right) =\text{dim}_{\mathbb{R}}\left( \widehat{M}_{\mathcal{N}%
=8}^{5}\right) =81,  \label{c=nc-1}
\end{eqnarray}%
and the result (\ref{c=nc-1}) can be simply explained by noticing that the
Cartan decomposition pertaining to $\widehat{M}_{\mathcal{N}=8}^{5}$ is
given by (\ref{second}) with $n=4$, which thus yields that the generators of
$\widehat{M}_{\mathcal{N}=8}^{5}$ fit into the irrep. $\left( \mathbf{Adj}%
_{SU(2)},\mathbf{\Lambda }_{0}^{2}\right) =\left( \mathbf{3},\mathbf{27}%
\right) $ of $SO\left( 3\right) _{J}\times USp(8)$, of total real dimension $%
81$.
\end{enumerate}

\subsection{\label{J3-O-Subsec}$\mathfrak{J}_{3}^{\mathbb{O}}$}

The case pertaining to the rank-$3$ Euclidean Jordan algebra over the normed
division algebra of octonions $\mathbb{O}$ has an interpretation as \textit{%
minimal} supergravity ($8$ supersymmetries), namely octonionic (also named
exceptional) \textit{magical} Maxwell-Einstein supergravity \cite{GST}.
Coupling is allowed to two types of matter multiplets, namely vector and
hyper multiplets.

An important difference with the case of $\mathfrak{J}_{3}^{\mathbb{O}_{s}}$
treated above is the presence of an hypermultiplet sector, which is
independent on the dimension $D=3,4,5,6$ in which the theory is considered.
The presence of such a $D$-independent hypersector enhances the gravity $%
\mathcal{R}$-symmetry of $\mathcal{N}=4$, $D=3$ $\mathfrak{J}_{3}^{\mathbb{O}%
}$-related (\textit{exceptional}) supergravity from the quaternionic $%
SU(2)_{H}$ (related to the $c$-map of the $D=4$ vector multiplets' scalar
manifold) to $SU(2)_{H}\times SU(2)^{\prime }$:%
\begin{equation}
SU(2)_{H}\longrightarrow SU(2)_{H}\times SU(2)^{\prime }\sim SO(4).
\label{SO(4)}
\end{equation}%
The maximal compact subalgebra (\textit{mcs}) of $\mathfrak{qconf}\left(
\mathfrak{J}_{3}^{\mathbb{O}}\right) =\mathfrak{e}_{8\left( -24\right) }$
and $\mathfrak{str}_{0}\left( \mathfrak{J}_{3}^{\mathbb{O}}\right) =%
\mathfrak{e}_{6\left( -26\right) }$ respectively reads%
\begin{equation}
mcs\left( \mathfrak{e}_{8\left( -24\right) }\right) =\mathfrak{e}_{7\left(
-133\right) }\oplus \mathfrak{su}(2)_{H};~mcs\left( \mathfrak{e}_{6\left(
-26\right) }\right) =\mathfrak{f}_{4\left( -52\right) },  \label{mcs-1}
\end{equation}%
but the corresponding relevant maximal non-symmetric embedding must also
include the $\mathfrak{su}(2)^{\prime }$ from the $D$-independent
hypersector (recall (\ref{pre-3-mcs})):%
\begin{eqnarray}
\mathfrak{e}_{7\left( -133\right) }\oplus \mathfrak{so}(4) &\sim &\mathfrak{e%
}_{7\left( -133\right) }\oplus \mathfrak{su}(2)_{H}\oplus \mathfrak{su}%
(2)^{\prime }  \notag \\
&=&\mathfrak{f}_{4\left( -52\right) }\oplus \mathfrak{su}(2)_{\mathfrak{e}%
_{7}}\oplus \mathfrak{su}(2)_{H}\oplus \mathfrak{su}(2)^{\prime }  \notag \\
&&\oplus ~\mathbf{26}\times \mathbf{3}\times \mathbf{1}\times \mathbf{1;}
\label{embb-1-1} \\
&&  \notag \\
E_{7\left( -133\right) }\times SO\left( 4\right) &\sim &E_{7\left(
-133\right) }\times SU\left( 2\right) _{H}\times SU\left( 2\right) ^{\prime }
\notag \\
&\supset &F_{4\left( -52\right) }\times SU\left( 2\right) _{E_{7}}\times
SU\left( 2\right) _{H}\times SU\left( 2\right) ^{\prime };  \label{embb-2-1}
\\
&&  \notag \\
\left( \mathbf{133,1,1}\right) +\left( \mathbf{1},\mathbf{3},\mathbf{1}%
\right) +\left( \mathbf{1},\mathbf{1},\mathbf{3}\right) &=&\left( \mathbf{52}%
,\mathbf{1,1,1}\right) +\left( \mathbf{1},\mathbf{3},\mathbf{1},\mathbf{1}%
\right)  \notag \\
&&+\left( \mathbf{26},\mathbf{3,1,1}\right) +\left( \mathbf{1},\mathbf{1},%
\mathbf{3,1}\right) +\left( \mathbf{1},\mathbf{1},\mathbf{1,3}\right) ,
\label{embb-2-2}
\end{eqnarray}%
where $\mathbf{26}$ is the fundamental irrep. of $F_{4\left( -52\right) }$.
As mentioned, $SU(2)_{H}\times SU(2)^{\prime }\sim SO(4)$ (\ref{SO(4)}) and $%
SU(2)^{\prime }\sim USp\left( 2\right) $ are the $\mathcal{R}$-symmetry of $%
\mathcal{N}=4$, $D=3$ \textit{exceptional} and of its uplift to $D=5$%
\footnote{%
This is a \textit{unified} $\mathcal{N}=2$ theory, namely all vectors sit in
an \textit{irreducible} representation of the $D=5$ $U$-duality group (in
this case, $\mathbf{27}$ of $E_{6(-26)}$; see \textit{e.g.} \cite{GZ-1}).},
respectively. The group $SU\left( 2\right) _{E_{7}}$ is the one commuting
with $F_{4\left( -52\right) }$ in the maximal non-symmetric embedding%
\begin{equation}
E_{7\left( -133\right) }\supset F_{4\left( -52\right) }\times SU\left(
2\right) _{E_{7}},
\end{equation}%
determining (\ref{embb-2-1})-(\ref{embb-2-2}).

Clearly, it holds that%
\begin{eqnarray}
\mathfrak{su}(2)^{\prime }\cap \mathfrak{e}_{8\left( -24\right) }
&=&\emptyset \Rightarrow \mathfrak{su}(2)^{\prime }\cap \mathfrak{su}%
(2)_{J}=\emptyset ;~\mathfrak{su}(2)^{\prime }\cap \mathfrak{sl}(3,\mathbb{R}%
)=\emptyset ; \\
\mathfrak{su}(2)_{\mathfrak{e}_{7}}\oplus \mathfrak{su}(2)_{H} &\nsubseteq &%
\mathfrak{sl}(3,\mathbb{R}).
\end{eqnarray}%
As for the case of $\mathfrak{J}_{3}^{\mathbb{O}_{s}}$ treated above (recall
(\ref{principal-1})), the $D=5$ Ehlers Lie algebra $\mathfrak{sl}(3,\mathbb{R%
})$ admits the massless spin algebra $\mathfrak{su}(2)_{J}$ as maximal
compact subalgebra. Due to the different multiplet structure, this latter is
defined in a slightly more involved way with respect to the $\mathfrak{J}%
_{3}^{\mathbb{O}_{s}}$-related maximally supersymmetric case treated above.

The branchings corresponding to the maximal symmetric embeddings (\ref{mcs-1}%
) read%
\begin{eqnarray}
E_{8\left( -24\right) } &\supset &E_{7\left( -133\right) }\times SU\left(
2\right) _{H}:\mathbf{248}=\left( \mathbf{133,1}\right) +\left( \mathbf{1},%
\mathbf{3}\right) +\left( \mathbf{56,2}\right) \mathbf{;}  \label{a-bis-1} \\
E_{6\left( -26\right) } &\supset &F_{4\left( -52\right) }:\mathbf{78}=%
\mathbf{52}+\mathbf{26,}  \label{a-bis-2}
\end{eqnarray}%
where $\left( \mathbf{56,2}\right) $ is the bi-fundamental irrep. of $%
E_{7\left( -133\right) }\times SU\left( 2\right) _{H}$, in which the
generators of the rank-$4$ symmetric quaternionic scalar manifold $\frac{%
E_{8\left( -24\right) }}{E_{7\left( -133\right) }\times SU\left( 2\right)
_{H}}$ of $\mathcal{N}=4$, $D=3$ exceptional supergravity sit. In (\ref%
{a-bis-2}), the generators of the rank-$2$ symmetric real special scalar
manifold $\frac{E_{6\left( -26\right) }}{F_{4\left( -52\right) }}$ of $%
\mathcal{N}=2$, $D=5$ exceptional supergravity sit in the fundamental irrep.
$\mathbf{26}$ of $F_{4\left( -52\right) }$. Thus, under (\ref{embb-1-1})-(%
\ref{embb-2-1}), it is worth considering also the following branching:%
\begin{eqnarray}
E_{7\left( -133\right) }\times SU\left( 2\right) _{H}\times SU\left(
2\right) ^{\prime } &\supset &F_{4\left( -52\right) }\times SU\left(
2\right) _{E_{7}}\times SU\left( 2\right) _{H}\times SU\left( 2\right)
^{\prime };  \notag \\
\left( \mathbf{56,2,1}\right) &=&\left( \mathbf{1,4,2,1}\right) +\left(
\mathbf{26},\mathbf{2},\mathbf{2},\mathbf{1}\right) .  \label{embb-3-1}
\end{eqnarray}%
\medskip

Some remarks are in order.

\begin{enumerate}
\item In this theory, the massless spin group $SU\left( 2\right) _{J}$ in $%
D=5$ can be identified with the \textit{diagonal} $SU(2)$ maximally and
symmetrically embedded into $SU\left( 2\right) _{E_{7}}\times SU\left(
2\right) _{H}$:%
\begin{equation}
SU\left( 2\right) _{J}\subset _{d}SU\left( 2\right) _{E_{7}}\times SU\left(
2\right) _{H},  \label{spin-SU(2)-2}
\end{equation}%
such that (\ref{embb-3-1}) can be completed to the following chain:%
\begin{eqnarray}
E_{7\left( -133\right) }\times SU\left( 2\right) _{H}\times SU\left(
2\right) ^{\prime } &\supset &F_{4\left( -52\right) }\times SU\left(
2\right) _{E_{7}}\times SU\left( 2\right) _{H}\times SU\left( 2\right)
^{\prime }  \notag \\
&\supset &F_{4\left( -52\right) }\times SU\left( 2\right) _{J}\times
SU\left( 2\right) ^{\prime };  \label{chain-1} \\
&&  \notag \\
\left( \mathbf{56,2,1}\right) &=&\left( \mathbf{1,4,2,1}\right) +\left(
\mathbf{26},\mathbf{2},\mathbf{2},\mathbf{1}\right)  \notag \\
&=&\left( \mathbf{1,5,1}\right) +\left( \mathbf{1,3,1}\right) +\left(
\mathbf{26},\mathbf{3},\mathbf{1}\right) +\left( \mathbf{26},\mathbf{1},%
\mathbf{1}\right) .  \label{chain-2}
\end{eqnarray}%
Indeed, the decomposition (\ref{chain-1}) corresponds to the massless
bosonic spectrum of $\mathcal{N}=2$, $D=5$ exceptional supergravity ($112$
states\footnote{%
In absence of $D$-independent hypermultiplets (as assumed throughout this
paper for the theories with $8$ local supersymmetries).}): $1$ graviton and $%
1$ graviphoton from the gravity multiplet, and $26$ vectors and $26$ scalars
from the $26$ vector multiplets. At the level of massless spectrum, the
action of supersymmetry amounts to the following exchange of irreps.:
\begin{equation}
E_{7\left( -133\right) }\times SU\left( 2\right) _{H}\times SU\left(
2\right) ^{\prime }:\underset{B}{\left( \mathbf{56,2,1}\right) }%
~\longleftrightarrow ~\underset{F}{\left( \mathbf{56,1,2}\right) }.
\label{susy-1-1}
\end{equation}%
Indeed, under (\ref{chain-1}), $\left( \mathbf{56,1,2}\right) $ decomposes
as follows:%
\begin{equation}
\left( \mathbf{56,1,2}\right) =\left( \mathbf{1},\mathbf{4},\mathbf{1},%
\mathbf{2}\right) +\left( \mathbf{26},\mathbf{2},\mathbf{1},\mathbf{2}%
\right) =\left( \mathbf{1},\mathbf{4},\mathbf{2}\right) +\left( \mathbf{26},%
\mathbf{2},\mathbf{2}\right) ,
\end{equation}%
thus reproducing the massless fermionic spectrum of $\mathcal{N}=2$, $D=5$
exceptional supergravity ($112$ states): $1$ $SU\left( 2\right) ^{\prime }$%
-doublet of gravitinos, and $26$ $SU\left( 2\right) ^{\prime }$-doublets of
gauginos from the $26$ vector multiplets. Note that, consistently, bosons
are $\mathcal{R}$-symmetry $SU\left( 2\right) ^{\prime }$-singlets, whereas
fermions fit into $SU\left( 2\right) ^{\prime }$-doublets. Note how the
change of the kind of octonions ($J_{3}^{\mathbb{O}_{s}}$ \textit{versus} $%
J_{3}^{\mathbb{O}}$) on which the bosonic theory is constructed affects its
supersymmetrization as well as the relevant irreps. and the number of
resulting massless states\footnote{%
In general, the number of massless bosonic states of the theory (in $D=3$ as
well as in any dimension) is given by the (real) dimension of the irrep. of
the $mcs(G_{\mathcal{N}}^{3})$ occurring in the Cartan decomposition of the $%
D=3$ scalar manifold. In supersymmetric theories, the numbers of bosonic and
fermionic states coincide.} : in (\ref{susy-1-1}), the chiral spinor irreps.
$\mathbf{128}$ and $\mathbf{128}^{\prime }$ of the maximal Clifford algebra $%
SO(16)$ of (\ref{susy-1}) are not replaced by the tri-fundamental $\left(
\mathbf{56},\mathbf{2},\mathbf{2}\right) $ of $E_{7\left( -133\right)
}\times SU\left( 2\right) _{H}\times SU\left( 2\right) ^{\prime }$, but
rather by the bi-fundamental $\left( \mathbf{56},\mathbf{2}\right) $ of $%
E_{7\left( -133\right) }\times SU\left( 2\right) _{H}$ (for bosons) and by
the bi-fundamental $\left( \mathbf{56},\mathbf{2}\right) $ of $E_{7\left(
-133\right) }\times SU\left( 2\right) ^{\prime }$ (for fermions).

\item As for the case of $\mathfrak{J}_{3}^{\mathbb{O}}$ treated above, $%
SU\left( 2\right) _{J}$, which commutes with $F_{4\left( -52\right) }\times
SU\left( 2\right) ^{\prime }$ inside $E_{7\left( -133\right) }\times
SU\left( 2\right) _{H}\times SU\left( 2\right) ^{\prime }$ (recall (\ref%
{chain-1}), is the Kostant \textit{\textquotedblleft principal"} $SU(2)$ (%
\ref{principal-1}) maximally embedded into the Ehlers group $SL\left( 3,%
\mathbb{R}\right) $ group. Therefore, as for $\mathfrak{J}_{3}^{\mathbb{O}}$%
, the split form $SL\left( 3,\mathbb{R}\right) $ of the \textit{Jordan-pair}
$SU(3)$ maximally enhances the $D=5$ massless spin group:%
\begin{equation}
SL\left( 3,\mathbb{R}\right) _{\text{Ehlers}}\cap \left[ SU\left( 2\right)
_{E_{7}}\times SU\left( 2\right) _{H}\right] =SU\left( 2\right) _{J}.
\end{equation}

\item As a consequence of the $\mathfrak{J}_{3}^{\mathbb{O}}$-related
embedding in (\ref{q=8}) and of the embedding (\ref{chain-1}), the following
(non-maximal, non-symmetric) manifold embedding holds:%
\begin{equation}
\frac{E_{8\left( -24\right) }}{E_{7\left( -133\right) }\times SU\left(
2\right) _{H}}\supset \frac{E_{6\left( -26\right) }}{F_{4\left( -52\right) }}%
\times \frac{SL\left( 3,\mathbb{R}\right) _{\text{Ehlers}}}{SU\left(
2\right) _{J}}.  \label{manifold-embedding-1}
\end{equation}%
This has the trivial interpretation of embedding of the scalar manifold of $%
\mathcal{N}=2$, $D=5$ theory into the scalar manifold of the theory
dimensionally reduced to $D=3$ dimensions.

\item As resulting from the above treatment, the main difference between the
$\mathfrak{J}_{3}^{\mathbb{O}_{s}}$ and $\mathfrak{J}_{3}^{\mathbb{O}}$
cases resides in the $D$-independent hypersector. In the former case,
pertaining to maximal supergravity, such a sector is forbidden by
supersymmetry. In the latter case, pertaining to minimal supergravity, such
a sector must be present for physical consistency; as mentioned above, this
hypersector is insensitive to dimensional reductions, and it is thus
independent on the number $D=3,4,5,6$ of space-time dimensions in which the
theory with $8$ supersymmetries is defined\footnote{%
We recall that the geometry of the hyperscalars in supergravity is
quaternionic (and thus Einstein); see \textit{e.g.} \cite{Bagger-Witten}.}.
In Lorentzian space-time signatures (which we consider throughout this
paper), it introduces a $D$-independent $SU(2)^{\prime }\sim USp(2)$ $%
\mathcal{R}$-symmetry, which enhances to $SO(4)\sim SU(2)_{H}\times
SU(2)^{\prime }$ in $D=3$. Note that $SU(2)^{\prime }$ is present also in
absence of $D$-independent hypermultiplets (as we assume throughout this
paper), in which case it is promoted to a \textit{global} symmetry of the
theory \cite{FScZ}. Moreover, in order to analyze the massless spectrum of
the theory, such a $D$-independent hypersector does not need to be
specified. By confining ourselves \textit{e.g.} to symmetric
hypermultiplets' scalar manifolds, they read%
\begin{equation}
\frac{\mathcal{G}_{3}}{\mathcal{H}_{3}\times SU\left( 2\right) ^{\prime }},
\label{hyper-coset}
\end{equation}%
where $\mathcal{H}_{3}\times SU\left( 2\right) ^{\prime }$ is the $mcs$ of
the $D$-independent hypersector global symmetry $\mathcal{G}_{3}$. The coset
(\ref{hyper-coset}) is not necessarily the $c$-map \cite{CFG} of the $D=4$
special K\"{a}hler vector multiplets' scalar manifold, as instead is the
quaternionic manifold obtained as space-like KK reduction from the latter
manifold (this parametrizes the scalar degrees of freedom of the genuinely $%
D=3$ hypersector). In the $\mathfrak{J}_{3}^{\mathbb{O}}$-related
exceptional theory under consideration, the $D=4$ vector multiplets' scalar
manifold is symmetric, and so is its $c$-map:%
\begin{equation}
\underset{D=4}{\frac{E_{7\left( -25\right) }}{E_{6\left( -78\right) }\times
U(1)}}\overset{c}{\longrightarrow }\underset{D=3}{\frac{E_{8\left(
-24\right) }}{E_{7\left( -133\right) }\times SU(2)_{H}}}.  \label{c-map}
\end{equation}%
For completeness, we recall that the $\mathfrak{J}_{3}^{\mathbb{O}_{s}}$
(maximally supersymmetric) analogue of (\ref{c-map}) reads%
\begin{equation}
\underset{D=4}{\frac{E_{7\left( 7\right) }}{SU(8)}}\overset{c_{m}}{%
\longrightarrow }\underset{D=3}{\frac{E_{8\left( 8\right) }}{SO(16)}},
\label{c-map-1}
\end{equation}%
where $c_{m}$ is the \textit{\textquotedblleft maximal"} analogue of $c$%
-map, and $\frac{E_{7\left( 7\right) }}{SU(8)}$ is the rank-$7$ scalar
symmetric coset of $\mathcal{N}=8$, $D=4$ maximal supergravity.

\item In $\mathcal{N}=2$, $D=5$ \textit{exceptional} supergravity, (\ref{M}%
)-(\ref{c=nc}) respectively specify to
\begin{eqnarray}
M_{\mathcal{N}=2,J_{3}^{\mathbb{O}}}^{5} &\equiv &\frac{E_{8(-24)}}{%
E_{6(-26)}\times SL\left( 3,\mathbb{R}\right) _{\text{Ehlers}}}; \\
\widehat{M}_{\mathcal{N}=2,J_{3}^{\mathbb{O}}}^{5} &\equiv &\frac{%
E_{7(-133)}\times SU(2)_{H}}{USp(8)\times SU\left( 2\right) _{J}}; \\
c\left( M_{\mathcal{N}=2,J_{3}^{\mathbb{O}}}^{5}\right) &=&nc\left( M_{%
\mathcal{N}=2,J_{3}^{\mathbb{O}}}^{5}\right) =\text{dim}_{\mathbb{R}}\left(
\widehat{M}_{\mathcal{N}=2,J_{3}^{\mathbb{O}}}^{5}\right) =81.
\label{c=nc-2}
\end{eqnarray}%
Disregarding $SU\left( 2\right) ^{\prime }$, the result (\ref{c=nc-2}) can
be explained by noticing that the Cartan decomposition pertaining to $%
\widehat{M}_{\mathcal{N}=2,J_{3}^{\mathbb{O}}}^{5}$ is given by further
branching (\ref{embb-2-2}) with respect to (\ref{spin-SU(2)-2}):%
\begin{eqnarray}
&&%
\begin{array}{l}
E_{7\left( -133\right) }\times SU\left( 2\right) _{H}\supset F_{4\left(
-52\right) }\times SU\left( 2\right) _{E_{7}}\times SU\left( 2\right)
_{H}\supset _{d}F_{4\left( -52\right) }\times SU(2)_{J} \\
\end{array}
\\
&&%
\begin{array}{l}
\left( \mathbf{133,1}\right) +\left( \mathbf{1},\mathbf{3}\right) \\
=\left( \mathbf{52},\mathbf{1,1}\right) +\left( \mathbf{1},\mathbf{3},%
\mathbf{1}\right) +\left( \mathbf{26},\mathbf{3,1}\right) +\left( \mathbf{1},%
\mathbf{1},\mathbf{3}\right) , \\
=\left( \mathbf{52},\mathbf{1}\right) +\left( \mathbf{1},\mathbf{3}\right)
+\left( \mathbf{26},\mathbf{3}\right) +\left( \mathbf{1},\mathbf{3}\right) ,%
\end{array}%
\end{eqnarray}%
which thus yields that the generators of $\widehat{M}_{\mathcal{N}=2,J_{3}^{%
\mathbb{O}}}^{5}$ fit into the sum of irreps. $\left( \mathbf{26},\mathbf{3}%
\right) +\left( \mathbf{1},\mathbf{3}\right) $ of $F_{4\left( -52\right)
}\times SU(2)_{J}$, of total real dimension $81$. Note that this is the same
dimension obtained in the case of $J_{3}^{\mathbb{O}_{s}}$, but with a
different covariant decomposition.; in particular, the presence of the $%
F_{4\left( -52\right) }$-singlet $\left( \mathbf{1},\mathbf{3}\right) $ (%
\textit{graviphoton}) is related to the fact that the theory has $8$ local
supersymmetries.
\end{enumerate}

\section{\label{q=4-J3-H-Sec}$q=4,$ $\mathfrak{J}_{3}^{\mathbb{H}}$
\textquotedblleft Twin" Theories}

As evident from the treatment above, the presence or absence of a $D$%
-independent hypersector is implied by the physical (supergravity)
interpretation of the model under consideration. From a group theoretical
perspective, of course one could have added an extra $SU(2)^{\prime }$ also
in the treatment of $\mathfrak{J}_{3}^{\mathbb{O}_{s}}$, or disregarded the $%
D$-independent hypersector in the treatment of $\mathfrak{J}_{3}^{\mathbb{O}%
} $. However, in both cases one would have failed to reproduce the massless
spectrum of the corresponding supergravity theory.

In some special cases, dubbed \textit{"(bosonic) twin"} theories, the $D$%
-independent hypersector can or cannot be considered, and in both instances
the resulting supergravity theory (of course with different number of local
supersymmetries) is physically meaningful. Indeed, "twin" theories share the
very same bosonic sector, which is however supersymmetrized in (\textit{at
least}) two different ways \cite{ADF-1,ADF-2,Gnecchi-1,Samtleben-twin}.

A nice example of "twin" theories is provided by the $q=4$ case of $%
\mathfrak{J}_{3}^{\mathbb{H}}$ \cite{GST} which we will now analyze.

From (\ref{2}) and Table 1, $q=4$ corresponds to%
\begin{equation}
\mathfrak{e}_{7\left( -133\right) }=\mathfrak{su}\left( 3\right) \oplus
\mathfrak{su}\left( 6\right) \oplus \mathbf{3}\times \overline{\mathbf{15}}%
\oplus \overline{\mathbf{3}}\times \mathbf{15},
\end{equation}%
or, at compact group level:%
\begin{eqnarray}
E_{7\left( -133\right) } &\supset &SU\left( 3\right) \times SU\left(
6\right) ; \\
\mathbf{133} &=&\left( \mathbf{8},\mathbf{1}\right) +\left( \mathbf{1},%
\mathbf{35}\right) +\left( \mathbf{3},\overline{\mathbf{15}}\right) +\left(
\overline{\mathbf{3}},\mathbf{15}\right) ,
\end{eqnarray}%
where $\mathbf{15}$ is the rank-$2$ antisymmetric irrep. of $SU\left(
6\right) $.

By confining ourselves to \textit{Euclidean} rank-$3$ simple Jordan
algebras, two possibility arise:%
\begin{equation}
q=4:\left\{
\begin{array}{l}
\mathfrak{J}_{3}^{\mathbb{H}}:\left\{
\begin{array}{l}
\mathfrak{e}_{7\left( -5\right) }=\mathfrak{sl}\left( 3,\mathbb{R}\right)
\oplus \mathfrak{su}^{\ast }\left( 6\right) \oplus \mathbf{3}\times \mathbf{%
15}^{\prime }\oplus \mathbf{3}^{\prime }\times \mathbf{15}, \\
~ \\
E_{7\left( -5\right) }\supset SL\left( 3,\mathbb{R}\right) \times SU^{\ast
}\left( 6\right) ;%
\end{array}%
\right. \\
\\
\mathfrak{J}_{3}^{\mathbb{H}_{s}}:\left\{
\begin{array}{l}
\mathfrak{e}_{7\left( 7\right) }=\mathfrak{sl}\left( 3,\mathbb{R}\right)
\oplus \mathfrak{sl}\left( 6,\mathbb{R}\right) \oplus \mathbf{3}\times
\mathbf{15}^{\prime }\oplus \mathbf{3}^{\prime }\times \mathbf{15}, \\
~ \\
E_{7\left( 7\right) }\supset SL\left( 3,\mathbb{R}\right) \times SL\left( 3,%
\mathbb{R}\right) .%
\end{array}%
\right.%
\end{array}%
\right.  \label{q=4}
\end{equation}%
As mentioned, we will here consider the case $\mathfrak{J}_{3}^{\mathbb{H}}$%
, relevant for \textquotedblleft (bosonic) twin" theories.

\subsection{\label{24-susys}$24$ Supersymmetries}

Let us start by considering the physical interpretation of the $\mathfrak{J}%
_{3}^{\mathbb{H}}$-related model as theory with $24$ local supersymmetries;
namely, in the framework under consideration, $\mathcal{N}=6$, $D=5$
supergravity and its dimensional reduction ($\mathcal{N}=12$) to $D=3$
dimensions. They are \textquotedblleft pure" theories : no matter coupling
is allowed.

The \textit{mcs} of $\mathfrak{qconf}\left( \mathfrak{J}_{3}^{\mathbb{H}%
}\right) =\mathfrak{e}_{7\left( -5\right) }$ and $\mathfrak{str}_{0}\left(
\mathfrak{J}_{3}^{\mathbb{H}}\right) =\mathfrak{su}^{\ast }\left( 6\right) $
respectively reads%
\begin{equation}
mcs\left( \mathfrak{e}_{7\left( -5\right) }\right) =\mathfrak{so}\left(
12\right) \oplus \mathfrak{su}(2)_{(H)};~mcs\left( \mathfrak{su}^{\ast
}\left( 6\right) \right) =\mathfrak{usp}\left( 6\right) ,  \label{mcs-bis}
\end{equation}%
and the corresponding relevant maximal non-symmetric embedding is%
\begin{eqnarray}
\mathfrak{so}\left( 12\right) \oplus \mathfrak{su}(2)_{(H)} &=&\mathfrak{usp}%
\left( 6\right) \oplus \mathfrak{su}(2)_{\mathfrak{so}(12)}\oplus \mathbf{3}%
\times \mathbf{14}\oplus su(2)_{(H)};  \label{embbb-1} \\
SO\left( 12\right) \times SU(2)_{(H)} &\supset &USp\left( 6\right) \times
SU(2)_{SO(12)}\times SU(2)_{(H)};  \label{embbb-2} \\
\left( \mathbf{66,1}\right) +\left( \mathbf{1},\mathbf{3}\right) &=&\left(
\mathbf{21},\mathbf{1,1}\right) +\left( \mathbf{1},\mathbf{3,1}\right)
+\left( \mathbf{14},\mathbf{3,1}\right) +\left( \mathbf{1},\mathbf{1},%
\mathbf{3}\right) ,
\end{eqnarray}%
where $\mathbf{14}$ is the rank-$2$ antisymmetric skew-traceless irrep. of $%
USp\left( 6\right) $. Note that $SO\left( 12\right) \times SU(2)_{(H)}$ and $%
USp\left( 6\right) $ are the $\mathcal{R}$-symmetry of $\mathcal{N}=12$, $%
D=3 $ and of $\mathcal{N}=6$, $D=5$ \textquotedblleft pure" supergravity,
respectively. Here, the subscript \textquotedblleft $(H)$" denotes the fact
that $SU(2)_{(H)}$ actually is the quaternionic $SU(2)$ in the physical
interpretation pertaining to $8$ local supersymmetries (see below). For
later convenience, by $SU\left( 2\right) _{SO(12)}$ we denote the group
commuting with $USp\left( 6\right) $ in the maximal non-symmetric embedding%
\begin{equation}
SO(12)\supset USp\left( 6\right) \times SU(2)_{SO(12)},  \label{x}
\end{equation}%
determining (\ref{embbb-2}).

On the other hand, the branchings corresponding to the maximal symmetric
embeddings (\ref{mcs-bis}) read%
\begin{eqnarray}
E_{7\left( -5\right) } &\supset &SO\left( 12\right) \times SU(2)_{(H)}:%
\mathbf{133}=\left( \mathbf{66,1}\right) +\left( \mathbf{1,3}\right) +\left(
\mathbf{32}^{\prime },\mathbf{2}\right) \mathbf{,}  \label{A-bis} \\
SU^{\ast }\left( 6\right) &\supset &USp\left( 6\right) :\mathbf{35}=\mathbf{%
21}+\mathbf{14,}  \label{A-tris}
\end{eqnarray}%
where $\mathbf{32}^{\prime }$ is one of the two chiral spinor irreps. of $%
SO\left( 12\right) $. The generators of the rank-$4$ quaternionic K\"{a}hler
symmetric scalar manifold $\frac{E_{7\left( -5\right) }}{SO\left( 12\right)
\times SU(2)_{(H)}}$ of $\mathcal{N}=12$, $D=3$ supergravity sit in the $%
\left( \mathbf{32}^{\prime },\mathbf{2}\right) $. On the other hand, the
generators of the rank-$2$ real special symmetric scalar manifold $\frac{%
SU^{\ast }\left( 6\right) }{USp\left( 6\right) }$ of $\mathcal{N}=6$, $D=5$
supergravity sit in the $\mathbf{14}$ of $USp(6)$. Thus, under (\ref{embbb-1}%
)-(\ref{embbb-2}), it is worth considering also the following branchings:%
\begin{eqnarray}
SO\left( 12\right) \times SU(2)_{(H)} &\supset &USp\left( 6\right) \times
SU(2)_{SO(12)}\times SU(2)_{(H)};  \label{embbbb} \\
\left( \mathbf{12,1}\right) &=&\left( \mathbf{6},\mathbf{2,1}\right) ;
\notag \\
\left( \mathbf{32},\mathbf{2}\right) &=&\left( \mathbf{14}^{\prime },\mathbf{%
1,2}\right) +\left( \mathbf{6},\mathbf{3},\mathbf{2}\right) ;  \notag \\
\left( \mathbf{32}^{\prime },\mathbf{2}\right) &=&\left( \mathbf{14},\mathbf{%
2,2}\right) +\left( \mathbf{1},\mathbf{4},\mathbf{2}\right) ;  \notag
\end{eqnarray}%
where $\mathbf{14}^{\prime }$ is the rank-$3$ antisymmetric skew-traceless
irrep. of $USp\left( 6\right) $, and $\mathbf{32}$ is the other chiral
spinor irrep. of $SO\left( 12\right) $, conjugate to $\mathbf{32}^{\prime }$%
.\medskip

Some remarks are in order.

\begin{enumerate}
\item Branching (\ref{embbb-1}) (or (\ref{embbbb})) is consistent with the
identification of the massless $D=5$ spin group with the diagonal $SU(2)$
embedded into $SU(2)_{SO(12)}\times SU(2)_{(H)}$:%
\begin{equation}
SU\left( 2\right) _{J}\subset _{d}SU\left( 2\right) _{SO(12)}\times
SU(2)_{(H)}.  \label{xx}
\end{equation}%
Thus, (\ref{embbb-1}) (or (\ref{embbbb})) can be completed to the following
chain:%
\begin{eqnarray}
SO\left( 12\right) \times SU(2)_{(H)} &\supset &USp\left( 6\right) \times
SU(2)_{SO(12)}\times SU(2)_{(H)}\supset USp\left( 6\right) \times SU\left(
2\right) _{J};  \notag \\
&&  \label{cchain} \\
\left( \mathbf{66,1}\right) +\left( \mathbf{1},\mathbf{3}\right) &=&\left(
\mathbf{21},\mathbf{1,1}\right) +\left( \mathbf{1},\mathbf{3,1}\right)
+\left( \mathbf{14},\mathbf{3,1}\right) +\left( \mathbf{1},\mathbf{1},%
\mathbf{3}\right)  \notag \\
&=&\left( \mathbf{21},\mathbf{1}\right) +\left( \mathbf{1},\mathbf{3}\right)
+\left( \mathbf{14+1},\mathbf{3}\right) ; \\
\left( \mathbf{32},\mathbf{2}\right) &=&\left( \mathbf{14}^{\prime },\mathbf{%
1,2}\right) +\left( \mathbf{6},\mathbf{3},\mathbf{2}\right) =\left( \mathbf{%
14}^{\prime },\mathbf{2}\right) +\left( \mathbf{6},\mathbf{4}\right) +\left(
\mathbf{6},\mathbf{2}\right) ;  \label{ff-1} \\
\left( \mathbf{32}^{\prime },\mathbf{2}\right) &=&\left( \mathbf{14},\mathbf{%
2,2}\right) +\left( \mathbf{1},\mathbf{4},\mathbf{2}\right) =\left( \mathbf{%
14},\mathbf{3}\right) +\left( \mathbf{14},\mathbf{1}\right) +\left( \mathbf{1%
},\mathbf{5}\right) +\left( \mathbf{1},\mathbf{3}\right) .  \label{bb-1}
\end{eqnarray}%
Note that the $\left( 3q+3\right) _{q=4}=15$-dimensional rep. of $USp\left(
6\right) $ is reducible as $\mathbf{14+1}$; as recently discussed \textit{%
e.g.} in \cite{super-Ehlers-1}, this is a peculiarity of the $\mathcal{N}=6$
theory, and allows for a different (\textquotedblleft twin")
supersymmetrization with only $8$ local supersymmetries (see below). The
decomposition (\ref{ff-1}) corresponds to the massless fermionic spectrum of
$\mathcal{N}=6$, $D=5$ supergravity ($64$ states): $14$ $+6$ spin $1/2$
fermions, and $6$ gravitinos. On the other hand, the decomposition (\ref%
{bb-1}) corresponds to the massless bosonic spectrum ($64$ states): $1$
graviton, $14+1$ graviphotons, and $14$ scalar fields. Thus, at the level of
massless spectrum, the action of supersymmetry amounts to the following
exchange of irreps.:%
\begin{equation}
SO\left( 12\right) \times SU(2)_{(H)}:\underset{F}{\left( \mathbf{32},%
\mathbf{2}\right) }~\longleftrightarrow ~\underset{B}{\left( \mathbf{32}%
^{\prime },\mathbf{2}\right) }.  \label{xxx}
\end{equation}

\item As for the cases of $\mathfrak{J}_{3}^{\mathbb{O}}$ and $\mathfrak{J}%
_{3}^{\mathbb{O}_{s}}$ treated above, and as holding true in general, $%
SU\left( 2\right) _{J}$, which commutes with $USp\left( 6\right) $ inside $%
SO\left( 12\right) \times SU(2)_{(H)}$ (recall (\ref{cchain}), is the
Kostant \textit{\textquotedblleft principal"} $SU(2)$ (\ref{principal-1})
into the $D=5$ Ehlers $SL\left( 3,\mathbb{R}\right) $:%
\begin{equation}
SL\left( 3,\mathbb{R}\right) _{\text{Ehlers}}\cap \left[ SU\left( 2\right)
_{SO(12)}\times SU(2)_{(H)}\right] =SU\left( 2\right) _{J}.
\end{equation}

\item As a consequence of the $\mathfrak{J}_{3}^{\mathbb{H}}$-related
embedding in (\ref{q=4}) and of the embedding (\ref{embbb-1}), the following
(non-maximal, non-symmetric) manifold embedding holds:%
\begin{equation}
\frac{E_{7\left( -5\right) }}{SO\left( 12\right) \times SU(2)_{(H)}}\supset
\frac{SU^{\ast }\left( 6\right) }{USp\left( 6\right) }\times \frac{SL\left(
3,\mathbb{R}\right) _{\text{Ehlers}}}{SU\left( 2\right) _{J}}.
\end{equation}%
As above, this has the trivial interpretation of embedding of the scalar
manifold of $\mathcal{N}=6$, $D=5$ theory into the scalar manifold of the
corresponding theory reduced to $D=3$.

\item Decompositions (\ref{ff-1}) and (\ref{bb-1}) of $\left( \mathbf{32},%
\mathbf{2}\right) $ and its conjugate $\left( \mathbf{32}^{\prime },\mathbf{2%
}\right) $ under the first embedding of (\ref{cchain}), which are consistent
with the space-time spin-statistics, are not the usual ones, as reported
\textit{e.g.} in \cite{Slansky} and \cite{Patera}. In fact, the first
embedding of (\ref{cchain}) is nothing but the case $n=3$ of the embedding
pattern discussed at point 4 of Subsec. \ref{J3-Os-Subsec}. In particular,
the case $n=3$ of (\ref{Emb}) splits into a \textquotedblleft standard"
embedding (as \textit{e.g.} reported in \cite{Slansky} and in \cite{Patera})
pertains to, say, $USp\left( 6\right) _{+}$, and it reads%
\begin{eqnarray}
SO\left( 12\right) &\supset &SU\left( 2\right) _{SO(12)}\times USp\left(
6\right) _{+}; \\
\mathbf{32} &=&\left( \mathbf{2,14}\right) +\left( \mathbf{4,1}\right) ; \\
\mathbf{32}^{\prime } &=&\left( \mathbf{1,14}^{\prime }\right) +\left(
\mathbf{3,6}\right) ,
\end{eqnarray}%
as well as into a \textquotedblleft non-standard" embedding, pertaining to $%
USp\left( 6\right) _{-}$, which is indeed given by the first step of (\ref%
{cchain}) and (\ref{ff-1})-(\ref{bb-1}). It is immediate to realize that the
role of the conjugate semi-spinor irreps. $\mathbf{32}$ and $\mathbf{32}%
^{\prime }$ of $SO(12)$ is interchanged in the \textquotedblleft standard"
and \textquotedblleft non-standard" embeddings, or equivalently, when
decomposed with respect to the maximal (singular) subalgebras $USp\left(
6\right) _{+}$ and $USp\left( 6\right) _{-}$.

\item In $\mathcal{N}=6$, $D=5$ supergravity, (\ref{M})-(\ref{c=nc})
respectively specify to%
\begin{eqnarray}
M_{\mathcal{N}=6}^{5} &\equiv &\frac{E_{7(-5)}}{SU^{\ast }\left( 6\right)
\times SL\left( 3,\mathbb{R}\right) _{\text{Ehlers}}}; \\
\widehat{M}_{\mathcal{N}=6}^{5} &\equiv &\frac{SO\left( 12\right) \times
SU(2)_{H}}{USp(6)\times SU\left( 2\right) _{J}}; \\
c\left( M_{\mathcal{N}=6}^{5}\right) &=&nc\left( M_{\mathcal{N}%
=6}^{5}\right) =\text{dim}_{\mathbb{R}}\left( \widehat{M}_{\mathcal{N}%
=6}^{5}\right) =45.  \label{t}
\end{eqnarray}%
The result (\ref{t}) has been explained in \cite{super-Ehlers-1} in terms of
the Cartan decomposition of $\widehat{M}_{\mathcal{N}=6}^{5}$.
\end{enumerate}

\subsection{\label{8-susys}$8$ Supersymmetries}

Let us proceed to considering the physical interpretation of the $\mathfrak{J%
}_{3}^{\mathbb{H}}$-related bosonic model as bosonic sector of a minimal
supergravity theory ($8$ local supersymmetries); in the framework under
consideration, involving \textit{Jordan pairs}, this corresponds to $%
\mathcal{N}=2$, $D=5$ quaternionic \textit{magical} Maxwell-Einstein
supergravity\footnote{%
This is a \textit{unified} $\mathcal{N}=2$ theory : all vectors sit in the $%
\mathbf{15}$ irrep. of $SU^{\ast }(6)$.} \cite{GST}, and its dimensional
reduction to $D=3$. Matter coupling is allowed through two types of
multiplets, namely vector and hyper multiplets.

A crucial difference with the case pertaining to $24$ supersymmetries
treated in previous Subsection is the presence of an hypermultiplet sector
which is independent on the dimension $D=3,4,5,6$ in which the
quarter-minimal theory is considered. Such a $D$-independent hypersector
enhances the $\mathcal{R}$-symmetry of $\mathcal{N}=4$, $D=3$ magical
quaternionic supergravity from the quaternionic $SU(2)_{H}$ (related to the $%
c$-map of the $D=4$ vector multiplets' scalar manifold) to $SU(2)_{H}\times
SU(2)^{\prime }$, as given by (\ref{SO(4)}).

Therefore, the corresponding relevant maximal non-symmetric embedding must
include the $\mathfrak{su}(2)^{\prime }$ algebra from the $D$-independent
hypersector, also when hypermultiplets are actually absent (in this case, $%
\mathfrak{su}(2)^{\prime }$ is a global symmetry):%
\begin{eqnarray}
\mathfrak{so}\left( 12\right) \oplus so\left( 4\right) &\sim &\mathfrak{so}%
\left( 12\right) \oplus \mathfrak{su}(2)_{H}\oplus \mathfrak{su}(2)^{\prime }
\notag \\
&=&\mathfrak{usp}\left( 6\right) \oplus \mathfrak{su}(2)_{\mathfrak{so}%
(12)}\oplus \mathbf{3}\times \mathbf{14}\oplus su(2)_{(H)}\oplus \mathfrak{su%
}(2)^{\prime };  \label{embbb-1-bis} \\
&&  \notag \\
SO\left( 12\right) \times SO(4) &\sim &SO\left( 12\right) \times
SU(2)_{H}\times SU(2)^{\prime }  \notag \\
&\supset &USp\left( 6\right) \times SU\left( 2\right) _{SO(12)}\times
SU(2)_{H}\times SU(2)^{\prime };  \label{embbb-2-bis} \\
&&  \notag \\
\left( \mathbf{66,1,1}\right) +\left( \mathbf{1},\mathbf{3,1}\right) +\left(
\mathbf{1},\mathbf{1,3}\right) &=&\left( \mathbf{21},\mathbf{1,1,1}\right)
+\left( \mathbf{1},\mathbf{3,1,1}\right)  \notag \\
&&+\left( \mathbf{14},\mathbf{3,1,1}\right) +\left( \mathbf{1},\mathbf{1},%
\mathbf{3,1}\right) +\left( \mathbf{1},\mathbf{1},\mathbf{1,3}\right) .
\end{eqnarray}%
As mentioned, $SU(2)_{H}\times SU(2)^{\prime }\sim SO(4)$ (\ref{SO(4)}) and $%
SU(2)^{\prime }\sim USp\left( 2\right) $ are the $\mathcal{R}$-symmetry of
the magical quaternionic theory in $D=3$ and $D=5$, respectively. $SO\left(
12\right) $ and $USp\left( 6\right) $ are to be interpreted as the
corresponding \textit{Clifford vacuum symmetry} in $D=3$ and $D=5$, encoding
the further degeneracy due matter vector multiplets.

It holds that%
\begin{eqnarray}
\mathfrak{su}(2)^{\prime }\cap \mathfrak{e}_{7\left( -5\right) }
&=&\emptyset \Rightarrow \mathfrak{su}(2)^{\prime }\cap \mathfrak{su}%
(2)_{J}=\emptyset ;~\mathfrak{su}(2)^{\prime }\cap \mathfrak{sl}(3,\mathbb{R}%
)=\emptyset ; \\
\mathfrak{su}(2)_{\mathfrak{so}(12)}\oplus \mathfrak{su}(2)_{H} &\nsubseteq &%
\mathfrak{sl}(3,\mathbb{R}).
\end{eqnarray}%
The $D=5$ Ehlers Lie algebra $\mathfrak{sl}(3,\mathbb{R})$ admits the
massless spin algebra $\mathfrak{su}(2)_{J}$ as maximal compact subalgebra.

Clearly, the branchings (\ref{A-bis}) and (\ref{A-tris}) hold in this case,
as well; however, now $\frac{E_{7\left( -5\right) }}{SO\left( 12\right)
\times SU(2)_{H}}$, whose generators sit into the $\left( \mathbf{32}%
^{\prime },\mathbf{2}\right) $ of $SO(12)\times SU(2)_{H}$, pertains to the
bosonic sector of $\mathcal{N}=4$, $D=3$ $\mathfrak{J}_{3}^{\mathbb{H}}$%
-related magical supergravity. Analogously, $\frac{SU^{\ast }\left( 6\right)
}{USp\left( 6\right) }$ is now to be considered as the rank-$2$ real special
symmetric scalar manifold of the corresponding magical Maxwell-Einstein
theory in $D=5$.

Thus, under (\ref{embb-1-1})-(\ref{embb-2-1}), it is worth considering also
the following branching:%
\begin{eqnarray}
E_{7\left( -133\right) }\times SU\left( 2\right) _{H}\times SU\left(
2\right) ^{\prime } &\supset &F_{4\left( -52\right) }\times SU\left(
2\right) _{E_{7}}\times SU\left( 2\right) _{H}\times SU\left( 2\right)
^{\prime };  \notag \\
\left( \mathbf{56,2,1}\right) &=&\left( \mathbf{1,4,2,1}\right) +\left(
\mathbf{26},\mathbf{2},\mathbf{2},\mathbf{1}\right) .
\end{eqnarray}

As within the supersymmetrization with $24$ local supersymmetries, also in
this case the massless $D=5$ spin group can be identified with the diagonal $%
SU(2)$ into $SU(2)_{SO(12)}\times SU(2)_{H}$, as given by (\ref{xx}). Thus, (%
\ref{embbb-2-bis}) can be completed to the following chain:%
\begin{eqnarray}
SO\left( 12\right) \times SO(4) &\sim &SO\left( 12\right) \times
SU(2)_{H}\times SU(2)^{\prime }  \notag \\
&\supset &USp\left( 6\right) \times SU\left( 2\right) _{SO(12)}\times
SU(2)_{H}\times SU(2)^{\prime }  \notag \\
&\supset &USp\left( 6\right) \times SU\left( 2\right) _{J}\times
SU(2)^{\prime };  \label{cchain-1} \\
\left( \mathbf{66,1,1}\right) +\left( \mathbf{1},\mathbf{3,1}\right) +\left(
\mathbf{1},\mathbf{1,3}\right) &=&\left( \mathbf{21},\mathbf{1,1,1}\right)
+\left( \mathbf{1},\mathbf{3,1,1}\right)  \notag \\
&&+\left( \mathbf{14},\mathbf{3,1,1}\right) +\left( \mathbf{1},\mathbf{1},%
\mathbf{3,1}\right) +\left( \mathbf{1},\mathbf{1},\mathbf{1,3}\right)  \notag
\\
&=&\left( \mathbf{21},\mathbf{1,1}\right) +\left( \mathbf{1},\mathbf{3,1}%
\right) +\left( \mathbf{14+1},\mathbf{3,1}\right) +\left( \mathbf{1},\mathbf{%
1,3}\right) ; \\
&&  \notag \\
\left( \mathbf{32}^{\prime },\mathbf{2,1}\right) &=&\left( \mathbf{14},%
\mathbf{2,2,1}\right) +\left( \mathbf{1},\mathbf{4},\mathbf{2,1}\right)
\notag \\
&=&\left( \mathbf{14},\mathbf{3,1}\right) +\left( \mathbf{14},\mathbf{1,1}%
\right) +\left( \mathbf{1},\mathbf{5,1}\right) +\left( \mathbf{1},\mathbf{3,1%
}\right) .  \label{bb-1-bis}
\end{eqnarray}%
The decomposition (\ref{bb-1-bis}) corresponds to the massless bosonic
spectrum of $\mathcal{N}=2$, $D=5$ magical quaternionic supergravity :
consistent with the fact that this theory is the \textquotedblleft bosonic
twin" of the $\mathcal{N}=6$, $D=5$ \textquotedblleft pure" supergravity,
they share the very same bosonic spectrum ($64$ states): $1$ graviton, $14+1$
vectors (in the $\mathcal{N}=2$ case, this splitting distinguishes between
the graviphoton and the $14$ vectors from the vector multiplets), and $14$
real scalar fields (in the $\mathcal{N}=2$ case, all belonging to the $14$
vector multiplets). Such states fit into%
\begin{eqnarray}
\mathcal{N} &=&6~(24\text{ susys}):\left( \mathbf{32}^{\prime },\mathbf{2}%
\right) ~\text{of~}SO\left( 12\right) \times SU(2)_{(H)}; \\
\mathcal{N} &=&2~(8~\text{susys}):\left( \mathbf{32}^{\prime },\mathbf{2,1}%
\right) ~\text{of~}SO\left( 12\right) \times SO(4)\sim SO\left( 12\right)
\times SU(2)_{H}\times SU(2)^{\prime }.
\end{eqnarray}%
However, the two theories have different fermionic sector; thus,
consistently, the massless fermionic spectrum of $\mathcal{N}=2$, $D=5$
magical quaternionic supergravity is not given by $\left( \mathbf{32},%
\mathbf{2,1}\right) $, but rather by $\left( \mathbf{32}^{\prime },\mathbf{%
1,2}\right) $, of $SO\left( 12\right) \times SU(2)_{H}\times SU(2)^{\prime }$%
. Indeed, under (\ref{cchain-1}), such an irrep. decomposes as follows:%
\begin{equation}
\left( \mathbf{32}^{\prime },\mathbf{1,2}\right) =\left( \mathbf{14},\mathbf{%
2,1,2}\right) +\left( \mathbf{1},\mathbf{4},\mathbf{1,2}\right) =\left(
\mathbf{14},\mathbf{2,2}\right) +\left( \mathbf{1},\mathbf{4,2}\right) ,
\end{equation}%
thus corresponding to $14$ $SU(2)^{\prime }$-doublets of gauginos (from the $%
14$ vector multiplets), and $1$ $SU(2)^{\prime }$-doublet of gravitinos.
Thus, at the level of massless spectrum, in the minimal interpretation the
action of supersymmetry amounts to the following exchange of irreps.%
\footnote{%
Consistently with the branching properties%
\begin{eqnarray*}
SO(12) &\supset &USp(6)\times SU(2)_{SO(12)}, \\
\mathbf{32} &=&(\mathbf{6},\mathbf{3})+\left( \mathbf{14}^{\prime },\mathbf{1%
}\right) ,
\end{eqnarray*}%
the irrep. $\left( \mathbf{32},\mathbf{2,1}\right) $ of $SO\left( 12\right)
\times SU(2)_{H}\times SU(2)^{\prime }$ does \textit{not} occur as
(massless) bosonic or fermionic representation in the $D=5$ theory.}:%
\begin{equation}
SO\left( 12\right) \times SU(2)_{H}\times SU(2)^{\prime }:\underset{F}{%
\left( \mathbf{32}^{\prime },\mathbf{1,2}\right) }~\longleftrightarrow ~%
\underset{B}{\left( \mathbf{32}^{\prime },\mathbf{2,1}\right) },
\end{equation}%
to be contrasted with its analogue (\ref{xxx}), holding in presence of $24$
local supersymmetries. Note that, consistently, bosons are $\mathcal{R}$%
-symmetry $SU\left( 2\right) ^{\prime }$-singlets, whereas fermions fit into
$SU\left( 2\right) ^{\prime }$-doublets.

\textit{Mutatis mutandis}, the very same considerations made in Subsec. \ref%
{24-susys} (in particular, the ones at points 4 and 5 therein) also hold in
this case. Due to the $8$-supersymmetries interpretation, from the remarks
made at point 4 of Subsec. \ref{J3-O-Subsec}, the $D=3$ manifold $\frac{%
E_{7\left( -5\right) }}{SO(12)\times SU(2)_{H}}$ can here be regarded as the
following $c$-map image \cite{CFG}:%
\begin{equation}
\underset{D=4}{\frac{SO^{\ast }\left( 12\right) }{SU(6)\times U(1)}}\overset{%
c}{\longrightarrow }\underset{D=4}{\frac{E_{7\left( -5\right) }}{%
SO(12)\times SU(2)_{H}}},
\end{equation}%
where the coset on the l.h.s. is the symmetric rank-$3$ special K\"{a}hler
vector multiplets' scalar manifold of the magical $\mathcal{N}=2$, $D=4$
quaternionic supergravity.

\section{\label{Simple-List}\textit{Simple} \textit{Jordan Pair} Embeddings}

In the present Section, we list and briefly analyze the relevant \textit{%
non-compact} real forms (\ref{1}) of the \textit{compact} \textit{Jordan pair%
} embeddings (\ref{2}) (listed in Table 1) \cite{Truini-1} pertaining to
\textit{simple} Euclidean rank-$3$ Jordan algebras.

We will thus briefly reconsider the cases $q=8$ $\mathfrak{J}_{3}^{\mathbb{O}%
_{s}}$ and $\mathfrak{J}_{3}^{\mathbb{O}}$, and $q=4$ $\mathfrak{J}_{3}^{%
\mathbb{H}}$ (recall Footnote 1 for the cases $q=-1$ and $q=-4/3$), but we
will not mention the peculiar (non-simple but triality-symmetric) case $q=0$
($J_{3}=\mathbb{R}\oplus \mathbb{R}\oplus \mathbb{R}$), which deserves a
separate treatment, given in Subsubsec. \ref{STU}.

\subsection{$q=8$}

\begin{itemize}
\item $\mathfrak{J}_{3}^{\mathbb{O}}$ ($\mathfrak{qconf}\left( \mathfrak{J}%
_{3}^{\mathbb{O}}\right) =\mathfrak{e}_{8\left( -24\right) }$; $\mathfrak{str%
}_{0}\left( \mathfrak{J}_{3}^{\mathbb{O}}\right) =\mathfrak{e}_{6\left(
-26\right) }\sim \mathfrak{sl}(3,\mathbb{O})$):%
\begin{eqnarray}
\mathfrak{e}_{8\left( -24\right) } &=&\mathfrak{sl}\left( 3,\mathbb{R}%
\right) \oplus \mathfrak{e}_{6\left( -26\right) }\oplus \mathbf{3}\times
\mathbf{27}\oplus \mathbf{3}^{\prime }\times \mathbf{27}^{\prime }; \\
E_{8(-24)} &\supset &SL(3,\mathbb{R})\times E_{6(-26)}; \\
\mathbf{248} &=&\left( \mathbf{8},\mathbf{1}\right) +\left( \mathbf{1},%
\mathbf{78}\right) +\left( \mathbf{3},\mathbf{27}\right) +\left( \mathbf{3}%
^{\prime },\mathbf{27}^{\prime }\right) ; \\
\mathfrak{J}_{3}^{\mathbb{O}} &:&E_{6(-26)}\supset F_{4(-52)}:\mathbf{27}=%
\mathbf{26}+\mathbf{1.}  \label{ssss}
\end{eqnarray}
In (\ref{ssss}), $\mathbf{27}$ and $\mathbf{26}$ respectively are the
fundamental irreps. of $E_{6(-26)}$ and of its maximal compact subgroup $%
F_{4(-52)}$. Physical (supergravity) interpretation : minimal theory ($8$
supersymmetries, with $D$-independent hypersector). See Subsec. \ref%
{J3-O-Subsec}.

\item $\mathfrak{J}_{3}^{\mathbb{O}_{s}}$ ($\mathfrak{qconf}\left( \mathfrak{%
J}_{3}^{\mathbb{O}_{s}}\right) =\mathfrak{e}_{8\left( 8\right) }$; $%
\mathfrak{str}_{0}\left( \mathfrak{J}_{3}^{\mathbb{O}_{s}}\right) =\mathfrak{%
e}_{6\left( 6\right) }\sim \mathfrak{sl}(3,\mathbb{O}_{s})$):%
\begin{eqnarray}
\mathfrak{e}_{8\left( 8\right) } &=&\mathfrak{sl}\left( 3,\mathbb{R}\right)
\oplus \mathfrak{e}_{6\left( 6\right) }\oplus \mathbf{3}\times \mathbf{27}%
\oplus \mathbf{3}^{\prime }\times \mathbf{27}^{\prime }; \\
E_{8(8)} &\supset &SL(3,\mathbb{R})\times E_{6(6)}; \\
\mathbf{248} &=&\left( \mathbf{8},\mathbf{1}\right) +\left( \mathbf{1},%
\mathbf{78}\right) +\left( \mathbf{3},\mathbf{27}\right) +\left( \mathbf{3}%
^{\prime },\mathbf{27}^{\prime }\right) ; \\
\mathfrak{J}_{3}^{\mathbb{O}_{s}} &:&E_{6(6)}\supset USp(8):\mathbf{27}=%
\mathbf{27.}  \label{sssss}
\end{eqnarray}%
In (\ref{sssss}), $\mathbf{27}$ is the fundamental irrep. of $E_{6(6)}$,
which becomes the rank-$2$ antisymmetric skew-traceless irrep. of its
maximal compact subgroup $USp(8)$. Physical (supergravity) interpretation :
maximal theory ($32$ supersymmetries, without $D$-independent hypersector).
See Subsec. \ref{J3-Os-Subsec}.
\end{itemize}

\subsection{$q=4$}

\begin{itemize}
\item $\mathfrak{J}_{3}^{\mathbb{H}}$ ($\mathfrak{qconf}\left( \mathfrak{J}%
_{3}^{\mathbb{H}}\right) =\mathfrak{e}_{7\left( -5\right) }$; $\mathfrak{str}%
_{0}\left( \mathfrak{J}_{3}^{\mathbb{H}}\right) =\mathfrak{su}^{\ast }\left(
6\right) \sim \mathfrak{sl}(3,\mathbb{H})$; $\mathbf{14}$ is the rank-$2$
antisymmetric skew-traceless of $USp(6)$):%
\begin{eqnarray}
\mathfrak{e}_{7\left( -5\right) } &=&\mathfrak{sl}\left( 3,\mathbb{R}\right)
\oplus \mathfrak{su}^{\ast }\left( 6\right) \oplus \mathbf{3}\times \mathbf{%
15}^{\prime }\oplus \mathbf{3}^{\prime }\times \mathbf{15}; \\
E_{7(-5)} &\supset &SL(3,\mathbb{R})\times SU^{\ast }(6); \\
\mathbf{133} &=&\left( \mathbf{8},\mathbf{1}\right) +\left( \mathbf{1},%
\mathbf{35}\right) +\left( \mathbf{3},\mathbf{15}^{\prime }\right) +\left(
\mathbf{3}^{\prime },\mathbf{15}\right) ; \\
\mathfrak{J}_{3}^{\mathbb{H}} &:&SU^{\ast }\left( 6\right) \supset USp(6):%
\mathbf{15}^{\prime }=\mathbf{14}+\mathbf{1.}  \label{zz}
\end{eqnarray}%
In (\ref{zz}), $\mathbf{15}^{\prime }$ and $\mathbf{14}$ respectively are
the (contravariant) rank-$2$ antisymmetric irrep. of $SU^{\ast }\left(
6\right) $ and the rank-$2$ antisymmetric skew-traceless irrep. of its
maximal compact subgroup $USp(6)$. Physical (supergravity) interpretation :
\textit{either} minimal theory ($8$ supersymmetries, with $D$-independent
hypersector), \textit{or} \textquotedblleft pure" theory with 24
supersymmetries (without $D$-independent hypersector) : this is indeed
example of a pair of \textquotedblleft (bosonic) twin" theories; see Sec. %
\ref{q=4-J3-H-Sec}.

\item $\mathfrak{J}_{3}^{\mathbb{H}_{s}}$ ($\mathfrak{qconf}\left( \mathfrak{%
J}_{3}^{\mathbb{H}_{s}}\right) =\mathfrak{e}_{7\left( 7\right) }=\mathfrak{%
conf}\left( \mathfrak{J}_{3}^{\mathbb{O}_{s}}\right) $; $\mathfrak{str}%
_{0}\left( \mathfrak{J}_{3}^{\mathbb{H}_{s}}\right) =\mathfrak{sl}(6,\mathbb{%
R})\sim \mathfrak{sl}(3,\mathbb{H}_{s})$):%
\begin{eqnarray}
\mathfrak{e}_{7\left( 7\right) } &=&\mathfrak{sl}\left( 3,\mathbb{R}\right)
\oplus \mathfrak{sl}(6,\mathbb{R})\oplus \mathbf{3}\times \mathbf{15}%
^{\prime }\oplus \mathbf{3}^{\prime }\times \mathbf{15}; \\
E_{7(7)} &\supset &SL(3,\mathbb{R})\times SL(6,\mathbb{R}); \\
\mathbf{133} &=&\left( \mathbf{8},\mathbf{1}\right) +\left( \mathbf{1},%
\mathbf{35}\right) +\left( \mathbf{3},\mathbf{15}^{\prime }\right) +\left(
\mathbf{3}^{\prime },\mathbf{15}\right) ; \\
\mathfrak{J}_{3}^{\mathbb{H}_{s}} &:&SL(6,\mathbb{R})\supset SO(6):\mathbf{15%
}^{\prime }=\mathbf{15.}  \label{sss}
\end{eqnarray}%
In (\ref{sss}), $\mathbf{15}^{\prime }$ is the (contravariant) rank-$2$
antisymmetric of $SL(6,\mathbb{R})$, which becomes the adjoint of its
maximal compact subgroup $SO(6)\sim SU(4)$. Physical interpretation :
\textit{non-supersymmetric} theory ($D$-independent hypersector irrelevant);
in fact, $E_{7(7)}$ can be the global symmetry of a non-linear scalar sigma
model coupled to gravity in $D=3$ dimensions (see \textit{e.g.} \cite{BGM}).
\end{itemize}

\subsection{$q=2$}

\begin{itemize}
\item $\mathfrak{J}_{3}^{\mathbb{C}}$ ($\mathfrak{qconf}\left( \mathfrak{J}%
_{3}^{\mathbb{C}}\right) =\mathfrak{e}_{6\left( 2\right) }$; $\mathfrak{str}%
_{0}\left( \mathfrak{J}_{3}^{\mathbb{C}}\right) =\mathfrak{sl}\left( 3,%
\mathbb{C}\right) $):%
\begin{eqnarray}
\mathfrak{e}_{6\left( 2\right) } &=&\mathfrak{sl}\left( 3,\mathbb{R}\right)
\oplus \mathfrak{sl}\left( 3,\mathbb{C}\right) \oplus \mathbf{3}\times
\left( \mathbf{3,}\overline{\mathbf{3}}\right) \oplus \mathbf{3}^{\prime
}\times \left( \overline{\mathbf{3}}\mathbf{,3}\right) ; \\
E_{6(2)} &\supset &SL(3,\mathbb{R})\times SL(3,\mathbb{C}); \\
\mathbf{78} &=&\left( \mathbf{8},\mathbf{1,1}\right) +\left( \mathbf{1},%
\mathbf{8,1}\right) +\left( \mathbf{1,1,8}\right) +\left( \mathbf{3,3,}%
\overline{\mathbf{3}}\right) +\left( \mathbf{3}^{\prime }\mathbf{,}\overline{%
\mathbf{3}}\mathbf{,3}\right) ; \\
\mathfrak{J}_{3}^{\mathbb{C}} &:&SL(3,\mathbb{C})\supset SU(3):\left(
\mathbf{3,}\overline{\mathbf{3}}\right) =\mathbf{8}+\mathbf{1.}
\end{eqnarray}%
Physical (supergravity) interpretation : minimal theory ($8$
supersymmetries, with $D$-independent hypersector). It is worth recalling
here that, from the theory of extremal black hole attractors, another
maximal non-symmetric embedding is known (see \textit{e.g.} App. of \cite%
{AFMT-1}):%
\begin{equation}
E_{6(2)}\supset SU(2,1)\times SU(2,1)\times SU(2,1).
\end{equation}

\item $\mathfrak{J}_{3}^{\mathbb{C}_{s}}$ ($\mathfrak{qconf}\left( \mathfrak{%
J}_{3}^{\mathbb{C}_{s}}\right) =\mathfrak{e}_{6\left( 6\right) }=\mathfrak{%
str}_{0}\left( \mathfrak{J}_{3}^{\mathbb{O}_{s}}\right) $; $\mathfrak{str}%
_{0}\left( \mathfrak{J}_{3}^{\mathbb{C}_{s}}\right) =\mathfrak{sl}(3,\mathbb{%
R})\oplus \mathfrak{sl}(3,\mathbb{R})\sim \mathfrak{sl}(3,\mathbb{C}_{s})$):%
\begin{eqnarray}
\mathfrak{e}_{6\left( 6\right) } &=&\mathfrak{sl}\left( 3,\mathbb{R}\right)
\oplus \mathfrak{sl}(3,\mathbb{R})_{I}\oplus \mathfrak{sl}(3,\mathbb{R}%
)_{II}\oplus \mathbf{3}\times \left( \mathbf{3,3}^{\prime }\right) \oplus
\mathbf{3}^{\prime }\times \left( \mathbf{3}^{\prime }\mathbf{,3}\right) ; \\
E_{6(6)} &\supset &SL(3,\mathbb{R})\times SL(3,\mathbb{R})_{I}\times SL(3,%
\mathbb{R})_{II}; \\
\mathbf{78} &=&\left( \mathbf{8},\mathbf{1,1}\right) +\left( \mathbf{1},%
\mathbf{8,1}\right) +\left( \mathbf{1,1,8}\right) +\left( \mathbf{3,3,3}%
^{\prime }\right) +\left( \mathbf{3}^{\prime }\mathbf{,3}^{\prime }\mathbf{,3%
}\right) ;  \label{not-tr} \\
\mathfrak{J}_{3}^{\mathbb{C}_{s}} &:&SL(3,\mathbb{R})_{I}\times SL(3,\mathbb{%
R})_{II}\supset SO(3)\times SO(3):\left( \mathbf{3,3}^{\prime }\right)
=\left( \mathbf{3,3}\right) \mathbf{.}  \label{s}
\end{eqnarray}%
Note that (\ref{not-tr}) does not give rise to a triality-symmetric
decomposition. Moreover, (\ref{s}) is a (double) maximal symmetric principal
embedding, of the same kind of the maximal enhancement of the $D=5$ spin
group $SU(2)_{J}$ into $SL(3,\mathbb{R})_{\text{Ehlers}}$ (\textit{cfr.}
\textit{e.g.} (\ref{principal-1})). Physical interpretation : \textit{%
non-supersymmetric} theory ($D$-independent hypersector irrelevant); in
fact, $E_{6(6)}$ can be a global symmetry of a non-linear scalar sigma model
coupled to $D=3$ gravity (see \textit{e.g.} \cite{BGM}).
\end{itemize}

\subsection{$q=1$}

\begin{itemize}
\item $\mathfrak{J}_{3}^{\mathbb{R}}$ ($\mathfrak{qconf}\left( \mathfrak{J}%
_{3}^{\mathbb{R}}\right) =\mathfrak{f}_{4\left( 4\right) }$; $\mathfrak{str}%
_{0}\left( \mathfrak{J}_{3}^{\mathbb{R}}\right) =\mathfrak{sl}\left( 3,%
\mathbb{R}\right) $):%
\begin{eqnarray}
\mathfrak{f}_{4\left( 4\right) } &=&\mathfrak{sl}\left( 3,\mathbb{R}\right)
\oplus \mathfrak{sl}\left( 3,\mathbb{R}\right) _{I}\oplus \mathbf{3}\times
\mathbf{6}^{\prime }\oplus \mathbf{3}^{\prime }\times \mathbf{6}; \\
F_{4(4)} &\supset &SL(3,\mathbb{R})\times SL(3,\mathbb{R})_{I}; \\
\mathbf{52} &=&\left( \mathbf{8},\mathbf{1}\right) +\left( \mathbf{1},%
\mathbf{8}\right) +\left( \mathbf{3,6}^{\prime }\right) +\left( \mathbf{3}%
^{\prime }\mathbf{,6}\right) ; \\
\mathfrak{J}_{3}^{\mathbb{R}} &:&SL(3,\mathbb{R})_{I}\supset SO(3):\mathbf{6}%
^{\prime }=\mathbf{5}+\mathbf{1.}  \label{s-1}
\end{eqnarray}%
In (\ref{s-1}), $\mathbf{6}^{\prime }$ is the (contravariant) rank-$2$
symmetric of $SL(3,\mathbb{R})_{I}$. As for $\mathfrak{J}_{3}^{\mathbb{C}%
_{s}}$, (\ref{s-1}) is a maximal symmetric principal embedding, of the same
kind of the maximal enhancement of the $D=5$ massless spin group $SU(2)_{J}$
into $SL(3,\mathbb{R})_{\text{Ehlers}}$ (\textit{cfr.} \textit{e.g.} (\ref%
{principal-1})). Physical (supergravity) interpretation : minimal theory ($8$
supersymmetries, with $D$-independent hypersector). It is worth recalling
here that, from the theory of attractors in $D=5$, $\mathfrak{J}_{3}^{%
\mathbb{C}}$-related magical supergravity, another maximal non-symmetric
embedding is known (see \textit{e.g.} App. of \cite{AFMT-1}):%
\begin{equation}
F_{4(4)}\supset SU(2,1)\times SU(2,1).
\end{equation}
\end{itemize}

\subsection{$q=-2/3$}

\begin{itemize}
\item $\mathbb{R}$ ($\mathfrak{qconf}\left( \mathbb{R}\right) =\mathfrak{g}%
_{2\left( 2\right) }$; $\mathfrak{str}_{0}\left( \mathbb{R}\right) =%
\mathfrak{\varnothing }$, and thus no non-trivial Jordan algebra
representation):%
\begin{eqnarray}
\mathfrak{g}_{2\left( 2\right) } &=&\mathfrak{sl}\left( 3,\mathbb{R}\right)
\oplus \mathbf{3}\oplus \mathbf{3}^{\prime }; \\
G_{2(2)} &\supset &SL(3,\mathbb{R});  \label{ss-1} \\
\mathbf{14} &=&\mathbf{8}+\mathbf{3}+\mathbf{3}^{\prime };  \label{ss-2} \\
\mathbb{R} &:&\text{Id}:\mathbf{1}=\mathbf{1.}
\end{eqnarray}%
For more on the maximal non-symmetric embedding (\ref{ss-1})-(\ref{ss-2}),
see \textit{e.g.} App. \cite{Compere}, and Refs. therein, as well as \cite%
{super-Ehlers-1}). Physical (supergravity) interpretation : minimal theory ($%
8$ supersymmetries, with $D$-independent hypersector), named $T^{3}$ model
in $D=4$; see also \cite{Miz}. It is worth recalling here that, from the
theory of $c$-map in supergravity (namely, from the \textit{universal
hypermultiplet} as $c$-map of \textquotedblleft pure" $\mathcal{N}=2$, $D=4$
supergravity \cite{CFG}), another maximal non-symmetric embedding is known:%
\begin{equation}
G_{2(2)}\supset SU(2,1).
\end{equation}
\end{itemize}

\section{\label{Semi-Simple-General}\textit{Jordan Pairs} : the \textit{%
Semi-Simple} Case}

We are now going to extend the treatment of \textit{Jordan pair} embeddings,
introduced in Sec. \ref{Simple-General} for \textit{simple} rank-$3$
Euclidean Jordan algebras, to \textit{semi-simple} rank-$3$ Euclidean Jordan
algebras, having the following structure \cite{JVNW}:%
\begin{equation}
\mathfrak{J}_{3}^{m,n}\equiv \mathbb{R}\oplus \mathbf{\Gamma }_{m-1,n-1},
\label{semi-simple-class}
\end{equation}%
where $\Gamma _{m-1,n-1}$ is the \textit{simple} rank-$2$\ Euclidean Jordan
algebra given by the Clifford algebra of $O\left( m-1,n-1\right) $, with%
\begin{equation}
\mathfrak{str}_{0}\left( \mathbf{\Gamma }_{m-1,n-1}\right) =\mathfrak{so}%
\left( m-1,n-1\right) .  \label{y-1}
\end{equation}%
The relevant cases for supergravity corresponds to:

\begin{itemize}
\item $m=2$, pertaining to an infinite sequence of models with $8$ local
supersymmetries (coupled to $n$ vector multiplets, namely $1$ dilatonic and $%
n-1$ non-dilatonic vector multiplets, in $D=5$), based on%
\begin{equation}
8~\text{susys}:\mathfrak{J}_{3}^{2,n}\equiv \mathbb{R}\oplus \mathbf{\Gamma }%
_{1,n-1};  \label{N=2}
\end{equation}

\item $m=6$, pertaining to half-maximal supergravity ($16$ supersymmetries),
coupled to $n-1$ matter (vector) multiplets in $D=5$, based on%
\begin{equation}
16~\text{susys}:\mathfrak{J}_{3}^{6,n}\equiv \mathbb{R}\oplus \mathbf{\Gamma
}_{5,n-1}.  \label{N=4}
\end{equation}
\end{itemize}

In general, it holds that%
\begin{eqnarray}
\mathfrak{L}\left( \mathfrak{J}_{3}^{m,n}\right) &\equiv &\mathfrak{qconf}%
\left( \mathfrak{J}_{3}^{m,n}\right) =\mathfrak{so}\left( m+2,n+2\right) ;
\label{gr-1} \\
\mathfrak{conf}\left( \mathfrak{J}_{3}^{m,n}\right) &=&\mathfrak{aut}\left(
\mathfrak{F}\left( \mathfrak{J}_{3}^{m,n}\right) \right) =\mathfrak{sl}(2,%
\mathbb{R})\oplus \mathfrak{so}\left( m,n\right) ;  \label{gr-2} \\
\mathfrak{str}_{0}\left( \mathfrak{J}_{3}^{m,n}\right) &=&\mathfrak{so}%
(1,1)\oplus \mathfrak{so}\left( m-1,n-1\right) =\mathfrak{so}(1,1)\oplus
\mathfrak{str}_{0}\left( \mathbf{\Gamma }_{m-1,n-1}\right) ,  \label{gr-3}
\end{eqnarray}%
where $\mathfrak{F}\left( \mathfrak{J}_{3}^{m,n}\right) $ denotes the
\textit{Freudenthal triple system} constructed over $\mathfrak{J}_{3}^{m,n}$
(see \textit{e.g.} \cite{G-Lects,GP-1}, and Refs. therein).

Furthermore, for (\ref{semi-simple-class}) one can define an
\textquotedblleft effective" parameter $q_{eff}\left( m,n\right) $ as
follows:%
\begin{equation}
3q_{eff}\left( m,n\right) +4=m+n\Leftrightarrow q_{eff}\left( m,n\right) =%
\frac{m+n-4}{3},  \label{q-eff}
\end{equation}%
such that the \textquotedblleft effective" dimension of $\mathfrak{J}%
_{3}^{m,n}$ relevant for the \textit{semi-simple} generalization of the
\textit{Jordan pair} embeddings (\ref{1}) and (\ref{2}) reads%
\begin{equation}
3q_{eff}\left( m,n\right) +3=m+n-1.  \label{eff-dim}
\end{equation}

In the following treatment, we will focus on the aforementioned cases $m=2$
and $m=6$, relevant for supergravity (a general treatment for (\ref%
{semi-simple-class}) can be given by a straightforward generalization%
\footnote{%
It is easily realized that $\mathfrak{J}_{3}^{m,n}\sim \mathfrak{J}%
_{3}^{n,m} $ as vector space isomorphism. Actually, for $\left( m,n\right)
=\left( 2,6\right) $, this entails a pair of \textquotedblleft (bosonic)
twin" theories, whose treatment in $D=5$ in terms of \textit{Jordan pairs}
is given in Sec. \ref{Semi-Simple-Twin}.}).

\subsection{\label{Semi-Simple-N=2}$\mathfrak{J}_{3}^{2,n}\equiv \mathbb{R}%
\oplus \mathbf{\Gamma }_{1,n-1}$}

An important difference between the \textit{simple} Jordan algebras treated
in\ Secs. \ref{Simple-General}-\ref{Simple-List} and the \textit{semi-simple}
Jordan algebras (\ref{semi-simple-class}) is the fact that $\mathfrak{sl}%
\left( 3,\mathbb{R}\right) \oplus \mathfrak{str}_{0}\left( \mathfrak{J}%
_{3}^{m,n}\right) $ is \textit{not} maximally embedded into $\mathfrak{L}%
\left( \mathfrak{J}_{3}^{m,n}\right) $, but rather it can be embedded by a
\textit{two-step} chain of maximal symmetric embeddings.

In the case $m=2$ under consideration, this chain reads as follows ($%
\mathfrak{so}(3,3)\sim sl(4,\mathbb{R})$):%
\begin{eqnarray}
\mathfrak{so}(4,n+2) &\supset &\mathfrak{so}(3,3)\oplus \mathfrak{so}%
(1,n-1)\oplus \mathbf{6}\times \mathbf{n}  \notag \\
&\supset &\mathfrak{sl}(3,\mathbb{R})\oplus \mathfrak{so}(1,n-1)\oplus
\mathfrak{so}(1,1)\oplus \mathbf{3}\times \left( \mathbf{n}_{2}+\mathbf{1}%
_{-4}\right) \oplus \mathbf{3}^{\prime }\times \left( \mathbf{n}_{-2}+%
\mathbf{1}_{4}\right) ,  \label{ccchain}
\end{eqnarray}%
or, at group level:%
\begin{eqnarray}
SO(4,n+2) &\supset &SO(3,3)\times SO(1,n-1)  \notag \\
&\supset &SL(3,\mathbb{R})\times SO(1,n-1)\times SO(1,1);
\label{ccchain-2-pre} \\
&&  \notag \\
\mathbf{Adj}_{SO(4,n+2)} &=&\mathbf{Adj}_{SO(3,3)}+\mathbf{Adj}%
_{SO(1,n-1)}+\left( \mathbf{6},\mathbf{n}\right)  \notag \\
&=&\mathbf{Adj}_{SL(3,\mathbb{R})}+\mathbf{Adj}_{SO(1,1)}+\mathbf{Adj}%
_{SO(1,n-1)}  \notag \\
&&+\left( \mathbf{3},\mathbf{n}_{2}+\mathbf{1}_{-4}\right) +\left( \mathbf{3}%
^{\prime },\mathbf{n}_{-2}+\mathbf{1}_{4}\right) ,  \label{ccchain-2}
\end{eqnarray}%
where the subscripts denote $SO(1,1)$-weights.

Note that, according to (\ref{q-eff}) and (\ref{eff-dim}), the
\textquotedblleft effective" dimension of $\mathfrak{J}_{3}^{2,n}$ is $n+1$,
and the corresponding representation is \textit{reducible} with respect to $%
Str_{0}\left( \mathfrak{J}_{3}^{2,n}\right) =SO(1,1)\times SO(1,n-1)$, as
given by (\ref{ccchain-2}):%
\begin{equation}
\mathbf{n+1}=\mathbf{n}_{2}+\mathbf{1}_{-4}.  \label{mm}
\end{equation}%
Thus, the $\mathfrak{J}_{3}^{2,n}$-related $\mathcal{N}=2$ theory in $D=5$
is \textit{not unified} \cite{GZ-1}; this is another difference with respect
to simple Jordan algebras, in which $\mathfrak{J}_{3}^{q}$ fits into an
\textit{irreducible} representation of $Str_{0}\left( \mathfrak{J}%
_{3}^{q}\right) $ itself, and therefore the corresponding $D=5$ theory is
\textit{unified}. In (\ref{mm}) (modulo redefinitions of the $SO(1,1)$
weights), $\mathbf{n}_{2}$ corresponds to the graviphoton and the $n-1$
matter vectors (respectively with positive and negative signature in $%
SO(1,n-1)$), whereas $\mathbf{1}_{-4}$ pertains to the vector from the
dilatonic vector multiplet.

The second line of (\ref{ccchain}) provides the extension of (\ref{1}) to $%
\mathfrak{J}_{3}^{2,n}$ (\ref{N=2}). Its compact counterpart (which extends (%
\ref{2}), and thus the results of \cite{Truini-1}) reads%
\begin{eqnarray}
\mathfrak{so}(n+6) &\supset &\mathfrak{so}(6)\oplus \mathfrak{so}(n)\oplus
\mathbf{6}\times \mathbf{n}  \notag \\
&\supset &\mathfrak{su}(3)\oplus \mathfrak{so}(n)\oplus \mathfrak{u}%
(1)\oplus \mathbf{3}\times \left( \mathbf{n}_{2}+\mathbf{1}_{-4}\right)
\oplus \overline{\mathbf{3}}\times \left( \mathbf{n}_{-2}+\mathbf{1}%
_{4}\right) ,  \label{ccchain-1-c}
\end{eqnarray}%
with subscripts here denoting $U(1)$-charges ($\mathfrak{so}(6)\sim su(4)$).
(Suitable real, non-compact forms of) orthogonal Lie algebras can be
characterized as quasi-conformal algebras of suitable \textit{semi-simple}
Euclidean Jordan algebras of rank $3$ \cite{GP-1}.

At group level, the algebraic decompositions (\ref{ccchain}) and (\ref%
{ccchain-1-c}) respectively correspond to the second line of (\ref%
{ccchain-2-pre}), which can be summarized by the non-maximal non-symmetric
embedding
\begin{equation}
QConf\left( \mathfrak{J}_{3}^{2,n}\right) \supset SL\left( 3,\mathbb{R}%
\right) \times Str_{0}\left( \mathfrak{J}_{3}^{2,n}\right) ,
\end{equation}%
and by its compact counterpart:%
\begin{equation}
SO\left( n+6\right) \supset SU(3)\times SO(n)\Leftrightarrow QConf_{c}\left(
\mathfrak{J}_{3}^{2,n}\right) \supset SU\left( 3\right) \times
Str_{0,c}\left( \mathfrak{J}_{3}^{2,n}\right) .
\end{equation}%
\bigskip

As mentioned, $\mathfrak{J}_{3}^{2,n}$ is related to minimal supergravity ($%
8 $ supersymmetries): therefore, matter coupling is allowed, in terms of two
types of matter multiplets, namely vector and hyper multiplets, and a $D$%
-independent hypermultiplet sector must be considered.

The \textit{mcs} of $\mathfrak{qconf}\left( \mathfrak{J}_{3}^{2,n}\right) =%
\mathfrak{so}\left( 4,n+2\right) $ and $\mathfrak{str}_{0}\left( \mathfrak{J}%
_{3}^{2,n}\right) =\mathfrak{so}(1,1)\oplus \mathfrak{so}\left( 1,n-1\right)
$ respectively read\footnote{%
The presence of an \textquotedblleft extra" commuting $SU(2)$ in (\ref{xxxx}%
) and (\ref{xxxxx}) can ultimately be traced back to the fact that for $m=2$
$SO(m+2)=SO(4)\sim SU(2)\times SU(2)_{H}$.}%
\begin{eqnarray}
mcs\left( \mathfrak{so}\left( 4,n+2\right) \right) &=&\mathfrak{so}%
(n+2)\oplus \mathfrak{so}(4)\sim \mathfrak{so}(n+2)\oplus \mathfrak{su}%
(2)\oplus \mathfrak{su}(2)_{H};  \label{xxxx} \\
mcs\left( \mathfrak{so}(1,1)\oplus \mathfrak{so}\left( 1,n-1\right) \right)
&=&\mathfrak{so}\left( n-1\right) ,  \label{xxxx-2}
\end{eqnarray}%
corresponding to the following maximal symmetric embeddings at group level:%
\begin{eqnarray}
SO(4,n+2) &\supset &SO(n+2)\times SO(4)\sim SO(n+2)\times SU(2)\times
SU(2)_{H};  \label{xxxxx} \\
\mathbf{Adj}_{SO(4,n+2)} &=&\mathbf{Adj}_{SO(n+2)}+\mathbf{Adj}%
_{SO(4)}+\left( \mathbf{n+2},\mathbf{4}\right)  \notag \\
&=&\mathbf{Adj}_{SO(n+2)}+\mathbf{Adj}_{SU(2)}+\mathbf{Adj}%
_{SU(2)_{H}}+\left( \mathbf{n+2},\mathbf{2},\mathbf{2}\right) ;
\label{xxxxxx}
\end{eqnarray}%
\begin{eqnarray}
SO(1,1)\times SO(1,n-1) &\supset &SO(n-1);  \label{xxxxx-2} \\
\mathbf{1}_{0}+\mathbf{Adj}_{SO(1,n-1),0} &=&\mathbf{1}+\mathbf{Adj}%
_{SO(n-1)}+\left( \mathbf{n-1}\right) .  \label{xxxxxx-2}
\end{eqnarray}%
However, the relevant maximal embedding must include the $\mathfrak{su}%
(2)^{\prime }$ algebra from the $D$-independent hypersector, as well:%
\begin{eqnarray}
\mathfrak{so}(n+2)\oplus \mathfrak{so}(4)\oplus \mathfrak{su}(2)^{\prime }
&\sim &\mathfrak{so}(n+2)\oplus \mathfrak{su}(2)\oplus \mathfrak{su}%
(2)_{H}\oplus \mathfrak{su}(2)^{\prime }  \notag \\
&=&\mathfrak{so}(n-1)\oplus \mathfrak{su}(2)_{\mathfrak{so}(n+2)}\oplus
\mathfrak{su}(2)\oplus \mathfrak{su}(2)_{H}\oplus \mathfrak{su}(2)^{\prime }
\notag \\
&&\oplus \left( \mathbf{n-1}\right) \times \mathbf{3}\times \mathbf{1}\times
\mathbf{1\times \mathbf{1};}  \label{embbb-2-x} \\
&&  \notag \\
SO(n+2)\times SO(4)\times SU(2)^{\prime } &\sim &SO(n+2)\times SU(2)\times
SU(2)_{H}\times SU(2)^{\prime }  \notag \\
&\supset &SO(n-1)\times SU\left( 2\right) _{SO(n+2)}  \notag \\
&&\times SU(2)\times SU(2)_{H}\times SU(2)^{\prime };  \label{embbb-2-xx} \\
&&  \notag \\
\mathbf{Adj}_{SO(n+2)}+\mathbf{Adj}_{SO(4)}+\mathbf{Adj}_{SU(2)^{\prime }}
&=&\mathbf{Adj}_{SO(n+2)}+\mathbf{Adj}_{SU(2)}+\mathbf{Adj}_{SU(2)_{H}}+%
\mathbf{Adj}_{SU(2)^{\prime }}  \notag \\
&=&\mathbf{Adj}_{SO(n-1)}+\mathbf{Adj}_{SU(2)_{SO(n+2)}}  \notag \\
&&+\mathbf{Adj}_{SU(2)}+\mathbf{Adj}_{SU(2)_{H}}+\mathbf{Adj}_{SU(2)^{\prime
}}  \notag \\
&&+(\mathbf{n-1},\mathbf{3},\mathbf{1},\mathbf{1},\mathbf{1}).
\end{eqnarray}%
By $SU\left( 2\right) _{SO(n+2)}$, we here denote the group commuting with $%
SO(n-1)$ in the maximal symmetric embedding%
\begin{equation}
SO(n+2)\supset SO(n-1)\times SO(3)\sim SO(n-1)\times SU(2)_{SO(n+2)},
\end{equation}%
determining (\ref{embbb-2-x}). Note that, differently from its analogue for
\textit{simple} rank-$3$ Euclidean Jordan algebras, the maximal embedding (%
\ref{embbb-2-x})-(\ref{embbb-2-xx}) is symmetric.

Moreover, it is worth pointing out that the $\mathcal{R}$-symmetry of $%
\mathcal{N}=4$, $D=3$ $\mathfrak{J}_{3}^{2,n}$-related supergravity is
consistently enhanced from the quaternionic $SU(2)_{H}$ (related to the $c$%
-map \cite{CFG} of the $D=4$ vector multiplets' scalar manifold) to $%
SU(2)_{H}\times SU(2)^{\prime }$, as given by (\ref{SO(4)}). On the other
hand, $SU(2)^{\prime }\sim USp\left( 2\right) $ is the $\mathcal{R}$%
-symmetry of the corresponding $D=5$ ($\mathcal{N}=2$) uplift of the theory.

Clearly, it holds that%
\begin{eqnarray}
\mathfrak{su}(2)^{\prime }\cap \mathfrak{so}\left( 4,n+2\right) &=&\emptyset
\Rightarrow \mathfrak{su}(2)^{\prime }\cap \mathfrak{su}(2)_{J}=\emptyset ;~%
\mathfrak{su}(2)^{\prime }\cap \mathfrak{sl}(3,\mathbb{R})=\emptyset ; \\
\mathfrak{su}(2)_{\mathfrak{so}\left( n+2\right) }\oplus \mathfrak{su}%
(2)_{H} &\nsubseteq &\mathfrak{sl}(3,\mathbb{R}),
\end{eqnarray}%
where, as in general, the $D=5$ Ehlers Lie algebra $\mathfrak{sl}(3,\mathbb{R%
})$ admits the massless spin algebra $\mathfrak{su}(2)_{J}$ as maximal
compact subalgebra.

From the embedding (\ref{xxxx}), (\ref{xxxxx}) and (\ref{xxxxxx}), $\left(
\mathbf{n+2},\mathbf{2},\mathbf{2}\right) $ is the tri-fundamental irrep. of
$SO(n+2)\times SU(2)\times SU(2)_{H}$, in which the generators of the rank-$%
4 $ symmetric quaternionic scalar manifold $\frac{SO\left( 4,n+2\right) }{%
SO(n+2)\times SU(2)\times SU(2)_{H}}$ of $\mathcal{N}=4$, $D=3$ $\mathfrak{J}%
_{3}^{2,n}$-related supergravity sit. Furthermore, from the the embedding (%
\ref{xxxx-2}), (\ref{xxxxx-2}) and (\ref{xxxxxx-2}), $\mathbf{1}+\left(
\mathbf{n-1}\right) $ is the (singlet $+$ fundamental) irrep. of $SO(n-1)$,
in which the generators of the rank-$2$ symmetric real special scalar
manifold $SO(1,1)\times \frac{SO(1,n-1)}{SO(n-1)}$ of the corresponding
theory in $D=5$ sit. Thus, under (\ref{embbb-2-x})-(\ref{embbb-2-xx}), it is
worth considering also the following branching:%
\begin{eqnarray}
SO(n+2)\times SO(4)\times SU(2)^{\prime } &\supset &SO(n-1)\times SU\left(
2\right) _{SO(n+2)}\times SU(2)\times SU(2)_{H}\times SU(2)^{\prime };
\notag \\
\left( \mathbf{n+2},\mathbf{2},\mathbf{2,1}\right) &=&\left( \mathbf{n-1},%
\mathbf{1},\mathbf{2},\mathbf{2,1}\right) +\left( \mathbf{1},\mathbf{3},%
\mathbf{2},\mathbf{2,1}\right) .  \label{xy}
\end{eqnarray}%
\medskip

Some remarks are in order.

\begin{enumerate}
\item Differently from the treatment of \textit{simple} rank-$3$ Euclidean
Jordan algebras, in order to identify the $D=5$ massless \textit{spin group}
$SU\left( 2\right) _{J}$, a two-step procedure is to be performed: \textbf{%
1.1]} one introduces the diagonal $SU(2)_{I}$ into $SU\left( 2\right) \times
SU\left( 2\right) _{H}$:%
\begin{equation}
SU\left( 2\right) _{I}\subset _{d}SU\left( 2\right) \times SU\left( 2\right)
_{H},  \label{diag-I}
\end{equation}%
such that (\ref{xy}) can be completed to the following chain:%
\begin{eqnarray}
SO(n+2)\times SO(4)\times SU(2)^{\prime } &\supset &SO(n-1)\times SU\left(
2\right) _{SO(n+2)}\times SU(2)\times SU(2)_{H}\times SU(2)^{\prime }  \notag
\\
&\supset &SO(n-1)\times SU\left( 2\right) _{SO(n+2)}\times SU(2)_{I}\times
SU(2)^{\prime };  \label{xyz} \\
&&  \notag \\
\left( \mathbf{n+2},\mathbf{2},\mathbf{2,1}\right) &=&\left( \mathbf{n-1},%
\mathbf{1},\mathbf{2},\mathbf{2,1}\right) +\left( \mathbf{1},\mathbf{3},%
\mathbf{2},\mathbf{2,1}\right)  \notag \\
&=&\left( \mathbf{n-1},\mathbf{1},\mathbf{3,1}\right) +\left( \mathbf{n-1},%
\mathbf{1},\mathbf{1,1}\right) +\left( \mathbf{1},\mathbf{3},\mathbf{3,1}%
\right) +\left( \mathbf{1},\mathbf{3},\mathbf{1,1}\right) .  \notag \\
&&  \label{xyz-2}
\end{eqnarray}%
\textbf{1.2]} Then, $SU\left( 2\right) _{J}$ is identified with the diagonal
$SU(2)_{II}$ into $SU\left( 2\right) _{SO(n+2)}\times SU\left( 2\right) _{I}$%
:%
\begin{equation}
SU(2)_{J}\equiv SU\left( 2\right) _{II}\subset _{d}SU\left( 2\right)
_{SO(n+2)}\times SU\left( 2\right) _{I}.  \label{diag-II}
\end{equation}%
Indeed, the chain (\ref{xyz})-(\ref{xyz-2}) can be further completed as
follows:%
\begin{eqnarray}
SO(n+2)\times SO(4)\times SU(2)^{\prime } &\supset &SO(n-1)\times SU\left(
2\right) _{SO(n+2)}\times SU(2)\times SU(2)_{H}\times SU(2)^{\prime }  \notag
\\
&\supset &SO(n-1)\times SU\left( 2\right) _{SO(n+2)}\times SU(2)_{I}\times
SU(2)^{\prime }  \notag \\
&\supset &SO(n-1)\times SU\left( 2\right) _{J}\times SU(2)^{\prime };
\label{xyzt} \\
&&  \notag \\
\left( \mathbf{n+2},\mathbf{2},\mathbf{2,1}\right) &=&\left( \mathbf{n-1},%
\mathbf{1},\mathbf{2},\mathbf{2,1}\right) +\left( \mathbf{1},\mathbf{3},%
\mathbf{2},\mathbf{2,1}\right)  \notag \\
&=&\left( \mathbf{n-1},\mathbf{1},\mathbf{3,1}\right) +\left( \mathbf{n-1},%
\mathbf{1},\mathbf{1,1}\right) +\left( \mathbf{1},\mathbf{3},\mathbf{3,1}%
\right) +\left( \mathbf{1},\mathbf{3},\mathbf{1,1}\right)  \notag \\
&=&\left( \mathbf{n-1},\mathbf{3,1}\right) +\left( \mathbf{n-1},\mathbf{1,1}%
\right)  \notag \\
&&+\left( \mathbf{1},\mathbf{5,1}\right) +\left( \mathbf{1},\mathbf{3,1}%
\right) +\left( \mathbf{1},\mathbf{1,1}\right) +\left( \mathbf{1},\mathbf{3,1%
}\right) .  \label{xyzt-2}
\end{eqnarray}%
The decomposition (\ref{xyzt-2}) corresponds to the massless bosonic
spectrum of $\mathcal{N}=2$, $D=5$ $\mathfrak{J}_{3}^{2,n}$-related
supergravity ($4\left( n+2\right) $ states): $1$ graviton and $1$
graviphoton (from the gravity multiplet), $1$ dilatonic vector and $1$
dilaton (from the dilatonic vector multiplet), and $n-1$ vectors and $n-1$
(real) scalars from the $n-1$ non-dilatonic vector multiplets. At the level
of massless spectrum, the action of supersymmetry amounts to the following
exchange of irreps.:
\begin{equation}
SO(n+2)\times SU(2)\times SU(2)_{H}\times SU(2)^{\prime }:\underset{B}{%
\left( \mathbf{n+2},\mathbf{2},\mathbf{2,1}\right) }~\longleftrightarrow ~%
\underset{F}{\left( \mathbf{n+2},\mathbf{2},\mathbf{1,2}\right) }.
\end{equation}%
This can be realized by noticing that, under the chain of maximal symmetric (%
\ref{xyzt}), $\left( \mathbf{n+2},\mathbf{2},\mathbf{1,2}\right) $
decomposes as follows:%
\begin{eqnarray}
\left( \mathbf{n+2},\mathbf{2},\mathbf{1,2}\right) &=&\left( \mathbf{n-1},%
\mathbf{1},\mathbf{2},\mathbf{1,2}\right) +\left( \mathbf{1},\mathbf{3},%
\mathbf{2},\mathbf{1,2}\right)  \notag \\
&=&\left( \mathbf{n-1},\mathbf{1},\mathbf{2,2}\right) +\left( \mathbf{1},%
\mathbf{3},\mathbf{2,2}\right)  \notag \\
&=&\left( \mathbf{n-1},\mathbf{2,2}\right) +\left( \mathbf{1},\mathbf{4,2}%
\right) +\left( \mathbf{1},\mathbf{2,2}\right) ,  \label{xyzt-3}
\end{eqnarray}%
thus reproducing the massless fermionic spectrum of $\mathcal{N}=2$, $D=5$ $%
\mathfrak{J}_{3}^{2,n}$-related supergravity ($4(n+2)$ states): $1$ $%
SU\left( 2\right) ^{\prime }$-doublet of gravitinos, $1$ $SU\left( 2\right)
^{\prime }$-doublet of dilatonic gauginos, and $n-1$ $SU\left( 2\right)
^{\prime }$-doublets of gauginos from the $n-1$ non-dilatonic vector
multiplets. Note that, consistently, bosons are $\mathcal{R}$-symmetry $%
SU\left( 2\right) ^{\prime }$-singlets, whereas fermions fit into $SU\left(
2\right) ^{\prime }$-doublets.

\item As generally holding true also for the \textit{semi-simple} cases, $%
SU\left( 2\right) _{J}$, which commutes with $SO(n-1)\times SU(2)^{\prime }$
inside $SO(n+2)\times SU(2)\times SU(2)_{H}\times SU(2)^{\prime }$ (\textit{%
cfr.} (\ref{xyzt})), is the Kostant \textit{\textquotedblleft principal"} $%
SU(2)$ (\ref{principal-1}) into the $D=5$ Ehlers $SL\left( 3,\mathbb{R}%
\right) $:%
\begin{equation}
SL\left( 3,\mathbb{R}\right) \cap \left[ SU\left( 2\right) _{SO(n+2)}\times
SU(2)\times SU(2)_{H}\right] =SU\left( 2\right) _{J}.
\end{equation}

\item As a consequence of the chain of maximal symmetric embeddings (\ref%
{ccchain-2-pre}) and (\ref{xyzt}), the following (non-maximal,
non-symmetric) manifold embedding holds:%
\begin{equation}
\frac{SO\left( 4,n+2\right) }{SO(n+2)\times SU(2)\times SU(2)_{H}}\supset
SO(1,1)\times \frac{SO(1,n-1)}{SO(n-1)}\times \frac{SL\left( 3,\mathbb{R}%
\right) }{SU\left( 2\right) _{J}}.  \label{yy}
\end{equation}%
As above, this has the trivial interpretation of embedding of the scalar
manifold of the $D=5$ theory into the scalar manifold of the corresponding
theory reduced to $D=3$.

\item In $\mathcal{N}=2$, $D=5$ $\mathfrak{J}_{3}^{2,n}$-related
supergravity, (\ref{M})-(\ref{c=nc}) respectively specify to \cite%
{super-Ehlers-1}%
\begin{eqnarray}
M_{\mathcal{N}=2,\mathfrak{J}_{3}^{2,n}}^{5} &\equiv &\frac{SO\left(
4,n+2\right) }{SO(1,1)\times SO(1,n-1)\times SL\left( 3,\mathbb{R}\right) _{%
\text{Ehlers}}};  \label{pre-pre-qq} \\
\widehat{M}_{\mathcal{N}=2,\mathfrak{J}_{3}^{2,n}}^{5} &\equiv &\frac{%
SO\left( 4\right) \times SO(n+2)}{SO(n-1)\times SU\left( 2\right) _{J}};
\label{pre-qq} \\
c\left( M_{\mathcal{N}=2,\mathfrak{J}_{3}^{2,n}}^{5}\right) &=&nc\left( M_{%
\mathcal{N}=2,\mathfrak{J}_{3}^{2,n}}^{5}\right) =\text{dim}_{\mathbb{R}%
}\left( \widehat{M}_{\mathcal{N}=2,\mathfrak{J}_{3}^{2,n}}^{5}\right)
=3n+3=9\left( q_{eff}\left( 2,n\right) +1\right) ,  \label{qq}
\end{eqnarray}%
where in (\ref{qq}) the definition (\ref{q-eff}) has been recalled.
\end{enumerate}

\subsubsection{\label{STU}$q=0$, $\mathfrak{J}_{3}^{2,2}$ (STU model)}

Within the class $\mathfrak{J}_{3}^{2,n}$ (\ref{N=2}), the $n=2$ element%
\begin{equation}
\mathfrak{J}_{3}^{2,2}\equiv \mathbb{R}\oplus \mathbf{\Gamma }_{1,1}\sim
\mathbb{R}\oplus \mathbb{R}\oplus \mathbb{R}  \label{N=2-STU}
\end{equation}%
deserves a more detailed analysis. After (\ref{gr-1})-(\ref{gr-3}), its
symmetry groups are%
\begin{eqnarray}
\mathfrak{L}\left( \mathfrak{J}_{3}^{2,2}\right) &\equiv &\mathfrak{qconf}%
\left( \mathfrak{J}_{3}^{2,2}\right) =\mathfrak{so}\left( 4,4\right) ;
\label{qconf-STU} \\
\mathfrak{conf}\left( \mathfrak{J}_{3}^{2,2}\right) &=&\mathfrak{aut}\left(
\mathfrak{F}\left( \mathfrak{J}_{3}^{2,2}\right) \right) =\mathfrak{sl}(2,%
\mathbb{R})\oplus \mathfrak{so}\left( 2,2\right)  \notag \\
&\sim &\mathfrak{sl}(2,\mathbb{R})\oplus \mathfrak{sl}(2,\mathbb{R})\oplus
\mathfrak{sl}(2,\mathbb{R});  \label{conf-STU} \\
\mathfrak{str}_{0}\left( \mathfrak{J}_{3}^{2,2}\right) &=&\mathfrak{so}%
(1,1)\oplus \mathfrak{so}\left( 1,1\right) =\mathfrak{so}(1,1)\oplus
\mathfrak{str}_{0}\left( \mathbf{\Gamma }_{1,1}\right) ,
\end{eqnarray}%
Furthermore, the corresponding \textquotedblleft effective" parameter $%
q_{eff}$ (\ref{q-eff}) vanishes%
\begin{equation}
\mathfrak{J}_{3}^{2,2}:q_{eff}\left( 2,2\right) =0,  \label{q-eff-STU}
\end{equation}%
such that the \textquotedblleft effective" dimension (\ref{eff-dim}) of $%
\mathfrak{J}_{3}^{2,2}$ takes value $3$.

The rank-$3$ Euclidean \textit{semi-simple} Jordan algebra $\mathfrak{J}%
_{3}^{2,2}$ (\ref{N=2-STU}) corresponds to the so-called $STU$ model \cite%
{STU-1,STU-2} of minimal supergravity ($8$ supersymmetries), whose \textit{%
triality symmetry} is related to the complete factorization of $\mathfrak{%
conf}\left( \mathfrak{J}_{3}^{2,2}\right) $ (\ref{conf-STU}). Furthermore,
the vanishing value (\ref{q-eff-STU}) of the \textquotedblleft effective"
parameter $q_{eff}$ yields the identification of $\mathfrak{J}_{3}^{2,2}$ as
a Jordan algebra pertaining to the $q=0$ element of the $q$-parametrized
\textit{\textquotedblleft exceptional sequence"} given by the second row of
Table 1 (see \textit{e.g.} \cite{LM-1}); indeed, $\mathfrak{L}\left(
\mathfrak{J}_{3}^{2,2}\right) =\mathfrak{so}\left( 4,4\right) $ (\ref%
{qconf-STU}) is a non-compact, real form (namely, the \textit{split} form)
of $\mathfrak{so}(8)$:%
\begin{equation}
\mathfrak{L}^{q=0}\equiv \mathfrak{L}\left( \mathfrak{J}_{3}^{2,2}\right) =%
\mathfrak{qconf}\left( \mathfrak{J}_{3}^{2,2}\right) =\mathfrak{so}(4,4).
\label{STU-STU}
\end{equation}

It is here worth observing that $\mathfrak{qconf}_{c}\left( \mathfrak{J}%
_{3}^{2,2}\right) =\mathfrak{L}_{c}^{q=0}=\mathfrak{so}(8)$ is the unique
classical Lie algebra in the \textit{\textquotedblleft exceptional sequence"}%
, besides the limit case of $\mathfrak{su}(3)$ (see also Footnote 2).\medskip

As pointed out above for the whole class $\mathfrak{J}_{3}^{2,n}$, $%
\mathfrak{sl}\left( 3,\mathbb{R}\right) \oplus \mathfrak{str}_{0}\left(
\mathfrak{J}_{3}^{2,2}\right) =\mathfrak{sl}\left( 3,\mathbb{R}\right)
\oplus \mathfrak{so}(1,1)\oplus \mathfrak{so}\left( 1,1\right) $ is not
maximally embedded into $\mathfrak{L}\left( \mathfrak{J}_{3}^{2,2}\right) =%
\mathfrak{so}\left( 4,4\right) $, but rather it can be embedded through a
two-step chain of maximal symmetric embedding (\textit{cfr.} (\ref{ccchain}%
)):%
\begin{eqnarray}
\mathfrak{so}(4,4) &\supset &\mathfrak{so}(3,3)_{0}\oplus \mathfrak{so}%
(1,1)_{0}\oplus \mathbf{6}_{2}\oplus \mathbf{6}_{-2}  \notag \\
&\supset &\mathfrak{sl}(3,\mathbb{R})_{0,0}\oplus \mathfrak{so}%
(1,1)_{0,0}\oplus \mathfrak{so}(1,1)_{0,0}  \notag \\
&&\oplus ~\mathbf{3}_{-4,0}\oplus \mathbf{3}_{4,0}^{\prime }\oplus \mathbf{3}%
_{2,2}\oplus \mathbf{3}_{-2,2}^{\prime }\oplus \mathbf{3}_{2,-2}\oplus
\mathbf{3}_{-2,-2}^{\prime },  \label{ccchain-STU}
\end{eqnarray}%
or, at group level:%
\begin{eqnarray}
SO(4,4) &\supset &SO(3,3)\times SO(1,1)\supset SL(3,\mathbb{R})\times
SO(1,1)\times SO(1,1);  \label{ccchain-pre-2-STU} \\
&&  \notag \\
\mathbf{28} &=&\mathbf{15}_{0}+\mathbf{1}_{0}+\mathbf{6}_{2}+\mathbf{6}_{-2}
\notag \\
&=&\mathbf{8}_{0,0}+\mathbf{1}_{0,0}+\mathbf{3}_{-4,0}+\mathbf{3}%
_{4,0}^{\prime }+\mathbf{1}_{0,0}+\mathbf{3}_{2,2}+\mathbf{3}_{-2,2}^{\prime
}+\mathbf{3}_{2,-2}+\mathbf{3}_{-2,-2}^{\prime }.  \label{ccchain-2-STU}
\end{eqnarray}%
By observing that in
\begin{eqnarray}
SL(3,\mathbb{R})\times SO(1,1) &:&\mathbf{6}_{2}+\mathbf{6}_{-2}=\left(
\mathbf{6},\mathbf{2}\right) ;  \label{pre-m} \\
SL(3,\mathbb{R})\times SO(1,1)\times SO(1,1) &:&\mathbf{3}_{-4,0}+\mathbf{3}%
_{2,-2}+\mathbf{3}_{2,2}=\left( \mathbf{3},\mathbf{2}_{2}+\mathbf{1}%
_{-4}\right) ,  \label{m}
\end{eqnarray}%
(\ref{ccchain-STU}) and (\ref{ccchain-2-STU}) can consistently be recast as
the $n=2$ case of the general expressions (\ref{ccchain}) and (\ref%
{ccchain-2}) (analogous formul\ae\ for the compact case hold). As mentioned,
the \textquotedblleft effective" dimension of $\mathfrak{J}_{3}^{2,2}$ is $%
n+1=3$, and the corresponding representation is \textit{reducible} as $%
\mathbf{2}_{2}+\mathbf{1}_{-4}$ (\ref{m}) with respect to $Str_{0}\left(
\mathfrak{J}_{3}^{2,2}\right) =SO(1,1)\times SO(1,1)$.

Thus, the uplift of $STU$ model to $D=5$, based on $\mathfrak{J}_{3}^{2,2}$,
is a \textit{non-unified} theory \cite{GZ-1}; in (\ref{m}) (modulo
redefinitions of the $SO(1,1)$ weights), $\mathbf{2}_{2}$ corresponds to the
graviphoton and the vector from the unique non-dilatonic vector multiplet
(respectively with positive and negative signature in $SO(1,1)$), whereas $%
\mathbf{1}_{-4}$ pertains to the vector from the dilatonic vector multiplet.
We also note that, within the class $\mathfrak{J}_{3}^{2,n}$, only for $%
\mathfrak{J}_{3}^{2,2}$ the total $SO(1,1)$-weight of the $3$-dimensional
representation of the Jordan algebra $\mathfrak{J}_{3}^{2,2}$ vanishes : $%
2\cdot 2-4=0$.\medskip

Concerning the massless spectrum of the $D=5$ uplift of the $STU$ model, the
analysis goes as the case $n=2$ of the general treatment for $\mathfrak{J}%
_{3}^{2,n}$ given in Subsec. \ref{Semi-Simple-N=2}; we briefly consider it
below (as implied by the interpretation with 8 local supersymmetries, a $D$%
-independent hypermultiplet sector must be considered).

The \textit{mcs} of $\mathfrak{qconf}\left( \mathfrak{J}_{3}^{2,2}\right) =%
\mathfrak{so}\left( 4,4\right) $ and $\mathfrak{str}_{0}\left( \mathfrak{J}%
_{3}^{2,2}\right) =\mathfrak{so}(1,1)\oplus \mathfrak{so}\left( 1,1\right) $
respectively reads%
\begin{eqnarray}
mcs\left( \mathfrak{so}\left( 4,4\right) \right) &=&\mathfrak{so}(4)\oplus
\mathfrak{so}(4)\sim \mathfrak{su}(2)\oplus \mathfrak{su}(2)\oplus \mathfrak{%
su}(2)\oplus \mathfrak{su}(2)_{H};  \label{xxxx-STU} \\
mcs\left( \mathfrak{so}(1,1)\oplus \mathfrak{so}\left( 1,1\right) \right) &=&%
\mathfrak{\emptyset }.  \label{xxxx-2-STU}
\end{eqnarray}%
(\ref{xxxx-STU}) corresponds to the following maximal symmetric embeddings
at group level:%
\begin{eqnarray}
SO(4,4) &\supset &SO(4)\times SO(4)\sim SU(2)\times SU(2)\times SU(2)\times
SU(2)_{H};  \label{xxxxx-STU} \\
\mathbf{28} &=&\left( \mathbf{6,1}\right) +\left( \mathbf{1},\mathbf{6}%
\right) +\left( \mathbf{4},\mathbf{4}\right)  \notag \\
&=&\left( \mathbf{3},\mathbf{1},\mathbf{1},\mathbf{1}\right) +(\mathbf{1},%
\mathbf{3},\mathbf{1},\mathbf{1})+(\mathbf{1},\mathbf{1},\mathbf{3},\mathbf{1%
})+(\mathbf{1},\mathbf{1},\mathbf{1},\mathbf{3})+(\mathbf{2},\mathbf{2},%
\mathbf{2},\mathbf{2});
\end{eqnarray}%
However, the relevant maximal embedding must include the $\mathfrak{su}%
(2)^{\prime }$ algebra from the $D$-independent hypersector, as well:%
\begin{eqnarray}
\mathfrak{so}(4)\oplus \mathfrak{so}(4)\oplus \mathfrak{su}(2)^{\prime }
&\sim &\mathfrak{su}(2)\oplus \mathfrak{su}(2)\oplus \mathfrak{su}(2)\oplus
\mathfrak{su}(2)_{H}\oplus \mathfrak{su}(2)^{\prime }  \notag \\
&=&\mathfrak{su}(2)_{\mathfrak{so}(4)}\oplus \mathfrak{su}(2)\oplus
\mathfrak{su}(2)_{H}\oplus \mathfrak{su}(2)^{\prime }  \notag \\
&&\oplus \mathbf{3}\times \mathbf{1}\times \mathbf{1\times \mathbf{1};}
\label{embbb-2-x-STU} \\
&&  \notag \\
SO(4)\times SO(4)\times SU(2)^{\prime } &\sim &SU(2)\times SU(2)\times
SU(2)\times SU(2)_{H}\times SU(2)^{\prime }  \notag \\
&\supset &SU\left( 2\right) _{SO(4)}\times SU(2)\times SU(2)_{H}\times
SU(2)^{\prime };  \label{embbb-2-xx-STU} \\
&&  \notag \\
\left( \mathbf{6,1,1}\right) +\left( \mathbf{1},\mathbf{6,1}\right) +\left(
\mathbf{1},\mathbf{1},\mathbf{3}\right) &=&\left( \mathbf{3},\mathbf{1},%
\mathbf{1},\mathbf{1},\mathbf{1}\right) +\left( \mathbf{1},\mathbf{3},%
\mathbf{1},\mathbf{1},\mathbf{1}\right)  \notag \\
&&+\left( \mathbf{1},\mathbf{1},\mathbf{3},\mathbf{1},\mathbf{1}\right)
+\left( \mathbf{1},\mathbf{1},\mathbf{1},\mathbf{3},\mathbf{1}\right)
+\left( \mathbf{1},\mathbf{1},\mathbf{1},\mathbf{1},\mathbf{3}\right)  \notag
\\
&=&\left( \mathbf{3},\mathbf{1},\mathbf{1},\mathbf{1}\right) +\left( \mathbf{%
3},\mathbf{1},\mathbf{1},\mathbf{1}\right)  \notag \\
&&+\left( \mathbf{1},\mathbf{3},\mathbf{1},\mathbf{1}\right) +\left( \mathbf{%
1},\mathbf{1},\mathbf{3},\mathbf{1}\right) +\left( \mathbf{1},\mathbf{1},%
\mathbf{1},\mathbf{3}\right) ,
\end{eqnarray}%
where $SU\left( 2\right) _{SO(4)}$ is diagonal into the first $SU(2)$'s on
the r.h.s. of the isomorphism in the first line of (\ref{embbb-2-xx-STU}):%
\begin{equation}
SU(2)_{SO(4)}\sim SO(3)\subset _{d}SU(2)\times SU(2)\sim SO(4).  \label{j}
\end{equation}

Consistently, the $\mathcal{R}$-symmetry of the $D=3$ dimensionally reduced $%
STU$ model is enhanced from the quaternionic $SU(2)_{H}$ (related to the $c$%
-map of the $D=4$ vector multiplets' scalar manifold) to $SO(4)\sim
SU(2)_{H}\times SU(2)^{\prime }$. On the other hand, $SU(2)^{\prime }\sim
USp\left( 2\right) $ is the $\mathcal{R}$-symmetry of the corresponding
uplifted theory in $D=5$.

Clearly, it holds that%
\begin{eqnarray}
\mathfrak{su}(2)^{\prime }\cap \mathfrak{so}\left( 4,4\right) &=&\emptyset
\Rightarrow \mathfrak{su}(2)^{\prime }\cap \mathfrak{su}(2)_{J}=\emptyset ;~%
\mathfrak{su}(2)^{\prime }\cap \mathfrak{sl}(3,\mathbb{R})=\emptyset ; \\
\mathfrak{su}(2)_{\mathfrak{so}\left( 4\right) }\oplus \mathfrak{su}(2)_{H}
&\nsubseteq &\mathfrak{sl}(3,\mathbb{R}),
\end{eqnarray}%
where, as in general, the $D=5$ Ehlers Lie algebra $\mathfrak{sl}(3,\mathbb{R%
})$ admits the massless spin algebra $\mathfrak{su}(2)_{J}$ as maximal
compact subalgebra.

From the embedding (\ref{xxxx-STU}) and (\ref{xxxxx-STU}), $\left( \mathbf{%
2,2},\mathbf{2},\mathbf{2}\right) $ is the quadri-fundamental irrep.%
\footnote{%
For application of $(\mathbf{2},\mathbf{2},\mathbf{2},\mathbf{2})$ irrep. to
the connection between QIT and supergravity, see \textit{e.g.} \cite%
{ICL-3,ICL-3-companion} (and Refs. therein).} of $SO(4)\times SU(2)\times
SU(2)_{H}\sim SU(2)\times SU(2)\times SU(2)\times SU(2)_{H}$, in which the
generators of the rank-$4$ symmetric quaternionic scalar manifold $\frac{%
SO\left( 4,4\right) }{SU(2)\times SU(2)\times SU(2)\times SU(2)_{H}}$ of the
$D=3$ dimensionally reduced $STU$ model sit. Trivially, the $n=2$ case of (%
\ref{xxxxx-2})-(\ref{xxxxxx-2}) yields that the generators of the rank-$2$
manifold $SO(1,1)\times SO(1,1)$ of the $D=5$ uplift of STU model sit in the
$\mathbf{1}+\mathbf{1}$. Under (\ref{embbb-2-x-STU})-(\ref{embbb-2-xx-STU}),
it is then worth considering also the following branching:%
\begin{eqnarray}
SO(4)\times SO(4)\times SU(2)^{\prime } &\sim &SU(2)\times SU(2)\times
SU(2)\times SU(2)_{H}\times SU(2)^{\prime }  \notag \\
&\supset &SU\left( 2\right) _{SO(4)}\times SU(2)\times SU(2)_{H}\times
SU(2)^{\prime }; \\
\left( \mathbf{2,2},\mathbf{2},\mathbf{2,1}\right) &=&\left( \mathbf{3},%
\mathbf{2},\mathbf{2,1}\right) +\left( \mathbf{1},\mathbf{2},\mathbf{2,1}%
\right) .  \label{xy-STU}
\end{eqnarray}

Some remarks are in order.\medskip

\begin{enumerate}
\item In order to identify the $D=5$ massless \textit{spin group} $SU\left(
2\right) _{J}$, a two-step procedure must be performed: \textbf{1.1]} one
introduces the diagonal $SU(2)_{I}$ into $SU\left( 2\right) \times SU\left(
2\right) _{H}$ (\textit{cfr.} (\ref{diag-I})), such that (\ref{xy-STU}) can
be completed to the following chain:%
\begin{eqnarray}
SU(2)\times SU(2)\times SU(2)\times SU(2)_{H}\times SU(2)^{\prime } &\supset
&SU\left( 2\right) _{SO(4)}\times SU(2)\times SU(2)_{H}\times SU(2)^{\prime }
\notag \\
&\supset &SU\left( 2\right) _{SO(4)}\times SU(2)_{I}\times SU(2)^{\prime };
\label{xyz-STU} \\
&&  \notag \\
\left( \mathbf{2,2},\mathbf{2},\mathbf{2,1}\right) &=&\left( \mathbf{3},%
\mathbf{2},\mathbf{2,1}\right) +\left( \mathbf{1},\mathbf{2},\mathbf{2,1}%
\right)  \notag \\
&=&\left( \mathbf{3},\mathbf{3,1}\right) +\left( \mathbf{3},\mathbf{1,1}%
\right) +\left( \mathbf{1},\mathbf{3,1}\right) +\left( \mathbf{1},\mathbf{1,1%
}\right) .  \notag \\
&&  \label{xyz-2-STU}
\end{eqnarray}%
\textbf{1.2]} Then, the $D=5$ massless \textit{spin group} $SU\left(
2\right) _{J}$ is identified with the diagonal $SU(2)_{II}$ into $SU\left(
2\right) _{SO(4)}\times SU\left( 2\right) _{I}$ (\textit{cfr.} (\ref{diag-II}%
)). Indeed, the chain (\ref{xyz-STU})-(\ref{xyz-2-STU}) can be further
completed as follows:%
\begin{eqnarray}
SU(2)\times SU(2)\times SU(2)\times SU(2)_{H}\times SU(2)^{\prime } &\supset
&SU\left( 2\right) _{SO(4)}\times SU(2)\times SU(2)_{H}\times SU(2)^{\prime }
\notag \\
&\supset &SU\left( 2\right) _{SO(4)}\times SU(2)_{I}\times SU(2)^{\prime }
\notag \\
&\supset &SU\left( 2\right) _{J}\times SU(2)^{\prime };  \label{xyzt-STU} \\
&&  \notag \\
\left( \mathbf{2,2},\mathbf{2},\mathbf{2,1}\right) &=&\left( \mathbf{3},%
\mathbf{2},\mathbf{2,1}\right) +\left( \mathbf{1},\mathbf{2},\mathbf{2,1}%
\right)  \notag \\
&=&\left( \mathbf{3},\mathbf{3,1}\right) +\left( \mathbf{3},\mathbf{1,1}%
\right) +\left( \mathbf{1},\mathbf{3,1}\right) +\left( \mathbf{1},\mathbf{1,1%
}\right)  \notag \\
&=&\left( \mathbf{5,1}\right) +\left( \mathbf{3,1}\right) +\left( \mathbf{1,1%
}\right) +\left( \mathbf{3,1}\right) +\left( \mathbf{3,1}\right) +\left(
\mathbf{1,1}\right) .  \notag \\
&&  \label{xyzt-2-STU}
\end{eqnarray}%
The decomposition (\ref{xyzt-2-STU}) corresponds to the massless bosonic
spectrum of the $D=5$ uplift of the $STU$ model ($16$ states): $1$ graviton
and $1$ graviphoton (from the gravity multiplet), $1$ dilatonic vector and $%
1 $ dilaton from the dilatonic vector multiplet, and $1$ vector and $1$
(real) scalar from the non-dilatonic vector multiplet. At the level of
massless spectrum, the action of supersymmetry amounts to the following
exchange of irreps.:
\begin{equation}
SO(n+2)\times SU(2)\times SU(2)_{H}\times SU(2)^{\prime }:\underset{B}{%
\left( \mathbf{2,2},\mathbf{2},\mathbf{2,1}\right) }~\longleftrightarrow ~%
\underset{F}{\left( \mathbf{2,2},\mathbf{2},\mathbf{1,2}\right) }.
\end{equation}%
This can be realized by noticing that, under the chain (\ref{xyzt-STU}) of
maximal symmetric embeddings, $\left( \mathbf{2,2},\mathbf{2},\mathbf{1,2}%
\right) $ decomposes as follows:%
\begin{equation}
\left( \mathbf{2,2},\mathbf{2},\mathbf{1,2}\right) =\left( \mathbf{3},%
\mathbf{2},\mathbf{1,2}\right) +\left( \mathbf{1},\mathbf{2},\mathbf{1,2}%
\right) =\left( \mathbf{3},\mathbf{2,2}\right) +\left( \mathbf{1},\mathbf{2,2%
}\right) =\left( \mathbf{2,2}\right) +\left( \mathbf{4,2}\right) +\left(
\mathbf{2,2}\right) ,
\end{equation}%
thus reproducing the massless fermionic spectrum of the theory ($16$
states): $1$ $SU\left( 2\right) ^{\prime }$-doublet of gravitinos, $1$ $%
SU\left( 2\right) ^{\prime }$-doublet of dilatonic gauginos, and $1$ $%
SU\left( 2\right) ^{\prime }$-doublet of gauginos from the non-dilatonic
vector multiplet. Note that, consistently, bosons are $\mathcal{R}$-symmetry
$SU\left( 2\right) ^{\prime }$-singlets, whereas fermions fit into $SU\left(
2\right) ^{\prime }$-doublets.

\item $SU\left( 2\right) _{J}$, which commutes with $SU(2)^{\prime }$ inside
$SU(2)\times SU(2)\times SU(2)\times SU(2)_{H}\times SU(2)^{\prime }$ (%
\textit{cfr.} (\ref{xyzt-STU})), is the Kostant \textquotedblleft principal"
$SU(2)$ (\ref{principal-1}) into the $D=5$ Ehlers group $SL\left( 3,\mathbb{R%
}\right) $:%
\begin{equation}
SL\left( 3,\mathbb{R}\right) \cap \left[ SU\left( 2\right) _{SO(4)}\times
SU(2)\times SU(2)_{H}\right] =SU\left( 2\right) _{J}.
\end{equation}

\item As a consequence of the chain of maximal symmetric embeddings (\ref%
{ccchain-pre-2-STU}) and (\ref{xyzt-STU}), the following (non-maximal,
non-symmetric) manifold embedding holds:%
\begin{equation}
\frac{SO\left( 4,4\right) }{SU(2)\times SU(2)\times SU(2)\times SU(2)_{H}}%
\supset SO(1,1)\times SO(1,1)\times \frac{SL\left( 3,\mathbb{R}\right) }{%
SU\left( 2\right) _{J}}.
\end{equation}%
This has the trivial interpretation of embedding of the scalar manifold of
the $D=5$ theory into the scalar manifold of the corresponding theory
reduced to $D=3$.

\item By setting $n=2$ in (\ref{pre-pre-qq})-(\ref{qq}), one obtains:%
\begin{eqnarray}
M_{\mathcal{N}=2,\mathfrak{J}_{3}^{2,2}}^{5} &\equiv &\frac{SO\left(
4,4\right) }{SO(1,1)\times SO(1,1)\times SL\left( 3,\mathbb{R}\right) _{%
\text{Ehlers}}}; \\
\widehat{M}_{\mathcal{N}=2,\mathfrak{J}_{3}^{2,2}}^{5} &\equiv &\frac{%
SO\left( 4\right) \times SO(4)}{SU\left( 2\right) _{J}}; \\
c\left( M_{\mathcal{N}=2,\mathfrak{J}_{3}^{2,2}}^{5}\right) &=&nc\left( M_{%
\mathcal{N}=2,\mathfrak{J}_{3}^{2,2}}^{5}\right) =\text{dim}_{\mathbb{R}%
}\left( \widehat{M}_{\mathcal{N}=2,\mathfrak{J}_{3}^{2,2}}^{5}\right) =9,
\label{qq-n=2}
\end{eqnarray}%
consistent with the vanishing of $q_{eff}$ for the STU model (\textit{cfr.} (%
\ref{q-eff-STU})).
\end{enumerate}

\subsection{\texttt{\ }\label{Semi-Simple-N=4}$\mathfrak{J}_{3}^{6,n}\equiv
\mathbb{R}\oplus \mathbf{\Gamma }_{5,n-1}$}

Let us now consider the class $\mathfrak{J}_{3}^{6,n}$ (\ref{N=4}) of
\textit{semi-simple} rank-$3$ Euclidean Jordan algebras. The relevant chain
of embeddings reads:
\begin{eqnarray}
\mathfrak{so}(8,n+2) &\supset &\mathfrak{so}(3,3)\oplus \mathfrak{so}%
(5,n-1)\oplus \mathbf{6}\times \left( \mathbf{n+4}\right)  \notag \\
&\supset &\mathfrak{sl}(3,\mathbb{R})\oplus \mathfrak{so}(5,n-1)\oplus
\mathfrak{so}(1,1)  \notag \\
&&\oplus ~\mathbf{3}\times \left( \left( \mathbf{n+4}\right) _{2}+\mathbf{1}%
_{-4}\right) \oplus \mathbf{3}^{\prime }\times \left( \left( \mathbf{n+4}%
\right) _{-2}+\mathbf{1}_{4}\right) ,  \label{ccchain-N=4}
\end{eqnarray}%
or, at group level:%
\begin{eqnarray}
SO(8,n+2) &\supset &SO(3,3)\times SO(5,n-1)  \notag \\
&\supset &SL(3,\mathbb{R})\times SO(5,n-1)\times SO(1,1);
\label{ccchain-2-pre-N=4} \\
&&  \notag \\
\mathbf{Adj}_{SO(8,n+2)} &=&\mathbf{Adj}_{SO(3,3)}+\mathbf{Adj}%
_{SO(5,n-1)}+\left( \mathbf{6},\mathbf{n+4}\right)  \notag \\
&=&\mathbf{Adj}_{SL(3,\mathbb{R})}+\mathbf{Adj}_{SO(1,1)}+\mathbf{Adj}%
_{SO(5,n-1)}  \notag \\
&&+\left( \mathbf{3},\left( \mathbf{n+4}\right) _{2}+\mathbf{1}_{-4}\right)
+\left( \mathbf{3}^{\prime },\left( \mathbf{n+4}\right) _{-2}+\mathbf{1}%
_{4}\right) ,  \label{ccchain-2-N=4}
\end{eqnarray}%
where the subscripts denote $SO(1,1)$-weights.

According to (\ref{q-eff}) and (\ref{eff-dim}), the \textquotedblleft
effective" dimension of $\mathfrak{J}_{3}^{6,n}$ is $n+5$, and the
corresponding Jordan algebra representation is \textit{reducible} with
respect to $Str_{0}\left( \mathfrak{J}_{3}^{6,n}\right) =SO(1,1)\times
SO(5,n-1)$, as given by (\ref{ccchain-N=4}) and (\ref{ccchain-2-N=4}):%
\begin{equation}
\mathbf{n+5}=\left( \mathbf{n+4}\right) _{2}+\mathbf{1}_{-4}.  \label{mmm}
\end{equation}%
Thus, the $\mathfrak{J}_{3}^{6,n}$-related $\mathcal{N}=4$ (half-maximal), $%
D=5$ supergravity is \textit{not unified}. In (\ref{mmm}) (modulo
redefinitions of the $SO(1,1)$ weights), $\left( \mathbf{n+4}\right) _{2}$
corresponds to the $5$ graviphotons and the $n-1$ matter vectors
(respectively with positive and negative signature in $SO(5,n-1)$), whereas $%
\mathbf{1}_{-4}$ pertains to the $2$-form in the gravity multiplet.

The second line of (\ref{ccchain-N=4}) can be regarded as the extension of (%
\ref{1}) to \textit{semi-simple} rank-$3$ Euclidean Jordan algebras $%
\mathfrak{J}_{3}^{6,n}$ (\ref{N=4}). Its compact counterpart (which
correspondingly generalizes (\ref{2})) reads%
\begin{eqnarray}
\mathfrak{so}(n+10) &\supset &\mathfrak{so}(6)\oplus \mathfrak{so}%
(n+4)\oplus \mathbf{6}\times \left( \mathbf{n+4}\right)  \notag \\
&\supset &\mathfrak{su}(3)\oplus \mathfrak{so}(n+4)\oplus \mathfrak{u}(1)
\notag \\
&&\oplus ~\mathbf{3}\times \left( \left( \mathbf{n+4}\right) _{2}+\mathbf{1}%
_{-4}\right) \oplus \overline{\mathbf{3}}\times \left( \left( \mathbf{n+4}%
\right) _{-2}+\mathbf{1}_{4}\right) ,  \label{ccchain-1-c-N=4}
\end{eqnarray}%
with subscripts here denoting $U(1)$-charges.

At group level, the algebraic decompositions (\ref{ccchain-N=4}) and (\ref%
{ccchain-1-c-N=4}) respectively correspond to the second line of (\ref%
{ccchain-2-pre-N=4}), which can be summarized by the non-maximal
non-symmetric embedding
\begin{equation}
QConf\left( \mathfrak{J}_{3}^{6,n}\right) \supset SL\left( 3,\mathbb{R}%
\right) \times Str_{0}\left( \mathfrak{J}_{3}^{6,n}\right) ,
\end{equation}%
and its compact counterpart:%
\begin{equation}
SO\left( n+10\right) \supset SU(3)\times SO(n+4)\Leftrightarrow
QConf_{c}\left( \mathfrak{J}_{3}^{6,n}\right) \supset SU\left( 3\right)
\times Str_{0,c}\left( \mathfrak{J}_{3}^{6,n}\right) .
\end{equation}%
\bigskip

As mentioned, $\mathfrak{J}_{3}^{6,n}$ pertains to \textit{half-maximal}
supergravity ($16$ supersymmetries): therefore, coupling is allowed to
matter (vector) multiplets, with no $D$-independent hypermultiplet sector.

The \textit{mcs} of $\mathfrak{qconf}\left( \mathfrak{J}_{3}^{6,n}\right) =%
\mathfrak{so}\left( 8,n+2\right) $ and $\mathfrak{str}_{0}\left( \mathfrak{J}%
_{3}^{6,n}\right) =\mathfrak{so}(1,1)\oplus \mathfrak{so}\left( 5,n-1\right)
$ respectively read%
\begin{eqnarray}
mcs\left( \mathfrak{so}\left( 8,n+2\right) \right) &=&\mathfrak{so}(8)\oplus
\mathfrak{so}(n+2);  \label{xxxx-N=4} \\
mcs\left( \mathfrak{so}(1,1)\oplus \mathfrak{so}\left( 5,n-1\right) \right)
&=&\mathfrak{so}(5)\oplus \mathfrak{so}\left( n-1\right) \sim \mathfrak{usp}%
(4)\oplus \mathfrak{so}\left( n-1\right) ,  \label{xxxx-2-N=4}
\end{eqnarray}%
corresponding to the following maximal symmetric embeddings at group level:%
\begin{eqnarray}
SO(8,n+2) &\supset &SO(8)\times SO(n+2);  \label{xxxxx-N=4} \\
\mathbf{Adj}_{SO(8,n+2)} &=&\mathbf{Adj}_{SO(n+2)}+\mathbf{Adj}%
_{SO(8)}+\left( \mathbf{8}_{v},\mathbf{n+2}\right) ;  \notag
\end{eqnarray}%
\begin{eqnarray}
SO(1,1)\times SO(5,n-1) &\supset &SO(5)\times SO(n-1)\sim USp(4)\times
SO(n-1);  \label{xxxxx-2-N=4} \\
\mathbf{1}_{0}+\mathbf{Adj}_{SO(5,n-1),0} &=&\mathbf{1}+\mathbf{Adj}%
_{SO(n-1)}+\mathbf{Adj}_{SO(5)}+\left( \mathbf{5,n-1}\right) ,
\label{xxxxxx-2-N=4}
\end{eqnarray}%
where $\mathbf{8}_{v}$ denotes the vector $\mathbf{8}$ irrep. of $SO(8)$.

From the embedding (\ref{xxxxx-N=4}), $\left( \mathbf{8}_{v},\mathbf{n+2}%
\right) $ is the bi-fundamental irrep. of $SO(8)\times SO(n+2)$, in which
the generators of the symmetric scalar manifold $\frac{SO\left( 8,n+2\right)
}{SO(8)\times SO(n+2)}$ of the $D=3$ half-maximal $\mathfrak{J}_{3}^{6,n}$%
-related supergravity sit. Furthermore, from the embedding (\ref{xxxxx-2-N=4}%
), the generators of the symmetric scalar manifold $SO(1,1)\times \frac{%
SO(5,n-1)}{USp(4)\times SO(n-1)}$ of the corresponding $D=5$ theory sit into
the $\left( \mathbf{1,1}\right) +\left( \mathbf{5,n-1}\right) $ of $%
SO(5)\times SO(n-1)\sim USp(4)\times SO(n-1)$. Thus, the relevant maximal
embedding reads\footnote{%
We use a different convention on the branchings of the $\mathbf{8}$'s of $%
SO(8)$ (with respect \textit{e.g.} to \cite{Slansky}), namely:%
\begin{eqnarray*}
SO(8) &\supset &SO(5)\times SO(3)\sim USp(4)\times SU(2)_{I}; \\
\mathbf{8}_{v} &=&\left( \mathbf{5},\mathbf{1}\right) +(\mathbf{1},\mathbf{3}%
); \\
\mathbf{8}_{s} &=&(\mathbf{4},\mathbf{2}); \\
\mathbf{8}_{c} &=&(\mathbf{4},\mathbf{2}).
\end{eqnarray*}%
This is one of the possible ones allowed by the $SO(8)$ \textit{triality},
and it can be regarded as the \textquotedblleft physical" one, in which the
vector $\mathbf{8}_{v}$ of $SO(8)$ decomposes into the fundamental (vector) $%
\mathbf{5}$ of $SO(5)$.}%
\begin{eqnarray}
\mathfrak{so}(8)\oplus \mathfrak{so}(n+2) &=&\mathfrak{so}(5)\oplus
\mathfrak{so}(n-1)\oplus \mathfrak{su}(2)_{I}\oplus \mathfrak{su}(2)_{II}
\notag \\
&&\oplus ~\mathbf{5\times 1\times 3\times 1}\oplus \mathbf{1}\times \left(
\mathbf{n-1}\right) \times \mathbf{1}\times \mathbf{3;}
\label{embbb-2-x-N=4} \\
&&  \notag \\
SO(8)\times SO(n+2) &\supset &SO(5)\times SO(n-1)\times SU(2)_{I}\times
SU(2)_{II};  \label{embbb-2-xx-N=4} \\
&&  \notag \\
\left( \mathbf{28,1}\right) +\mathbf{Adj}_{SO(n+2)} &=&\left( \mathbf{%
10,1,1,1}\right) +\mathbf{Adj}_{SO(n-1)}  \notag \\
&&+\left( \mathbf{1},\mathbf{1},\mathbf{3,1}\right) +\left( \mathbf{1},%
\mathbf{1},\mathbf{1},\mathbf{3}\right) +(\mathbf{5},\mathbf{1},\mathbf{3},%
\mathbf{1})+\left( \mathbf{1},\mathbf{n-1},\mathbf{1},\mathbf{3}\right) ; \\
&&  \notag \\
\left( \mathbf{8}_{v},\mathbf{n+2}\right) &=&\left( \mathbf{5,n-1,1,1}%
\right) +\left( \mathbf{5,1,1,3}\right) +\left( \mathbf{1,n-1,3,1}\right)
+\left( \mathbf{1,1,3,3}\right) ,
\end{eqnarray}%
where%
\begin{eqnarray}
SO(8) &\supset &SO(5)\times SO(3)\sim USp(4)\times SU(2)_{I};  \label{n-1} \\
\mathbf{8}_{v} &=&(\mathbf{5},\mathbf{1})+(\mathbf{1},\mathbf{3});  \notag \\
\mathbf{28} &=&(\mathbf{10},\mathbf{1})+(\mathbf{1},\mathbf{3})+(\mathbf{5},%
\mathbf{3});  \notag
\end{eqnarray}%
\begin{eqnarray}
SO(n+2) &\supset &SO(n-1)\times SO(3)\sim SO(n-1)\times SU(2)_{II};
\label{n-2} \\
\mathbf{n+2} &=&\left( \mathbf{n-1},\mathbf{1}\right) +(\mathbf{1},\mathbf{3}%
);  \notag \\
\mathbf{Adj}_{SO(n+2)} &=&\mathbf{Adj}_{SO(n-1)}+(\mathbf{1},\mathbf{3})+(%
\mathbf{n-1},\mathbf{3}).  \notag
\end{eqnarray}%
Note that, differently from the analogous formula for \textit{simple} rank-$%
3 $ Euclidean Jordan algebras, the maximal embedding (\ref{embbb-2-x-N=4})-(%
\ref{embbb-2-xx-N=4}) is symmetric, as (\ref{n-1}) and (\ref{n-2}) are.

As consistently yielded by (\ref{xxxxx-N=4}) and (\ref{xxxxx-2-N=4}), the $%
\mathcal{R}$-symmetry of $\mathcal{N}=8$, $D=3$ $\mathfrak{J}_{3}^{6,n}$%
-related half-maximal supergravity is $SO(8)$, whereas $SO(5)\sim USp\left(
4\right) $ is the $\mathcal{R}$-symmetry of the theory uplifted to $D=5$.

\begin{enumerate}
\item Differently from the semi-simple class $\mathfrak{J}_{3}^{2,n}$
treated above, the massless $D=5$ \textit{spin group} $SU\left( 2\right)
_{J} $ can be identified by a \textit{one-step} procedure, with the diagonal
$SU(2)$ into $SU(2)_{I}\times SU(2)_{II}$:%
\begin{equation}
SU\left( 2\right) _{J}\subset _{d}SU\left( 2\right) _{I}\times SU\left(
2\right) _{II}.
\end{equation}%
Thus, (\ref{embbb-2-xx-N=4}) can be completed to the following chain:%
\begin{eqnarray}
SO(8)\times SO(n+2) &\supset &SO(5)\times SO(n-1)\times SU(2)_{I}\times
SU(2)_{II}  \notag \\
&\supset &SO(5)\times SO(n-1)\times SU(2)_{J};  \label{xyz-N=4} \\
&&  \notag \\
\left( \mathbf{8}_{v},\mathbf{n+2}\right) &=&\left( \mathbf{5,n-1,1,1}%
\right) +\left( \mathbf{5,1,1,3}\right) +\left( \mathbf{1,n-1,3,1}\right)
+\left( \mathbf{1,1,3,3}\right)  \notag \\
&=&\left( \mathbf{5,n-1,1}\right) +\left( \mathbf{5,1,3}\right) +\left(
\mathbf{1,n-1,3}\right)  \notag \\
&&+\left( \mathbf{1,1,5}\right) +\left( \mathbf{1,1,3}\right) +\left(
\mathbf{1,1,1}\right) .  \label{xyz-2-N=4}
\end{eqnarray}%
Indeed, the decomposition (\ref{xyz-2-N=4}) corresponds to the massless
bosonic spectrum of $\mathcal{N}=4$, $D=5$ $\mathfrak{J}_{3}^{4,n}$-related
half-maximal supergravity ($8\left( n+2\right) $ states): $1$ graviton, $1$ $%
2$-form, $5$ graviphotons and $1$ real scalar from the gravity multiplet,
and $5\left( n-1\right) $ scalars and $n-1$ vectors from the $n-1$ matter
(vector) multiplets. At the level of massless spectrum, the action of
supersymmetry amounts to the following exchange of irreps.:
\begin{equation}
SO(8)\times SO(n+2):\underset{B}{\left( \mathbf{8}_{v},\mathbf{n+2}\right) }%
~\longleftrightarrow ~\underset{F}{\left( \mathbf{8}_{s},\mathbf{n+2}\right)
},
\end{equation}%
where $\mathbf{8}_{s}$ is the chiral spinor\footnote{%
Instead of $\mathbf{8}_{s}$, the conjugated chiral spinor $\mathbf{8}_{c}$
can equivalently be chosen, as well.} irrep. of $SO(8)$. This can be
realized by noticing that, under the chain (\ref{xyz-2-N=4}) of maximal
symmetric embeddings, $\left( \mathbf{8}_{s},\mathbf{n+2}\right) $
decomposes as follows:%
\begin{equation}
\left( \mathbf{8}_{s},\mathbf{n+2}\right) =\left( \mathbf{4,n-1,2,1}\right)
+\left( \mathbf{4,1,2,3}\right) =\left( \mathbf{4,n-1,2}\right) +\left(
\mathbf{4,1,4}\right) +\left( \mathbf{4,1,2}\right) ,
\end{equation}%
thus reproducing the massless fermionic spectrum of the theory ($8(n+2)$
states): $4$ gravitinos and $4$ spin $1/2$ fermions (from the gravity
multiplet), and $4\left( n-1\right) $ gauginos from the $n-1$ matter
(vector) multiplets.

\item $SU\left( 2\right) _{J}$, which commutes with $USp(4)\times SO(n-1)$
inside $SO(8)\times SO(n+2)$ (\textit{cfr.} (\ref{xyz-N=4})), is the Kostant
\textquotedblleft principal" $SU(2)$ (\ref{principal-1}) into the $D=5$
Ehlers group $SL\left( 3,\mathbb{R}\right) $:%
\begin{equation}
SL\left( 3,\mathbb{R}\right) \cap \left[ SU(2)_{I}\times SU(2)_{II}\right]
=SU\left( 2\right) _{J}.
\end{equation}

\item As a consequence of the chain of maximal symmetric embeddings (\ref%
{ccchain-2-pre-N=4}) and (\ref{xyz-N=4}), the following (non-maximal,
non-symmetric) manifold embedding holds:%
\begin{equation}
\frac{SO\left( 8,n+2\right) }{SO(8)\times SO(n+2)}\supset SO(1,1)\times
\frac{SO(5,n-1)}{USp(4)\times SO(n-1)}\times \frac{SL\left( 3,\mathbb{R}%
\right) }{SU\left( 2\right) _{J}}.  \label{zz-N=4}
\end{equation}%
As usual, this has the trivial interpretation of embedding of the scalar
manifold of the $D=5$ theory into the scalar manifold of the corresponding
theory reduced to $D=3$.

\item As resulting from the above treatment, and analogously to the
\textquotedblleft $8$ \textit{versus} $24$ supersymmetries" interpretation
of $\mathfrak{J}_{3}^{\mathbb{H}}$ discussed in Sec. \ref{q=4-J3-H-Sec}, the
main difference between the semi-simple classes $\mathfrak{J}_{3}^{6,n}$ (%
\ref{N=4}) and $\mathfrak{J}_{3}^{2,n}$ (\ref{N=2}) resides in the $D$%
-independent hypersector. In the former case, pertaining to \textit{%
half-maximal} ($16$ supersymmetries) supergravity, such a sector is
forbidden by supersymmetry; in the latter case, pertaining to \textit{minimal%
} ($8$ supersymmetries) supergravity, \textit{such a sector must be present
for physical consistency}; as mentioned above, this hypersector is
insensitive to dimensional reduction, and it is thus independent on the
number $D=3,4,5,6$ of space-time dimensions in which the theory with $8$
supersymmetries is defined. The very same comments made at point 4 of
Subsec. \ref{J3-O-Subsec} also hold in this case, with (\ref{c-map})
replaced by%
\begin{equation}
\underset{D=4}{SL(2,\mathbb{R})\times \frac{SO\left( 6,n\right) }{%
SO(6)\times SO(n)}}\overset{c_{hm}}{\longrightarrow }\underset{D=3}{\frac{%
SO(8,n+2)}{SO(8)\times SO(n+2)}},
\end{equation}%
where the coset on the l.h.s. is the vector multiplets' scalar manifold of
the $D=4$ half-maximal theory; it is symmetric, as is its image $\frac{%
SO(8,n+2)}{SO(8)\times SO(n+2)}$ through the \textit{\textquotedblleft
half-maximal"} analogue $c_{hm}$ of $c$-map.

\item In $\mathcal{N}=4$, $D=5$ $\mathfrak{J}_{3}^{6,n}$-related
supergravity, (\ref{M})-(\ref{c=nc}) respectively specify to\footnote{%
With respect to the parameter $m$ used in \cite{super-Ehlers-1} (see \textit{%
e.g.} Table 12 therein), we define $n=m-1$.} \cite{super-Ehlers-1}%
\begin{eqnarray}
M_{\mathcal{N}=4,\mathfrak{J}_{3}^{6,n}}^{5} &\equiv &\frac{SO\left(
8,n+2\right) }{SO(1,1)\times SO(5,n-1)\times SL\left( 3,\mathbb{R}\right) _{%
\text{Ehlers}}};  \label{pre-pre-qq-2} \\
\widehat{M}_{\mathcal{N}=4,\mathfrak{J}_{3}^{6,n}}^{5} &\equiv &\frac{%
SO\left( 8\right) \times SO(n+2)}{SO(5)\times SO(n-1)\times SU\left(
2\right) _{J}};  \label{pre-qq-2} \\
c\left( M_{\mathcal{N}=4,\mathfrak{J}_{3}^{6,n}}^{5}\right) &=&nc\left( M_{%
\mathcal{N}=4,\mathfrak{J}_{3}^{6,n}}^{5}\right) =\text{dim}_{\mathbb{R}%
}\left( \widehat{M}_{\mathcal{N}=4,\mathfrak{J}_{3}^{6,n}}^{5}\right) =3n+15,
\label{qq-2}
\end{eqnarray}%
where in (\ref{qq-2}) the definition (\ref{q-eff}) has been recalled.
\end{enumerate}

\section{\label{Semi-Simple-Twin}$\mathfrak{J}_{3}^{2,6}\sim \mathfrak{J}%
_{3}^{6,2}$ \textquotedblleft Twin" Theories}

As pointed out above, the presence or absence of a $D$-independent
hypersector is implied by the physical (supergravity) interpretation of the
model under consideration. In "(bosonic) twin" theories, sharing the very
same bosonic sector, the $D$-independent hypersector can or cannot be
considered, and in both cases the resulting supergravity theory (of course
with different supersymmetry properties) is physically meaningful.

Besides the case of $\mathfrak{J}_{3}^{\mathbb{H}}$-related "(bosonic) twin"
theories, treated in Sec. \ref{q=4-J3-H-Sec}, another example is provided by
the semi-simple rank-$3$ Euclidean Jordan algebra\footnote{%
After the treatment of \cite{Gnecchi-1} (see also Refs. therein), the cases
of $\mathfrak{J}_{3}^{\mathbb{H}}$ and $\mathfrak{J}_{3}^{2,6}\sim \mathfrak{%
J}_{3}^{6,2}$ are the unique cases of "(bosonic) twin" with symmetric scalar
manifolds and with an interpretation in terms of rank-$3$ Euclidean Jordan
algebras.}%
\begin{equation}
\mathfrak{J}_{3}^{2,6}\sim \mathfrak{J}_{3}^{6,2},
\end{equation}%
given by the element $n=6$ of the class $\mathfrak{J}_{3}^{2,n}$ (\ref{N=2}):%
\begin{equation}
8~\text{susys}:\mathfrak{J}_{3}^{2,6}\equiv \mathbb{R}\oplus \mathbf{\Gamma }%
_{1,5},
\end{equation}%
or by the element $n=6$ of the class $\mathfrak{J}_{3}^{6,n}$ (\ref{N=4}):%
\begin{equation}
16~\text{susys}:\mathfrak{J}_{3}^{6,2}\equiv \mathbb{R}\oplus \mathbf{\Gamma
}_{5,1}.
\end{equation}

The relevant symmetries read%
\begin{eqnarray}
\mathfrak{L}\left( \mathfrak{J}_{3}^{2,6}\right) &\equiv &\mathfrak{qconf}%
\left( \mathfrak{J}_{3}^{2,6}\right) =\mathfrak{so}\left( 4,8\right) ;
\label{gr-1-x} \\
\mathfrak{conf}\left( \mathfrak{J}_{3}^{2,6}\right) &=&\mathfrak{aut}\left(
\mathfrak{F}\left( \mathfrak{J}_{3}^{2,6}\right) \right) =\mathfrak{sl}(2,%
\mathbb{R})\oplus \mathfrak{so}\left( 2,6\right) ;  \label{gr-2-x} \\
\mathfrak{str}_{0}\left( \mathfrak{J}_{3}^{2,6}\right) &=&\mathfrak{so}%
(1,1)\oplus \mathfrak{so}\left( 1,5\right) =\mathfrak{so}(1,1)\oplus
\mathfrak{str}_{0}\left( \mathbf{\Gamma }_{1,5}\right) ,  \label{gr-3-x}
\end{eqnarray}%
where $\mathfrak{F}\left( \mathfrak{J}_{3}^{2,6}\right) $ denotes the
\textit{Freudenthal triple system} constructed over $\mathfrak{J}_{3}^{2,6}$%
. By recalling the definition (\ref{q-eff}), the corresponding
\textquotedblleft effective" parameter reads%
\begin{equation}
q_{eff}\left( 2,6\right) =\frac{4}{3},
\end{equation}%
such that the \textquotedblleft effective" dimension of $\mathfrak{J}%
_{3}^{2,6}$ is $7$.

\subsection{\label{Semi-Simple-Twin-N=4}$\mathfrak{J}_{3}^{6,2}$, $16$
Supersymmetries}

Let us start by considering $\mathfrak{J}_{3}^{6,2}$.

Consistently with Subsec. \ref{Semi-Simple-N=4}, it is associated to a
theory with $16$ supersymmetries, namely $\mathcal{N}=4$, $D=5$ half-maximal
supergravity coupled to $1$ matter (vector) multiplet, whose reduction to $%
D=3$ yields $\mathcal{N}=8$ supergravity coupled to $4$ matter multiplets.
No $D$-independent hypersector is allowed.

In this case, the relevant chain of embeddings reads ($\mathfrak{so}%
(3,3)\sim \mathfrak{sl}(4,\mathbb{R})$; $\mathfrak{so}(5,1)\sim \mathfrak{su}%
^{\ast }(4)$):
\begin{eqnarray}
\mathfrak{so}(8,4) &\supset &\mathfrak{so}(3,3)\oplus \mathfrak{so}%
(5,1)\oplus \mathbf{6}\times \mathbf{6}  \notag \\
&\supset &\mathfrak{sl}(3,\mathbb{R})\oplus \mathfrak{so}(5,1)\oplus
\mathfrak{so}(1,1)\oplus \mathbf{3}\times \left( \mathbf{6}_{2}+\mathbf{1}%
_{-4}\right) \oplus \mathbf{3}^{\prime }\times \left( \mathbf{6}_{-2}+%
\mathbf{1}_{4}\right) ,  \label{one}
\end{eqnarray}%
or, at group level:%
\begin{eqnarray}
SO(8,4) &\supset &SO(3,3)\times SO(5,1)  \notag \\
&\supset &SL(3,\mathbb{R})\times SO(5,1)\times SO(1,1);  \label{one-two} \\
&&  \notag \\
\mathbf{66} &=&\left( \mathbf{15,1}\right) +\left( \mathbf{1},\mathbf{15}%
\right) +\left( \mathbf{6},\mathbf{6}\right)  \notag \\
&=&\left( \mathbf{8},\mathbf{1}\right) _{0}+\left( \mathbf{1},\mathbf{1}%
\right) _{0}+\left( \mathbf{1},\mathbf{15}\right) _{0}+\left( \mathbf{3},%
\mathbf{6}_{2}+\mathbf{1}_{-4}\right) +\left( \mathbf{3}^{\prime },\mathbf{6}%
_{-2}+\mathbf{1}_{4}\right) ,  \label{two}
\end{eqnarray}%
where the subscripts denote $SO(1,1)$-weights.

The \textquotedblleft effective" dimension $7$ of $\mathfrak{J}_{3}^{6,2}$
is \textit{reducible} with respect to $Str_{0}\left( \mathfrak{J}%
_{3}^{6,2}\right) =SO(1,1)\times SO(5,1)\sim SO(1,1)\times SU^{\ast }\left(
4\right) $, as given by (\ref{one}) and (\ref{two}):%
\begin{equation}
\mathbf{7}=\mathbf{6}_{2}+\mathbf{1}_{-4},  \label{7}
\end{equation}%
yielding that the $\mathfrak{J}_{3}^{6,2}$-related $\mathcal{N}=4$, $D=5$
theory is \textit{non-unified}. In (\ref{7}) (modulo redefinitions of the $%
SO(1,1)$ weights), $\mathbf{6}_{2}$ corresponds to the $5$ graviphotons and
the unique matter vector (respectively with positive and negative signature
in $SO(5,1)$), whereas $\mathbf{1}_{-4}$ pertains to the $2$-form in the
gravity multiplet.

The \textit{mcs} of $\mathfrak{qconf}\left( \mathfrak{J}_{3}^{6,2}\right) =%
\mathfrak{so}\left( 8,4\right) $ and of $\mathfrak{str}_{0}\left( \mathfrak{J%
}_{3}^{6,2}\right) =\mathfrak{so}(1,1)\oplus \mathfrak{so}\left( 5,1\right) $
respectively read%
\begin{eqnarray}
mcs\left( \mathfrak{so}\left( 8,4\right) \right) &=&\mathfrak{so}(8)\oplus
\mathfrak{so}(4)\sim \mathfrak{so}(8)\oplus \mathfrak{su}(2)\times \mathfrak{%
su}(2)_{(H)};  \label{xxxx-x} \\
mcs\left( \mathfrak{so}(1,1)\oplus \mathfrak{so}\left( 5,1\right) \right) &=&%
\mathfrak{so}(5)\sim \mathfrak{usp}(4),  \label{xxxx-2-x}
\end{eqnarray}%
corresponding to the following maximal symmetric embeddings at group level:%
\begin{eqnarray}
SO(8,4) &\supset &SO(8)\times SO(4)\sim SO(8)\times SU(2)\times SU(2)_{(H)};
\label{xxxxx-x} \\
\mathbf{66} &=&\left( \mathbf{28,1}\right) +\left( \mathbf{1,6}\right)
+\left( \mathbf{8}_{v},\mathbf{4}\right)  \notag \\
&=&\left( \mathbf{28,1,1}\right) +\left( \mathbf{1,3,1}\right) +(\mathbf{1},%
\mathbf{1},\mathbf{3})+\left( \mathbf{8}_{v},\mathbf{2,2}\right) ;
\end{eqnarray}%
\begin{eqnarray}
SO(1,1)\times SO(5,1) &\supset &SO(5)\sim USp(4);  \label{xxxxx-2-x} \\
\mathbf{1}_{0}+\mathbf{15}_{0} &=&\mathbf{1}+\mathbf{10}+\mathbf{5}.
\label{xxxxxx-2-x}
\end{eqnarray}%
As for $\mathfrak{J}_{3}^{\mathbb{H}}$ treated in Subsec. \ref{24-susys},
the subscript \textquotedblleft $(H)$" denotes the fact that $SU(2)_{(H)}$
actually is the quaternionic $SU(2)$ connection in the physical
interpretation pertaining to $8$ local supersymmetries (see below).

From the embedding (\ref{xxxxx-x}), $\left( \mathbf{8}_{v},\mathbf{2,2}%
\right) $ is the tri-fundamental irrep. of $SO(8)\times SU(2)\times
SU(2)_{(H)}$, in which the generators of the symmetric scalar manifold $%
\frac{SO\left( 8,4\right) }{SO(8)\times SU(2)\times SU(2)_{(H)}}$ of $%
\mathcal{N}=8$, $D=3$ $\mathfrak{J}_{3}^{6,2}$-related supergravity sit. On
the other hand, (\ref{xxxxx-2-x}) yields that $\mathbf{1+5}$ is the
representation of $SO(5)$ in which the generators of the symmetric scalar
manifold $SO(1,1)\times \frac{SO(5,1)}{SO(5)}$ of the corresponding $D=5$
theory sit. Thus, the relevant maximal embedding reads%
\begin{eqnarray}
\mathfrak{so}(8)\oplus \mathfrak{so}(4) &\sim &\mathfrak{so}(8)\oplus
\mathfrak{su}(2)\oplus \mathfrak{su}(2)_{(H)}=\mathfrak{so}(5)\oplus
\mathfrak{su}(2)_{I}\oplus \mathfrak{su}(2)_{II}  \notag \\
&&\oplus ~\mathbf{5\times 3\times 1}\oplus \mathbf{1}\times \mathbf{1}\times
\mathbf{3;}  \label{embbb-2-x-x} \\
&&  \notag \\
SO(8)\times SO(4) &\sim &SO(8)\times SU(2)\times SU(2)_{(H)}  \notag \\
&\supset &SO(5)\times SU(2)_{I}\times SU(2)_{II};  \label{embbb-2-xx-x} \\
&&  \notag \\
\left( \mathbf{28,1,1}\right) +\left( \mathbf{1},\mathbf{3,1}\right) +\left(
\mathbf{1},\mathbf{1,3}\right) &=&\left( \mathbf{10,1,1}\right) +\left(
\mathbf{1},\mathbf{3,1}\right) +\left( \mathbf{1},\mathbf{1},\mathbf{3}%
\right) +(\mathbf{5},\mathbf{3},\mathbf{1})+\left( \mathbf{1},\mathbf{1},%
\mathbf{3}\right) ; \\
&&  \notag \\
\left( \mathbf{8}_{v},\mathbf{2,2}\right) &=&\left( \mathbf{5,1,1}\right)
+\left( \mathbf{5,1,3}\right) +\left( \mathbf{1,3,1}\right) +\left( \mathbf{%
1,3,3}\right) ,
\end{eqnarray}%
where (\ref{n-1}) holds, and $SU(2)_{II}$ is diagonal into $SU(2)\times
SU(2)_{(H)}$ (\textit{cfr.} (\ref{j}))
\begin{eqnarray}
SO(4) &\sim &SU(2)\times SU(2)_{(H)}\supset SO(3)\sim SU(2)_{II};
\label{n-2-x} \\
\mathbf{(2,2)} &=&\mathbf{3}+\mathbf{1};~~\left( \mathbf{3,1}\right) +\left(
\mathbf{1},\mathbf{3}\right) =\mathbf{3}+\mathbf{3}.  \notag
\end{eqnarray}%
Note that, differently from the analogous formula for \textit{simple} rank-$%
3 $ Euclidean Jordan algebras, the maximal embedding (\ref{embbb-2-x-x})-(%
\ref{embbb-2-xx-x}) is symmetric, as (\ref{n-1}) and (\ref{n-2-x}) are.

As consistently yielded by (\ref{xxxxx-x}) and (\ref{xxxxx-2-x}), the $%
\mathcal{R}$-symmetry of $\mathcal{N}=8$, $D=3$ $\mathfrak{J}_{3}^{6,2}$%
-related supergravity is $SO(8)$, whereas $SO(5)\sim USp\left( 4\right) $ is
the $\mathcal{R}$-symmetry of the same theory uplifted to $D=(5)$.

Some remarks are in order.\medskip

\begin{enumerate}
\item As pointed out in the analysis of $\mathfrak{J}_{3}^{6,n}$ in Subsec. %
\ref{Semi-Simple-N=4}, the massless $D=5$ \textit{spin group} $SU\left(
2\right) _{J}$ can be identified, by a \textit{one-step} procedure, with the
diagonal $SU(2)$ into $SU(2)_{I}\times SU(2)_{II}$:%
\begin{equation}
SU\left( 2\right) _{J}\subset _{d}SU\left( 2\right) _{I}\times SU\left(
2\right) _{II}.
\end{equation}%
Thus, (\ref{embbb-2-xx-x}) can be completed to the following chain:%
\begin{eqnarray}
SO(8)\times SO(4) &\sim &SO(8)\times SU(2)\times SU(2)_{(H)}\supset
SO(5)\times SU(2)_{I}\times SU(2)_{II}  \notag \\
&\supset &SO(5)\times SU(2)_{J};  \label{xyz-N=4-x} \\
&&  \notag \\
\left( \mathbf{8}_{v},\mathbf{4}\right) &=&\left( \mathbf{8}_{v},\mathbf{2,2}%
\right) =\left( \mathbf{5,1,1}\right) +\left( \mathbf{5,1,3}\right) +\left(
\mathbf{1,3,1}\right) +\left( \mathbf{1,3,3}\right)  \notag \\
&=&\left( \mathbf{5,1}\right) +\left( \mathbf{5,3}\right) +\left( \mathbf{1,3%
}\right) +\left( \mathbf{1,5}\right) +\left( \mathbf{1,3}\right) +\left(
\mathbf{1,1}\right) .  \label{xyz-N=4-2-x}
\end{eqnarray}%
Indeed, the decomposition (\ref{xyz-N=4-2-x}) corresponds to the massless
bosonic spectrum of $\mathcal{N}=4$, $D=5$ $\mathfrak{J}_{3}^{4,2}$-related
half-maximal supergravity ($32$ states): $1$ graviton, $1$ $2$-form, $5$
graviphotons and $1$ real scalar from the gravity multiplet, and $5$ real
scalars and $1$ vector from the unique matter (vector) multiplet. At the
level of massless spectrum, the action of supersymmetry amounts to the
following exchange of irreps.:
\begin{equation}
SO(8)\times SU(2)\times SU(2)_{(H)}:\underset{B}{\left( \mathbf{8}_{v},%
\mathbf{2,2}\right) }~\longleftrightarrow ~\underset{F}{\left( \mathbf{8}%
_{s},\mathbf{2,2}\right) },  \label{c}
\end{equation}%
where the the conjugated chiral spinor $\mathbf{8}_{c}$ can be equivalently
considered in place of $\mathbf{8}_{s}$, as well. This can be realized by
noticing that, under the chain (\ref{xyz-N=4-2-x}) of maximal symmetric
embeddings, $\left( \mathbf{8}_{s},\mathbf{2,2}\right) $ decomposes as
follows:%
\begin{equation}
\left( \mathbf{8}_{s},\mathbf{2,2}\right) =\left( \mathbf{4,2,1}\right)
+\left( \mathbf{4,2,3}\right) =\left( \mathbf{4,2}\right) +\left( \mathbf{4,4%
}\right) +\left( \mathbf{4,2}\right) ,
\end{equation}%
thus reproducing the massless fermionic spectrum of the theory ($32$
states): $4$ gravitinos and $4$ spin $1/2$ fermions (from gravity
multiplet), and $4$ gauginos from the unique matter (vector) multiplet.

\item $SU\left( 2\right) _{J}$, which commutes with $USp(4)$ inside $%
SO(8)\times SU(2)\times SU(2)_{(H)}$ (\textit{cfr.} (\ref{xyz-N=4-x})), is
the Kostant \textquotedblleft principal" $SU(2)$ (\ref{principal-1}) into
the $D=5$ Ehlers group $SL\left( 3,\mathbb{R}\right) $:%
\begin{equation}
SL\left( 3,\mathbb{R}\right) \cap \left[ SU(2)_{I}\times SU(2)_{II}\right]
=SU\left( 2\right) _{J}.
\end{equation}

\item As a consequence of the chain of maximal symmetric embeddings (\ref%
{one-two}) and (\ref{xyz-N=4-x}), the following (non-maximal, non-symmetric)
manifold embedding holds:%
\begin{equation}
\frac{SO\left( 8,4\right) }{SO(8)\times SU(2)\times SU(2)_{(H)}}\supset
SO(1,1)\times \frac{SO(5,1)}{SO(5)}\times \frac{SL\left( 3,\mathbb{R}\right)
}{SU\left( 2\right) _{J}}\sim SO(1,1)\times \frac{SU^{\ast }\left( 4\right)
}{USp(4)}\times \frac{SL\left( 3,\mathbb{R}\right) }{SU\left( 2\right) _{J}}.
\label{zz-N=4-x}
\end{equation}%
As usual, this has the trivial interpretation of embedding of the scalar
manifold of the $D=5$ theory into the scalar manifold of the corresponding
theory reduced to $D=3$.

\item In $\mathcal{N}=4$, $D=5$ $\mathfrak{J}_{3}^{6,2}$-related
supergravity, (\ref{pre-pre-qq-2})-(\ref{qq-2}) respectively specify to%
\begin{eqnarray}
M_{\mathcal{N}=4,\mathfrak{J}_{3}^{6,2}}^{5} &\equiv &\frac{SO\left(
8,4\right) }{SO(1,1)\times SO(5,1)\times SL\left( 3,\mathbb{R}\right) _{%
\text{Ehlers}}};  \label{pre-pre-qq-3} \\
\widehat{M}_{\mathcal{N}=4,\mathfrak{J}_{3}^{6,2}}^{5} &\equiv &\frac{%
SO\left( 8\right) \times SO(4)}{SO(5)\times SU\left( 2\right) _{J}};
\label{pre-qq-3} \\
c\left( M_{\mathcal{N}=4,\mathfrak{J}_{3}^{6,2}}^{5}\right) &=&nc\left( M_{%
\mathcal{N}=4,\mathfrak{J}_{3}^{6,2}}^{5}\right) =\text{dim}_{\mathbb{R}%
}\left( \widehat{M}_{\mathcal{N}=4,\mathfrak{J}_{3}^{6,2}}^{5}\right) =21.
\label{qq-3}
\end{eqnarray}
\end{enumerate}

\subsection{\label{Semi-Simple-Twin-N=2}$\mathfrak{J}_{3}^{2,6}$, $8$
Supersymmetries}

Let us now consider $\mathfrak{J}_{3}^{2,6}$.

Consistently with Subsec. \ref{Semi-Simple-N=2}, it is associated to a
supergravity model with $8$ supersymmetries, namely the $\mathfrak{J}%
_{3}^{2,6}$-based $\mathcal{N}=2$, $D=5$ supergravity coupled to $6$ vector
multiplets, and its dimensional reduction down to $D=3$, which is coupled to
$8$ matter multiplets. For physical consistency, a $D$-independent
hypermultiplet sector must be considered.

Clearly, the chains of embeddings (\ref{one}), (\ref{one-two}) and (\ref{two}%
) also hold in this case, along with the considerations on the reducibility
of the representation of $\mathfrak{J}_{3}^{2,6}$ with respect to $%
Str_{0}\left( \mathfrak{J}_{3}^{2,6}\right) =SO(1,1)\times SO(1,5)$ (\textit{%
cfr.} (\ref{7})). Furthermore, the embeddings (\ref{xxxx-x})-(\ref%
{xxxxxx-2-x}) also hold, with the brackets removed in the subscript
\textquotedblleft $(H)$".

(\ref{7}) is still true, but with a different interpretation, namely: the $%
\mathfrak{J}_{3}^{2,6}$-related $\mathcal{N}=2$, $D=5$ theory is \textit{%
non-unified}, with $\mathbf{6}_{2}$ corresponding to the graviphoton and the
vectors from the $5$ non-dilatonic vector multiplets (respectively with
positive and negative signature in $SO(1,5)$; notice the consistent flip of
signs with respect to the \textquotedblleft twin" $\mathfrak{J}_{3}^{6,2}$%
-related theory), whereas $\mathbf{1}_{-4}$ pertains to the vector from the
dilatonic vector multiplet.

However, the relevant maximal embedding must include the $\mathfrak{su}%
(2)^{\prime }$ algebra from the $D$-independent hypersector, as well:%
\begin{eqnarray}
\mathfrak{so}(8)\oplus \mathfrak{so}(4)\oplus \mathfrak{su}(2)^{\prime }
&\sim &\mathfrak{so}(8)\oplus \mathfrak{su}(2)\oplus \mathfrak{su}%
(2)_{H}\oplus \mathfrak{su}(2)^{\prime }  \notag \\
&=&\mathfrak{so}(5)\oplus \mathfrak{su}(2)_{\mathfrak{so}(8)}\oplus
\mathfrak{su}(2)\oplus \mathfrak{su}(2)_{H}\oplus \mathfrak{su}(2)^{\prime }
\notag \\
&&\oplus ~\mathbf{5}\times \mathbf{3}\times \mathbf{1}\times \mathbf{1\times
\mathbf{1};}  \label{embbb-2-x-y} \\
&&  \notag \\
SO(8)\times SO(4)\times SU(2)^{\prime } &\sim &SO(8)\times SU(2)\times
SU(2)_{H}\times SU(2)^{\prime }  \notag \\
&\supset &SO(5)\times SU\left( 2\right) _{SO(8)}\times SU(2)\times
SU(2)_{H}\times SU(2)^{\prime };  \label{embbb-2-xx-y} \\
&&  \notag \\
\left( \mathbf{28,1,1}\right) +\left( \mathbf{1},\mathbf{6,1}\right) +\left(
\mathbf{1},\mathbf{1},\mathbf{3}\right) &=&\left( \mathbf{28,1,1,1}\right)
+\left( \mathbf{1},\mathbf{3,1,1}\right) +\left( \mathbf{1},\mathbf{1,3,1}%
\right) +\left( \mathbf{1},\mathbf{1},\mathbf{1},\mathbf{3}\right)  \notag \\
&=&\left( \mathbf{10,1,1,1,1}\right) +\left( \mathbf{1,3,1,1,1}\right)
\notag \\
&&+\left( \mathbf{1},\mathbf{1},\mathbf{3,1,1}\right) +\left( \mathbf{1},%
\mathbf{1},\mathbf{1,3,1}\right)  \notag \\
&&+\left( \mathbf{1},\mathbf{1},\mathbf{1},\mathbf{1},\mathbf{3}\right) +(%
\mathbf{5},\mathbf{3},\mathbf{1},\mathbf{1},\mathbf{1}).
\end{eqnarray}%
As above, by $SU\left( 2\right) _{SO(8)}$ we denote the group commuting with
$SO(5)$ in the maximal symmetric embedding%
\begin{equation}
SO(8)\supset SO(5)\times SO(3)\sim USp(4)\times SU(2)_{SO(8)},
\end{equation}%
determining (\ref{embbb-2-x-y}). Note that, differently from the analogous
formula for \textit{simple} rank-$3$ Euclidean Jordan algebras, the maximal
embedding (\ref{embbb-2-x-y})-(\ref{embbb-2-xx-y}) is symmetric.

Note that the $\mathcal{R}$-symmetry of $\mathcal{N}=4$, $D=3$ $\mathfrak{J}%
_{3}^{2,6}$-related supergravity is consistently enhanced from the
quaternionic $SU(2)_{H}$ (related to the $c$-map of the $D=4$ vector
multiplets' scalar manifold) to $SU(2)_{H}\times SU(2)^{\prime }$, as given
by (\ref{SO(4)}). On the other hand, $SU(2)^{\prime }\sim USp\left( 2\right)
$ is the $\mathcal{R}$-symmetry of the theory in $D=5$.

Clearly, it holds that%
\begin{eqnarray}
\mathfrak{su}(2)^{\prime }\cap \mathfrak{so}\left( 4,8\right) &=&\emptyset
\Rightarrow \mathfrak{su}(2)^{\prime }\cap \mathfrak{su}(2)_{J}=\emptyset ;~%
\mathfrak{su}(2)^{\prime }\cap \mathfrak{sl}(3,\mathbb{R})=\emptyset ; \\
\mathfrak{su}(2)_{\mathfrak{so}\left( 8\right) }\oplus \mathfrak{su}(2)_{H}
&\nsubseteq &\mathfrak{sl}(3,\mathbb{R}),
\end{eqnarray}%
where, as in general, the $D=5$ Ehlers Lie algebra $\mathfrak{sl}(3,\mathbb{R%
})$ admits the massless spin algebra $\mathfrak{su}(2)_{J}$ as maximal
compact subalgebra.

From the embeddings considered above, $\left( \mathbf{8}_{v},\mathbf{2},%
\mathbf{2}\right) $ is the tri-fundamental irrep. of $SO(8)\times
SU(2)\times SU(2)_{H}$, in which the generators of the rank-$4$ symmetric
quaternionic scalar manifold $\frac{SO\left( 4,8\right) }{SO(8)\times
SU(2)\times SU(2)_{H}}$ of $\mathcal{N}=4$, $D=3$ $\mathfrak{J}_{3}^{2,6}$%
-related supergravity sit. Furthermore, from (\ref{xxxxxx-2-x}) $\mathbf{1}+%
\mathbf{5}$ is the representation of $SO(5)$ in which the generators of the
rank-$2$ symmetric real special scalar manifold $SO(1,1)\times \frac{SO(1,5)%
}{SO(5)}$ of the $D=5$ theory sit. Thus, under (\ref{embbb-2-x-y})-(\ref%
{embbb-2-xx-y}), it is worth considering also the following branching:%
\begin{eqnarray}
SO(8)\times SO(4)\times SU(2)^{\prime } &\sim &SO(8)\times SU(2)\times
SU(2)_{H}\times SU(2)^{\prime }  \notag \\
&\supset &SO(5)\times SU\left( 2\right) _{SO(8)}\times SU(2)\times
SU(2)_{H}\times SU(2)^{\prime }; \\
\left( \mathbf{8}_{v},\mathbf{2},\mathbf{2,1}\right) &=&\left( \mathbf{5},%
\mathbf{1},\mathbf{2},\mathbf{2,1}\right) +\left( \mathbf{1},\mathbf{3},%
\mathbf{2},\mathbf{2,1}\right) .  \label{xy-y}
\end{eqnarray}

\begin{enumerate}
\item As for the class $\mathfrak{J}_{3}^{2,n}$ (\ref{N=2}) treated in
Subsec. \ref{Semi-Simple-N=2}, and differently from the case of \textit{%
simple} rank-$3$ Euclidean Jordan algebras, in order to identify the
massless $D=5$ \textit{spin group} $SU\left( 2\right) _{J}$ a \textit{%
two-step} procedure must be performed: \textbf{1.1]} one introduces the
diagonal $SU(2)_{I}$ into $SU\left( 2\right) \times SU\left( 2\right) _{H}$,
as given by (\ref{diag-I}), such that (\ref{xy-y}) can be completed to the
following chain:%
\begin{eqnarray}
SO(8)\times SU(2)\times SU(2)_{H}\times SU(2)^{\prime } &\supset
&SO(5)\times SU\left( 2\right) _{SO(8)}\times SU(2)\times SU(2)_{H}\times
SU(2)^{\prime }  \notag \\
&\supset &SO(5)\times SU\left( 2\right) _{SO(8)}\times SU(2)_{I}\times
SU(2)^{\prime };  \label{xyz-y} \\
&&  \notag \\
\left( \mathbf{8}_{v},\mathbf{2},\mathbf{2,1}\right) &=&\left( \mathbf{5},%
\mathbf{1},\mathbf{2},\mathbf{2,1}\right) +\left( \mathbf{1},\mathbf{3},%
\mathbf{2},\mathbf{2,1}\right)  \notag \\
&=&\left( \mathbf{5},\mathbf{1},\mathbf{3,1}\right) +\left( \mathbf{5},%
\mathbf{1},\mathbf{1,1}\right) +\left( \mathbf{1},\mathbf{3},\mathbf{3,1}%
\right) +\left( \mathbf{1},\mathbf{3},\mathbf{1,1}\right) .  \notag \\
&&  \label{xyz-2-y}
\end{eqnarray}%
\textbf{1.2]} Then, the massless $D=5$ \textit{spin group} $SU\left(
2\right) _{J}$ can be identified with the diagonal $SU(2)_{II}$ into $%
SU\left( 2\right) _{SO(8)}\times SU\left( 2\right) _{I}$:%
\begin{equation}
SU(2)_{J}\equiv SU\left( 2\right) _{II}\subset _{d}SU\left( 2\right)
_{SO(8)}\times SU\left( 2\right) _{I},  \label{diag-II-y}
\end{equation}%
such that the chain (\ref{xyz-y})-(\ref{xyz-2-y}) can be further completed
as follows:%
\begin{eqnarray}
SO(8)\times SU(2)\times SU(2)_{H}\times SU(2)^{\prime } &\supset
&SO(5)\times SU\left( 2\right) _{SO(8)}\times SU(2)\times SU(2)_{H}\times
SU(2)^{\prime }  \notag \\
&\supset &SO(5)\times SU\left( 2\right) _{SO(8)}\times SU(2)_{I}\times
SU(2)^{\prime }  \notag \\
&\supset &SO(5)\times SU\left( 2\right) _{J}\times SU(2)^{\prime };
\label{xyzt-y} \\
&&  \notag \\
\left( \mathbf{8}_{v},\mathbf{2},\mathbf{2,1}\right) &=&\left( \mathbf{5},%
\mathbf{1},\mathbf{2},\mathbf{2,1}\right) +\left( \mathbf{1},\mathbf{3},%
\mathbf{2},\mathbf{2,1}\right)  \notag \\
&=&\left( \mathbf{5},\mathbf{1},\mathbf{3,1}\right) +\left( \mathbf{5},%
\mathbf{1},\mathbf{1,1}\right) +\left( \mathbf{1},\mathbf{3},\mathbf{3,1}%
\right) +\left( \mathbf{1},\mathbf{3},\mathbf{1,1}\right)  \notag \\
&=&\left( \mathbf{5},\mathbf{3,1}\right) +\left( \mathbf{5},\mathbf{1,1}%
\right)  \notag \\
&&+\left( \mathbf{1},\mathbf{5,1}\right) +\left( \mathbf{1},\mathbf{3,1}%
\right) +\left( \mathbf{1},\mathbf{1,1}\right) +\left( \mathbf{1},\mathbf{3,1%
}\right) .  \label{xyzt-2-y}
\end{eqnarray}%
The decomposition (\ref{xyzt-2-y}) corresponds to the massless bosonic
spectrum of $\mathcal{N}=2$, $D=5$ $\mathfrak{J}_{3}^{2,6}$-related
supergravity : consistent with the fact that this theory is the
\textquotedblleft bosonic twin" of the $\mathcal{N}=4$, $D=5$ $\mathfrak{J}%
_{3}^{6,2}$-related supergravity, they share the very same bosonic spectrum (%
$32$ states): $1$ graviton and $1$ graviphoton (belonging to the unique $%
\mathcal{N}=4$ vector multiplet) from the $\mathcal{N}=2$ gravity multiplet,
$1$ dilaton (corresponding to the scalar from the $\mathcal{N}=4$ gravity
multiplet) and $1$ dilatonic vector (which in the $\mathcal{N}=4$
interpretation corresponds to the $2$-form in the gravity multiplet) from
the $\mathcal{N}=2$ dilatonic vector multiplet, and $5$ vectors
(corresponding to the $5$ $\mathcal{N}=4$ graviphotons) and $5$ real scalars
(belonging to the $\mathcal{N}=4$ vector multiplet) from the $5$
non-dilatonic $\mathcal{N}=2$ vector multiplets. Such states fit into%
\begin{eqnarray}
\mathcal{N} &=&4~(16\text{ susys}):\left( \mathbf{8}_{v},\mathbf{2,2}\right)
~\text{of~}SO(8)\times SU(2)\times SU(2)_{(H)}; \\
\mathcal{N} &=&2~(8~\text{susys}):\left( \mathbf{8}_{v},\mathbf{2,2,1}%
\right) ~\text{of~}SO(8)\times SU(2)\times SU(2)_{H}\times SU(2)^{\prime }.
\end{eqnarray}%
On the other hand, the two theories consistently have different fermionic
sectors; thus, the massless fermionic spectrum of $\mathcal{N}=2$, $D=5$ $%
\mathfrak{J}_{3}^{2,6}$-related supergravity is not given by $\left( \mathbf{%
8}_{s},\mathbf{2,2}\right) $ (or $\left( \mathbf{8}_{c},\mathbf{2,2}\right) $%
), but rather by $\left( \mathbf{8}_{v},\mathbf{2,1,2}\right) $, of $%
SO(8)\times SU(2)\times SU(2)_{H}\times SU(2)^{\prime }$. This can be
realized by observing that, under(\ref{xyzt-2-y}), such an irrep. decomposes
as:%
\begin{eqnarray}
\left( \mathbf{8}_{v},\mathbf{2,1,2}\right) &=&\left( \mathbf{5},\mathbf{1},%
\mathbf{2},\mathbf{1,2}\right) +\left( \mathbf{1},\mathbf{3},\mathbf{2},%
\mathbf{1,2}\right) =\left( \mathbf{5},\mathbf{1},\mathbf{2,2}\right)
+\left( \mathbf{1},\mathbf{3},\mathbf{2,2}\right)  \notag \\
&=&\left( \mathbf{5},\mathbf{2,2}\right) +\left( \mathbf{1},\mathbf{4,2}%
\right) +\left( \mathbf{1},\mathbf{2,2}\right) ,
\end{eqnarray}%
thus corresponding to $5$ $SU(2)^{\prime }$-doublets of non-dilatonic
gauginos (from the $5$ non-dilatonic vector multiplets), $1$ $SU(2)^{\prime
} $-doublet of gravitinos, and $1$ $SU(2)^{\prime }$-doublet of dilatonic
gauginos. Thus, at the level of massless spectrum, in the minimal case the
action of supersymmetry amounts to the following exchange of irreps.%
\footnote{%
Consistently with the branching properties of $SO(8)$ mentioned in Footnote
15, the irrep. $\left( \mathbf{8}_{s},\mathbf{2,2,1}\right) $ (or $\left(
\mathbf{8}_{c},\mathbf{2,2,1}\right) $) of $SO(8)\times SU(2)\times
SU(2)_{H}\times SU(2)^{\prime }$ does \textit{not} occur as (massless)
bosonic or fermionic representation pertaining to the $D=5$ theory.}:%
\begin{equation}
SO(8)\times SU(2)\times SU(2)_{H}\times SU(2)^{\prime }:\underset{B}{\left(
\mathbf{8}_{v},\mathbf{2,2,1}\right) }~\longleftrightarrow ~\underset{B}{%
\left( \mathbf{8}_{v},\mathbf{2,1,2}\right) },
\end{equation}%
to be contrasted with its analogue (\ref{c}), holding in presence of $16$
local supersymmetries. Note that, consistently, in the minimal
interpretation bosons are $\mathcal{R}$-symmetry $SU\left( 2\right) ^{\prime
}$-singlets, whereas fermions fit into $SU\left( 2\right) ^{\prime }$%
-doublets.

\item $SU\left( 2\right) _{J}$, which commutes with $SO(5)\times
SU(2)^{\prime }$ inside $SO(8)\times SU(2)\times SU(2)_{H}\times
SU(2)^{\prime }$ (\textit{cfr.} (\ref{xyzt-y})), is the Kostant
\textquotedblleft principal" $SU(2)$ (\ref{principal-1}) maximally embedded
into the $D=5$ Ehlers group $SL\left( 3,\mathbb{R}\right) $:%
\begin{equation}
SL\left( 3,\mathbb{R}\right) \cap \left[ SU\left( 2\right) _{SO(8)}\times
SU(2)\times SU(2)_{H}\right] =SU\left( 2\right) _{J}.
\end{equation}

\item As a consequence of the chain of maximal symmetric embeddings (\ref%
{one-two}) and (\ref{xyzt-y}), the non-maximal, non-symmetric manifold
embedding (\ref{zz-N=4-x}) holds, with a different interpretation in terms
of $8$ supersymmetries.

\item As resulting from the above treatment, and analogously to the
\textquotedblleft $8$ susys versus $24$ susys" interpretation of $\mathfrak{J%
}_{3}^{\mathbb{H}}$, the main difference between $\mathfrak{J}_{3}^{6,2}$ ($%
16$ supersymmetries) and $\mathfrak{J}_{3}^{2,6}$ ($8$ supersymmetries)
resides in the $D$-independent hypersector. In the former case, pertaining
to half-maximal supergravity, such a sector is forbidden by supersymmetry.
In the latter case, pertaining to minimal supergravity, such a sector must
be present for physical consistency; as mentioned above, the hypersector is
insensitive to dimensional reductions, and it is thus independent on the
number $D=3,4,5,6$ of space-time dimensions in which the theory with $8$
supersymmetries is defined. The very same comments made at point 4 of
Subsec. \ref{J3-O-Subsec} also hold in this case, with (\ref{c-map})
replaced by%
\begin{equation}
\underset{D=4}{SL(2,\mathbb{R})\times \frac{SO\left( 6,2\right) }{%
SO(6)\times SO(2)}}\overset{c}{\longrightarrow }\underset{D=3}{\frac{SO(8,4)%
}{SO(8)\times SU(2)\times SU(2)_{H}}},
\end{equation}%
where the coset on the l.h.s. is the scalar manifold of the $\left(
\mathfrak{J}_{3}^{6,2}\sim \mathfrak{J}_{3}^{2,6}\right) $-related $\mathcal{%
N}=4$ (or $\mathcal{N}=2$), $D=4$ supergravity theory; it is symmetric, as
is its image $\frac{SO(8,4)}{SO(8)\times SU(2)\times SU(2)_{H}}$ through $c$%
-map \cite{CFG}.

\item The very same formul\ae\ (\ref{pre-pre-qq-3})-(\ref{qq-3}) also hold
in this case, but with the different interpretation (pertaining to $8$ local
supersymmetries) considered in this Subsection.
\end{enumerate}

\section{\label{Conclusion}Conclusion}

In the present investigation, we have spelled out the relation which exist
between the Ehlers group in five dimensions and the rank-$3$, Euclidean
(simple and semi-simple) Jordan algebra interpretation of supergravity
theories, whose $U$-duality symmetry is given by the reduced structure group
of the cubic norm the underlying Jordan algebra.

The massless spin (helicity) is enhanced to the Ehlers symmetry, and gets
further enlarged to the so-called \textit{super-Ehlers} symmetry \cite%
{super-Ehlers-1} by the inclusion of the $U$-duality, which consistently
encode the supermultiplet structure of the corresponding supergravity theory.

It is interesting to note that the general $D=5$ \textit{Jordan pair}
non-symmetric embedding (\ref{pre-2}) is maximal for theories based on
\textit{simple} Jordan algebras, such as $N=16$ and $12$ \textquotedblleft
pure" theories, as well as $N=4$ magical Maxwell-Einstein supergravities,
whereas it is non-maximal for the $N=8$ and $N=4$ matter coupled theories
based on \textit{semi-simple} Jordan algebras. However, (\ref{pre-2}) always
preserves the group rank, which is not related to supersymmetry but rather
to the underlying Jordan algebra.

\section*{Acknowledgements}

We would like to thank Piero Truini and Shannon McCurdy for useful
correspondence and discussions.

A.M. would like to thank the Department of Physics, University of California
at Berkeley, where this project was completed, for kind hospitality and
stimulating environment.

The work of S.F. has been supported by the ERC Advanced Grant no. 226455,
Supersymmetry, Quantum Gravity and Gauge Fields (SUPERFIELDS).

The work of B. Z. ~has been supported in part by the Director, Office of
Science, Office of High Energy and Nuclear Physics, Division of High Energy
Physics of the U.S. Department of Energy under Contract No.
DE-AC02-05CH11231, and in part by NSF grant 30964-13067-44PHHXM.

\end{document}